\documentclass[dvips]{acta}
\usepackage{supertabular,lscape,epsfig}
\usepackage{amssymb}
\usepackage{amsmath}

\long\def\symbolfootnote[#1]#2{\begingroup%
\def\thefootnote{\fnsymbol{footnote}}\footnote[#1]{#2}\endgroup}

\SetPages{0}{0}

\SetVol{61}{2011}

\begin{document}

\begin{Titlepage}
\Title{A study of contact binaries with large temperature differencies between components\symbolfootnote[2]{Based on observations obtained at:\newline
- the {\it Fort Ska{\l}a} Astronomical Observatory of the Jagiellonian University;\\
- the Mount Suhora Astronomical Observatory, Cracow Pedagogical University;\\
- the David Dunlap Observatory, University of Toronto;\\ 
- the South African Astronomical Observatory, with the 1.9, 0.75, and 0.5~m telescopes;\\ 
- the ESO-La Silla Observatory with the 2.2~m telescope, under programme ID 077.D-0789;\\ 
- the Las Campanas Observatory, Carnegie Institute of Washington with the 2.5~m Du Pont telescope.}}

\Author{Micha{\l} Siwak$^{1,2}$, Stanis{\l}aw Zola$^{1,3}$, Dorota Koziel-Wierzbowska$^1$}
{
$^1$Astronomical Observatory of the Jagiellonian University, ul. Orla 171, 30-244 Krak{\'o}w,\\
e-mail: siwak@oa.uj.edu.pl,\\
$^2$Department of Astronomy and Astrophysics, University of Toronto, 50 St.George Street, M5S 3H4 Toronto, Ontario,\\ 
e-mail: siwak@astro.utoronto.ca,\\
$^3$Mount Suhora Astronomical Observatory, Cracow Pedagogical University,\\
    ul. Pochor\c{a}{\.z}ych 2, 30-084 Krak{\'o}w}

\Received{25 11, 2010}

\end{Titlepage}

\Abstract{We present an extensive
analysis of new light and radial-velocity 
(RV) curves, as well as
high-quality broadening-function (BF)
profiles of twelve binary systems
for which a contact configuration
with large temperature differencies
between components has been reported
in the literature.  We find that six
systems (V1010~Oph, WZ~Cyg, VV~Cet,
DO~Cas, FS~Lup, V747~Cen) have
near-contact configurations.  For the
remaining systems (CX~Vir, FT~Lup,
BV~Eri, FO~Hya, CN~And, BX~And), our
solutions of the new observations 
once again
converge in a contact configuration
with large temperature differencies
between the components.  However,
the bright regions discovered in the
BFs for V747~Cen,
CX~Vir, FT~Lup, BV~Eri, FO~Hya, and
CN~And, and further attributed to hot
spots, shed new light on the physical
processes taking place between the
components and imply 
the possibility that the contact
configurations obtained from light-
and RV-curve modelling
are a spurious result.} {eclipsing
binaries, contact binaries, physical
parameters: individual: V1010~Oph,
WZ~Cyg, VV~Cet, DO~Cas, FS~Lup,
V747~Cen, CX~Vir, FT~Lup, BV~Eri,
FO~Hya, CN~And, BX~And}

\section{Introduction}
\label{intr}

Theoretical investigation of W~UMa-type
binary stars began with the work of
Kuiper (1941), who found that contact
binary stars with components belonging
to the Main Sequence (MS) cannot exist
in an equilibrium state (a
result known as the {\it Kuiper
paradox}).  Lucy (1968) has shown that
Kuiper's conclusion is valid only for contact
binary stars with radiative common
envelopes and unequal-mass components:
contact binaries consisting
of MS stars with common convective
envelopes can exist in hydrodynamical
and thermal equilibrium states if
CNO and p-p cycles occur
in the cores of both components.
According to Lucy's common-envelope
model, both stars in a contact binary
system should have equal effective
temperatures.  Although this result
resolved the Kuiper paradox,
Lucy's theoretical models of contact
binary stars with components differing
in their structure because of 
CNO and p-p cycle reactions             
covered only the blue end 
of the period-color diagram
(Eggen, 1967).

The small
masses of primary components obtained
from combined radial-velo- city 
(RV) and
light-curve modelling (Lucy, 1973) 
prompted Lucy (1976) and Flannery
(1976) to consider a situation with
both solar-type components harbouring 
p-p reactions.
It turned out that in such a
situation it is not possible to attain 
a thermally stable equilibrium. 
Instead, in a thermal time
scale of 10~milion~years, systems
will oscillate between contact
and semi-detached configurations.
Robertson \& Eggleton (1977) performed
more realistic computations, taking
into account an angular-momentum
loss during a system evolution.  As a
result, they obtained models of 
contact-binary systems with physical parameters
of components closely resembling those
obtained from light- and RV-curve
modelling.

According to the leading work on the
Thermal Relaxation Oscillation (TRO)
theory (Lucy, 1976; Flannery 1976),
the maximum temperature difference
($\Delta T$) between components of
solar-type contact-binary system
should not exceed 880--1000~K.
Similar values have been recently
obtained by Li et al.~(2004a, 2004b,
2005) and K{\"a}hler~(2004).  The 
marginal-contact configurations and
temperature differences between the
components in the 300--700~K range
obtained for RW~PsA, AK~Her, and
W~Crv by Lucy \& Wilson (1979) have
confirmed the theoretical findings.

Ka{\l}u{\.z}ny~(1983, 1985,
1986), Hilditch et al.~(1984), and
Hilditch~\&~King~(1988) have found
other contact systems exhibiting large
$\Delta T$ between their components. 
For some of these systems, light-curve
modelling resulted in temperature
differences that were considerably
larger than the theoretical limit
of about 1000~K.  Rucinski (1986)
and Eggleton (1996) proposed that
indications of 
contact binaries with such large
temperature differencies are in fact
artefacts caused by the application of
simple Roche lobe-based light curve
synthetic models, with disregard for
intensive accretion processes expected
to take place between components
during short and violent phases when
a contact configuration is broken or
re-established.

In Table~1, we present a sample of
close binary stars for which at least
one light-curve solution was previously
obtained using a light-curve synthesis
program based on the Roche model and
yielded a contact configuration with
a temperature difference between
the components larger than 1000~K.
However, as indicated in column~7 
of Table~1, almost all the
models are based on photometric
mass ratios, a limitation
which significantly
decreases the reliability of the
obtained configurations.  Closer
inspection reveals that in many cases
an O'Connel effect (i.e.,
occurrence of unequal maxima, 
most probably due to a combination of
effects introduced by spots on one or
both components) has not been taken
into account during the light-curve
modelling. Additionally, closer inspection 
reveals the 
applied albedo
and gravity-darkening coefficients
to have been far from their theoretical
values.  Moreover, the scatter visible in
the light curves was often too large
to permit a unique solution.  We refer
an interested reader to references
given in the last column of Table~1
for more historical information about
the targets investigated in this paper.

All the above has led us to the
conclusion that in order to get unique
configurations for these systems, one
must obtain high quality light and
RV curves (Section~2) and
must perform advanced modelling on this
data (Section~3).  Results obtained
from the modelling of individual
systems are described in Sections~4
and 5. 
Physical parameters of these systems 
are presented in Section~6.  
In Sections~4 and 5 we also present
a discovery of bright regions (hot
spots) on secondary components, made
possible by the outstanding
quality of the broadening functions 
(BFs) obtained for some systems. 
In Section~7, we discuss the implications 
of this discovery 
for results obtained classically (i.e., 
obtained from light- and RV-curve fits 
within the Roche-lobe model).

\MakeTable{lccccclr}{12.5cm}{The sample of contact binary systems with large 
temperature differencies between the components.}
%
%
%
%
{\hline
System &  $T_\mathrm{eff}^\mathrm{prim}$[K] & $T_\mathrm{eff}^\mathrm{sec}$[K]& $\Delta T$[K] & f(\%)&conf.&~~~q &  First author, year\\ \hline
GW Tau   &  8750  & 5620 & 3130 &  11  &C   &0.309 &Zhu+, 2006    \\
V747 Cen &  8500  & 4700 & 3800 &  4.3 &m-c &0.399 &Leung, 1974          \\
         &  8260  & 4640 &      &      &n-c &0.319 &Barone+, 1991  \\
DO Cas   &  7700  & 3850 & 3850 &  13  &C   &0.310 &Ka{\l}u{\.z}ny, 1985        \\
         &  7700  & 4000 &      &      &SD1 &0.325 &Ka{\l}u{\.z}ny, 1985        \\
         &  8500  & 4090 &      &      &SD1 &0.415 &Karimie+, 1985 \\
         &  9070  & 4830 & 4240 &  13  &C   &0.313 &Oh+, 1992      \\
         &  8300  & 4690 &      &      &SD1 &0.326 &Barone+, 1992  \\
V1010 Oph&  8200  & 5670 & 2530 &  18  &C   &0.489 &Leung+, 1977\\
         &  7500  & 5200 &      &      &SD1 &0.446 &Corcoran+., 1991\\
         &  8200  & 5700 &      &      &SD1 &0.448 &Niarchos+, 2003\\
VV Cet   &  8100  & 5900 & 2200 &  16  &C   &0.249 &Rahman, 2000         \\
BL And   &  7500  & 5370 & 2130 &  49  &C   &0.311 &Ka{\l}u{\.z}ny, 1985        \\
         &  7500  & 4830 &      &      &SD1 &0.377 &Zhu+, 2006    \\ \hline
WZ Cyg   &  7050  & 5750 & 1300 &  ?   &C   &0.540 &Rovithis+, 1999\\
FO Hya   &  6950  & 4260 & 2690 &  7   &m-c &0.552 &Candy+, 1997 \\
BV Eri   &  6850  & 5650 & 1200 &  20  &C   &0.252 &Gu, 1999        \\
FT Lup   &  7400  & 5100 & 2300 &  ?   &C   &0.400 &Mauder+, 1982 \\
         &  6700  & 4560 & 2140 &  ?   &C   &0.430$^{*}$&Hilditch+, 1984 \\
         &  6700  & 4800 & 1900 &  12  &C   &0.430$^{*}$&Ka{\l}u{\.z}ny, 1986      \\
         &  6700  & 3920 &      &      &SD1 &0.465 &Lipari+, 1986 \\
BX And   &  6500  & 4680 & 1820 &  6   &m-c &0.623 &Samec+, 1989      \\
         &  6800  & 4500 & 2300 &  8   &m-c &0.497$^{*}$&Bell+, 1990  \\
CX Vir   &  6500  & 4500 & 2000 &  0   &m-c &0.340$^{*}$&Hilditch+, 1988 \\
CN And   &  6500  & 5500 & 1000 &  ?   &m-c &~~~?  &Ka{\l}u{\.z}ny, 1983  \\
         &  6200  & 4680 & 1520 &  43  &C   &0.579 &Rafert, 1985   \\
         &  6500  & 5920 &      &      &SD1 &0.389$^{*}$&Van Hamme+, 2001\\
         &  6450  & 5375 &      &      &SD2 &0.371$^{*}$&Zola+, 2005     \\
         &  6160  & 4660 & 1500 &  17  &C   &0.390$^{*}$&Jassur+, 2006   \\
FS Lup   &  6000  & 4700 & 1300 &  33  &C   &0.510 &Milano+, 1987  \\  \hline
\multicolumn{8}{p{12cm}}{The 
line in the middle separates systems
with radiative envelopes (above) 
from systems with convective
envelopes (below).  The column
denoted ``conf.'' gives the configuration
derived for the system in question: 
C for 
contact; m-c for marginal contact; 
SD1 and SD2 for semi-detached systems, 
where it is, respectively, the 1st
or the 2nd component that fills its Roche
lobe; and n-c for 
near-contact.  The quantity $f$
is the {\it fill-out} parameter,
given only for contact
and marginal-contact
systems in cases in which we had
the necessary data to make a computation  
through the equation 
$f=(\Omega-\Omega_{L_1})/(\Omega_{L_1}-\Omega_{L_2})$ 
(where $\Omega_{L_1}$ and $\Omega_{L_2}$
are the Roche potentials at the $L_1$
and $L_2$ points, respectively, 
$\Omega$ being a common surface
potential obtained from the light-curve 
modelling).  We also show the
mass ratios $q$ obtained
from the light-curve solution, adding
an asterisk to mark cases in which
a spectroscopic mass ratio was used.} 
}

\section{Observations and data reduction}

\subsection{Spectroscopy}

Medium-resolution spectra of FS~Lup
were obtained at the South African
Astronomical Observatory (SAAO) in the
second half of April 2006,
with the 1.9~m
Radcliffe telescope equipped with
a slit spectrograph and the grating
``no.4'' (1200 l/mm).
With the 1.2~arcsec-wide slit projected
on 2.2 pixels of a CCD camera, this
results in effective resolving power
$R=8\,000$.  The star was observed in
two spectral regions: during the first
night the spectral window was centered
at 4400\AA, and during all other
nights at 5250\AA.

FT~Lup, CX~Vir, and V747~Cen were observed
in service mode at ESO-La
Silla during six nights of
April, May, and June 2006.  In this
facility the 2.2~m
MPG telescope equipped with the 
FEROS spectrograph provided high
resolution ($R=48\,000$) and very good
signal-to-noise spectra.

The targets reachable from the northern hemisphere
(WZ~Cyg, VV~Cet, BX~And, DO~Cas, and BV~Eri) 
were observed at the David 
Dunlap Observatory (DDO), a suburban facility
then owned by 
the University of Toronto,
during November and December of
2006 and February of 2007, 
with the DDO 1.9~m telescope and a slit spectrograph. 
Most of the data were obtained with the 2160~l/mm grating, but 
because of 
the relatively faint magnitude of VV~Cet and WZ~Cyg, during nights 
of poor seeing (a few arcsec), 
the observations were made with a lower-resolution 
(1200~l/mm) grating. 
With the 1.5-arcsec slit projected on 2.5 pixels, 
the effective resolving 
power was $R=19\,000$ and $R=11\,000$ 
for the 2160~l/mm and 1200~l/mm gratings, 
respectively.        

FO~Hya was observed at the Las Campanas
Observatory (LCO) in February 2009
with the 2.5~m Du Pont telescope, 
equipped with the classical slit
echelle spectrograph.  With the 1-arcsec slit,
this instrument assured spectral
resolution of $R=40\,000$.

The data gathered at the ESO-La~Silla
observatory were reduced in the
usual way with FEROS-DRS 
under ESO-MIDAS.
The data obtained at SAAO, DDO, and LCO
were reduced with IRAF. 

For RV measurements, we
applied the BF
method (Rucinski, 1992, 1999, 2002, 
and
Pribulla et al., 2009; 
the 2009 publication has the fullest
list of references). 
%
%
%
%
BFs were calculated on the basis
of RV standards from the list of 
Stefanik et al.~(1999), observed in all
the observatories on the same nights as
the main targets, and 
on the basis of synthetic spectra
from the SPECTRUM software
(Gray, 2001), using
Kurucz's (1993) atmosphere models for
MS stars with solar abundances.

RVs of light
centroids were measured by means of
synthetic rotational profiles for
spherical stars.  This choice assured
reasonable model-independence, while
still approximating very 
accurately the observed
profiles of elliptically disturbed
stars.  As primary and
secondary components of our targets
have very different spectral types,
the RVs of primary components were obtained
from BF profiles calculated using
the early-type RV standards, while RVs of
secondary components were obtained from
BF profiles calculated using  
late-type RV standards (Table~2).  We 
found that in most cases this approach
minimized scatter of RV 
curves and yielded almost identical
(within $1-2~km/s$) mean system
velocities $v_{\gamma}$ for the primary
and secondary components.  The BF
method gave larger semi-amplitudes
of the RV variations ($K_i$) than did
the cross-correlation function (CCF)
method previously applied for
RV measurements of CX~Vir (Hilditch \&
King, 1988), FT~Lup (Hilditch et al.,
1984) and BX~And (Bell et al., 1990),
thereby fully confirming the predictions of 
Rucinski (1999) that there would be
a tendency for 
systematic reduction of $K_i$ under the
CCF approach.

Table~3 contains preliminary orbital
parameters obtained from the new
RV curves by fitting the
sine-curve to the data in the phase
ranges 0.15--0.35 and 0.65--0.85.

\MakeTable{lcc}{12.5cm}{Spectral regions and RV-standard stars 
used for the RV determination.}
{\hline\hline
system   & used spectral range [\AA]       & RV standards \\ \hline
FS Lup   & 3920-4750 \& 4880-5640          & HD 102870 (F8V) \& HD 92588 (K1IV)\\
FT Lup   & 4378-5743 (H$_{\beta}$ excluded)& F2V \& K3V                        \\
CX Vir   & 4378-5743 (H$_{\beta}$ excluded)& F5V \& K0V + K3V                  \\
V747 Cen & 4905-5743                       & A5V \& K1V                        \\
FO Hya   & 4910-5560                       & F0V \& K0V                        \\  
V1010 Oph& 5074-5305                       & HD 128167 (F2V) \& HD 144571(G8V) \\
BV Eri   & 5074-5305                       & HD 128167 (F2V) \& HD 65588 (G8V) \\
BX And   & 5074-5305                       & HD 128167 (F2V) \& HD 3765 (K2V)  \\
DO Cas   & 5074-5305                       & A5V \& K1V                        \\
VV Cet   & 5074-5305                       & HD 222738 (F7V)                   \\
WZ Cyg   & 5074-5305           & HD 222738 (F7V) \& HD 3765 (K2V) \\ \hline\hline
}


\MakeTable{lrrrc}{12.5cm}{Preliminary orbital parameters obtained by  
sine-curve fitting to spectroscopic observations in the phase ranges: 
0.15-0.35 and 0.65-0.85.}
{\hline\hline
system   &$K_1~[km/s]$~~~~~~~&$K_2~[km/s]$~~~~~~~&$v_{\gamma}~[km/s]$&      q           \\ \hline
FO Hya   &  62.48 $\pm$ 0.97 & 253.20 $\pm$ 4.50 &  55.80 $\pm$ 0.90 & 0.247 $\pm$ 0.009\\
FS Lup   & 112.10 $\pm$ 0.87 & 236.36 $\pm$ 3.05 &  12.74 $\pm$ 0.84 & 0.474 $\pm$ 0.010\\
FT Lup   & 112.21 $\pm$ 0.37 & 239.22 $\pm$ 3.28 &  -2.22 $\pm$ 0.35 & 0.469 $\pm$ 0.008\\
CX Vir   &  75.64 $\pm$ 0.45 & 218.19 $\pm$ 1.63 & -11.47 $\pm$ 0.42 & 0.347 $\pm$ 0.005\\
V747 Cen &  78.30 $\pm$ 0.52 & 231.39 $\pm$ 3.42 & -18.57 $\pm$ 0.49 & 0.338 $\pm$ 0.007\\
V1010 Oph& 107.32 $\pm$ 0.46 & 226.32 $\pm$ 0.85 & -19.09 $\pm$ 0.44 & 0.474 $\pm$ 0.004\\
BV Eri   &  67.41 $\pm$ 0.48 & 250.81 $\pm$ 0.87 & -39.89 $\pm$ 0.48 & 0.269 $\pm$ 0.003\\
BX And   & 106.35 $\pm$ 0.61 & 233.58 $\pm$ 1.77 & -24.69 $\pm$ 0.90 & 0.455 $\pm$ 0.006\\
DO Cas   &  78.85 $\pm$ 0.75 & 259.01 $\pm$ 2.78 &  26.22 $\pm$ 0.25 & 0.304 $\pm$ 0.006\\
VV Cet   &  62.49 $\pm$ 1.51 & 220.23 $\pm$ 6.31 &  -0.01 $\pm$ 1.42 & 0.284 $\pm$ 0.015\\
WZ Cyg   & 130.91 $\pm$ 1.05 & 205.11 $\pm$ 3.54 &   0.98 $\pm$ 1.52 & 0.638 $\pm$ 0.016\\ 
\hline\hline
}

\subsection{Photometry}

New high-quality photometric data for
targets reachable
from the northern hemisphere were  
taken during the three consecutive
northern autumn-winter seasons 2004/2005,
2005/2006, and 2006/2007.  Most of the data
were obtained with the Carl-Zeiss
50~cm telescope at the Fort
Ska{\l}a Astronomical Observatory
of the Jagiellonian University.
This telescope was equipped with the
Photometrics S300 CCD camera with the
SITe SI003B, 1024x1024 pixels chip,
carrying a set of broadband BVRI (Bessell)
filters.  With a focal length of 6.7~m,
the field of view was 12~arcmin~x~12~arcmin.  
In this observatory, we took complete light
curves for WZ~Cyg, BX~And, GW~Tau, and
BL~And.

DO~Cas, VV~Cet, and CN~And
were observed at the Mount Suhora
Observatory of the Cracow Pedagogical
University using the Carl-Zeiss
60~cm telescope.  The first two stars
were observed with a three-channel
photometer equipped with the Hamamatsu
R2949 photomultipliers and broadband
Johnson BVRI filters.  CN And was
observed using the SBIG ST10XME CCD
camera (containing
the KODAK KAF-3200E/ME chip) and
wideband BVRI (Bessell) filters.

CX~Vir and FT~Lup were observed in
April 2005 at the SAAO with the
50~cm Boller~\&~Chivens telescope and a
single-channel modular photometer 
equipped with the Hamamatsu R943-02
GaAs photomultiplier and BV$(RI)_c$
Johnson-Cousins filters.  In March
2006, at the same observatory, we
observed V747~Cen with the 75~cm
Grubb-Parsons telescope, equipped with
the UCT CCD camera with the
EEV 576x420 pixel chip (binned 2x2)
and Johnson BVRI filters.  The same
telescope, but equipped with the
single-channel UCT Photometer
with the Hamamatsu RCA 31034A GaAs
photomultiplier and Johnson-Cousins
BV$(RI)_c$ filters, was used during
April and May 2006 for observations of
FS~Lup and FO~Hya.

Data gathered taken with CCD cameras
were reduced in the standard way
through ESO-MIDAS scripts. 
Aperture photometry with variable
aperture size was performed through 
Kopacki's programs (private
communication), based on the 
DAOPHOT~II package (Stetson, 1987).
Data obtained with single-channel
photometers were dead-time 
corrected using the procedure given in
the XLUCY manual (Balona, 2000).
Sky-background and comparison star
interpolations were applied with our
own software.  The data taken with the
three-channel photometer at the Mount
Suhora Observatory were reduced using
dedicated computer programs developed
by M. Dr{\'o}{\.z}d{\.z} (private
communication).  Standard corrections
for differential 
extinction and colour extinction
were applied to the new light curves. 
 
Light and RV curves
were phased with the most up-to-date
elements determined by Kreiner
(2005--2008, private communication).
These ephemerides were obtained
using the minima times included
already in the Kreiner (2004)
database, but supplemented with several
new times of minima determined from
our own observations.  The comparison
and check stars, as well as elements
applied for phases calculation, are
listed in Table~4.

As attempts to take RV curves of
GW~Tau and BL~And failed in bad
weather, these targets were excluded
from the current investigation.


\MakeTable{lcccc}{12.5cm}{Comparison and check stars applied for differential 
photometry. The two right panels contain elements used for phases calculation.}
{\hline\hline
system   & comparison star(s) & check star(s) & $M_0~[HJD]$ & $P~[d]$ \\ \hline
BX And  & mean of 3 stars & internal comp.& 2453350.23061& 0.61011240\\
WZ Cyg  & mean of 3 stars & internal comp.& 2454019.26728& 0.58446762\\
VV Cet  & GSC 4674 0662 & no check star   & 2453675.39228& 0.52239410 \\
CN And  & GSC 2787 1803 & GSC 2787 1891   & 2452500.12040& 0.46279081 \\
FO Hya  & GSC 6049 0074 & no check star   & 2453855.35760& 0.46955663 \\
CX Vir  & GSC 6137 0576 & GSC 6138 0644   & 2453470.50082& 0.74608003 \\
FT Lup  & GSC 7828 1478 & GSC 7829 0247   & 2453478.64697& 0.47008002 \\
FS Lup  & GSC 8292 1678 & GSC 8292 0914   & 2452500.08800& 0.38139970 \\
V747 Cen& GSC 8261 1863 & GSC 8261 0719   & 2453811.34417& 0.53719482 \\
\hline\hline
}

\section{Light- and RV-curve modelling: the method}

Light- and RV-curve
modelling was performed through 
the Wilson-Devinney (WD) code (Wilson,
1996), with a Monte Carlo search
method applied to locate a global
minimum in parameter space (Zola
et al., 2010).  An iterative procedure
of light- and RV-curve
modelling, introduced by Baran et~al.~(2004), 
was used during our
computations.  Errors of adjusted
parameters were calculated from the
search matrix at confidence level
of 90\%, as described by 
Kreiner~et~al.~(2003).  The values of a reduced
(weighted) $\chi^2$ were used to
determine the quality of the obtained fits.

The effective-temperature-versus-spectral-type
calibration of Harmanec
(1988) was used to determine the
effective temperatures of the primary
components ($T_\mathrm{eff}^\mathrm{prim}$) 
according
to their spectral types. (Table~5 gives
details.)  This parameter, like 
the spectroscopic
mass ratio $q_\mathrm{spec}$, was not
adjusted during light curve modelling.
The albedo $A$ and gravity-darkening 
coefficients $g$
of both stars were
assumed at the theoretical values,
i.e. $A=1,~g=1$ for stars with
radiative envelopes ($T_\mathrm{eff}>7200~\mathrm{K}$)
(von Ziepel, 1924), and $A=0.5,~g=0.32$
for stars with convective envelopes
($T_\mathrm{eff}<7200~\mathrm{K}$) (Lucy 1967, Rucinski
1969).  The square-root limb darkening
law and the coefficients published by
Dia{\'z}-Cordoves et al. (1995) and
Claret et al. (1995) were assumed as
a function of an effective temperature
of a star and the effective wavelength
of the filter.
As the choice between the detailed
and simple reflection-effect treatment
turned out to have no practical impact
on the obtained results, especially
on the obtained configurations, we
fitted our data under the assumption 
of a simple
reflection model ($mref=1$) with a single
reflection ($nref=1$).

During light curve modelling, we
adjusted the orbital inclination $i$,
the sec- ondary-compontent 
effective temperature
$T_\mathrm{eff}^\mathrm{sec}$,
the Roche potentials $\Omega$ and the
luminosities $L$ of both components,
the phase shift, and the third
light $l_3$.  Additionally, if an
O'Connel effect was visible in the
light curves, we adjusted the spot
parameters: its coordinates ($\phi,
\lambda$), a spot radius $r$, and the
ratio between the spot temperature and
the surrounding surface.  During the
RV-curve modelling, we
adjusted the orbital semi-major axis
$a$, the spectroscopic mass ratio
$q_\mathrm{spec}$, and the mean system velocity
$v_{\gamma}$.

As the result of the iterative
procedure of light- and RV-curve  
 modelling, we obtained two groups
of solutions. The first one comprises 
the evidently near-contact systems (V1010~Oph,
WZ~Cyg, VV~Cet, DO~Cas, FS~Lup, and
V747~Cen). The second comprises
the systems whose light curves were better
fitted by a contact configuration
model, again with large temperature
differences between the components 
(CX~Vir, FT~Lup, BV~Eri, FO~Hya, CN~And, 
and BX~And).


\MakeTable{lccc}{12.5cm}{General information concerning spectral classification 
of our targets.}
{\hline\hline
system  & method          & spectral type & author(s)                  \\ \hline
WZ Cyg  & spectral class. & F4V           & Og{\l}oza \& Rucinski (priv. comm.) \\
DO Cas  & spectral class. & A4V           & Og{\l}oza \& Rucinski (priv. comm.) \\
BV Eri  & spectral class. & F2V           & Og{\l}oza \& Rucinski (priv. comm.) \\
CN And  & spectral class. & F5V           & Rucinski et al. (2000)    \\
BX And  & spectral class. & F2V           & Bell et al. (1990)        \\
CX Vir  & spectral class. & F5V           & Hilditch \& King (1988)   \\
FT Lup  & spectral class. & F2V           & Hilditch et al. (1984)    \\
FO Hya  & B-V             & F0V           & Candy \& Candy (1997), this work     \\
FS Lup  & B-V             & G2V           & this work                 \\
V747 Cen& B-V             & A5V           & Chambliss (1970)          \\
VV Cet  & B-V             & A5V           & Rahman (2000), this work             \\
V1010 Oph & UV continuum  & A7V           & Corcoran et al. (1991)    \\ \hline\hline
}

\MakeTable{lllllll}{12.5cm}{The results obtained from light- and RV-curve modelling 
of {\it near-contact} sytems. 
\newline $^*$ - parameters fixed during light curve modelling (see Sections~3 and 4 for explanations).}
{\hline\hline
parameter          &  V1010 Oph   & WZ Cyg      & VV Cet     & DO Cas     & FS Lup     & V747 Cen    \\ \hline
$i[^o]$            &  86.14(37)   & 83.497(32)  & 81.89(13)  & 89.94(30)  & 82.64(23)  & 87.16(18)   \\
$T_\mathrm{eff}^\mathrm{prim}[K]$&  7500$^*$    & 6530$^*$    & 8150$^*$   & 8350$^*$   & 5860$^*$   & 8150$^*$    \\
$T_\mathrm{eff}^\mathrm{sec}[K]$ &  5132(11)    & 4932(2)     & 6252(7)    & 4297(26)   & 5130(5)    & 4275(16)    \\
$\Omega_\mathrm{prim}$    &  2.81773$^*$ & 3.1200(1)   & 2.43060$^*$& 2.4879(15) & 2.8181(5)  & 2.51001$^*$ \\
$\Omega_\mathrm{sec}$     &  2.8352(33)  & 3.1200(1)   & 2.4802(12) & 2.4751(1)  & 2.8443(39) & 2.5342(14)  \\
$q_\mathrm{spec}$         &  0.470(3)$^*$&0.631(36)$^*$&0.284(8)$^*$&0.304(5)$^*$&0.470(7)$^*$& 0.320(5)    \\
$v_{\gamma}[km/s]$ & -19.65(4)    & 1.05(88)    & -1.36(90)  & 26.22(30)  & 11.64(63)  & -18.55(46)  \\
$a[R_{\odot}]$     &  4.487(15)   & 4.039(27)   & 3.006(41)  & 4.587(32)  & 2.746(16)  & 3.522(21)   \\ \hline
$L_{1B}$           & 11.323(12)   & 10.4347(46) & 11.5776(67)& 12.3042(92)& 10.227(15) & 12.5055(64) \\
$L_{1V}$           & 10.960(14)   & 10.1042(44) & 11.2295(70)& 12.1172(99)&  9.978(16) & 12.2276(60) \\
$L_{1R}$           &  ---         & 9.9063(42)  &  ---       &  ---       &  9.805(16) & 11.9661(64) \\
$L_{1I}$           &  ---         & 9.4824(41)  &  ---       &  ---       &  9.586(16) & 11.7091(67) \\
$L_{2B}$           &  0.849(11)   & 1.5717(38)  & 1.0013(59) & 0.1278(64) &  2.404(16) &  0.1346(43) \\
$L_{2V}$           &  1.122(13)   & 1.8488(39)  & 1.2078(63) & 0.2092(88) &  2.589(15) &  0.2156(58) \\
$L_{2R}$           &  ---         & 2.1257(40)  &  ---       &  ---       &  2.717(15) &  0.3034(72) \\
$L_{2I}$           &  ---         & 2.5910(38)  &  ---       &  ---       &  2.918(15) &  0.4866(91) \\ \hline
$\phi[^o]$         &  90.0$^*$    &  ---        & 105.20(87) &  ---       &  93.7(2.0) &  90.0$^*$   \\
$\lambda[^o]$      &  79.5(1.5)   &  ---        & 0.4(1.5)   &  ---       &   1.80(12) &  41.0(1.2)  \\
$r[^o]$            &  19.7(1.4)   &  ---        & 14.744(97) &  ---       &  58.07(54) &  64.7(2.0)  \\
$T_\mathrm{spot}/T_{eff}^{sec}$&1.378(31)&  ---        &  ---       &  ---       &   ---      &  1.2094(40) \\ 
$T_\mathrm{spot}/T_{eff}^{prim}$&  ---   &  ---        & 0.253(79)  &  ---       &  0.9136(10)&   ---       \\ \hline\hline
}


\begin{figure}
\centerline{%
\begin{tabular}{c@{\hspace{1pc}}c}
\includegraphics[width=1.8in,angle=-90]{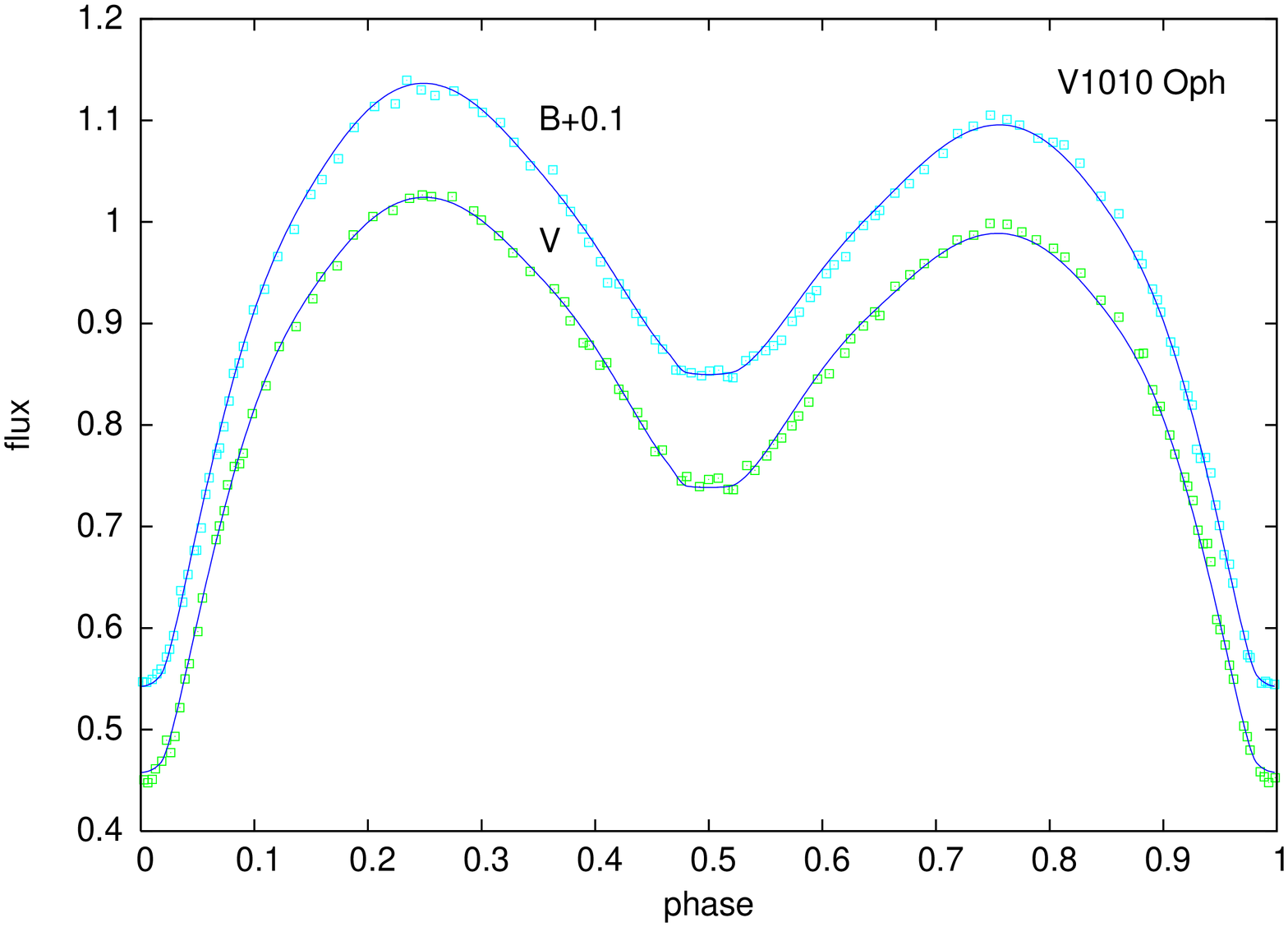}&
\includegraphics[width=1.8in,angle=-90]{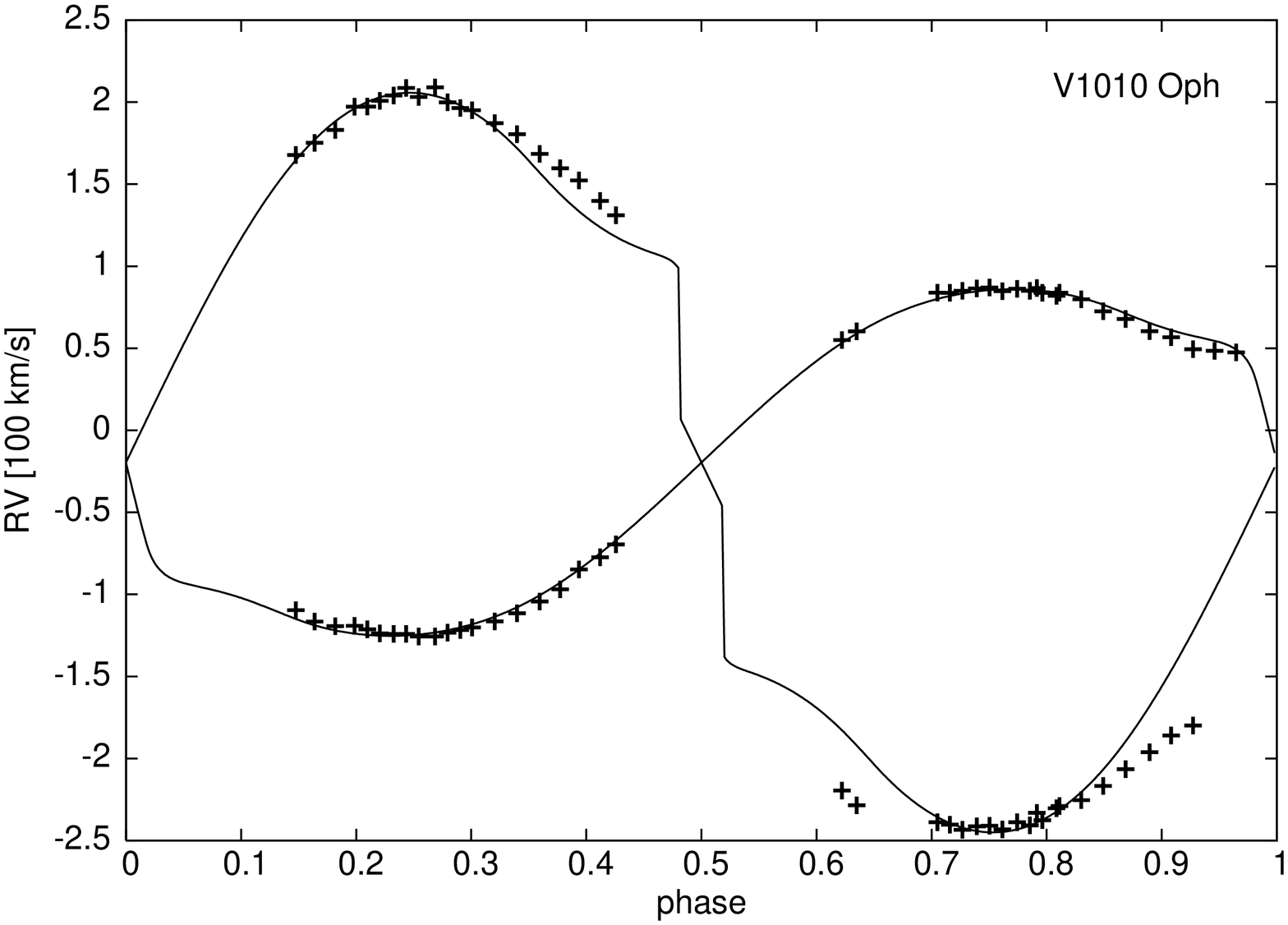} \\
\includegraphics[width=1.8in,angle=-90]{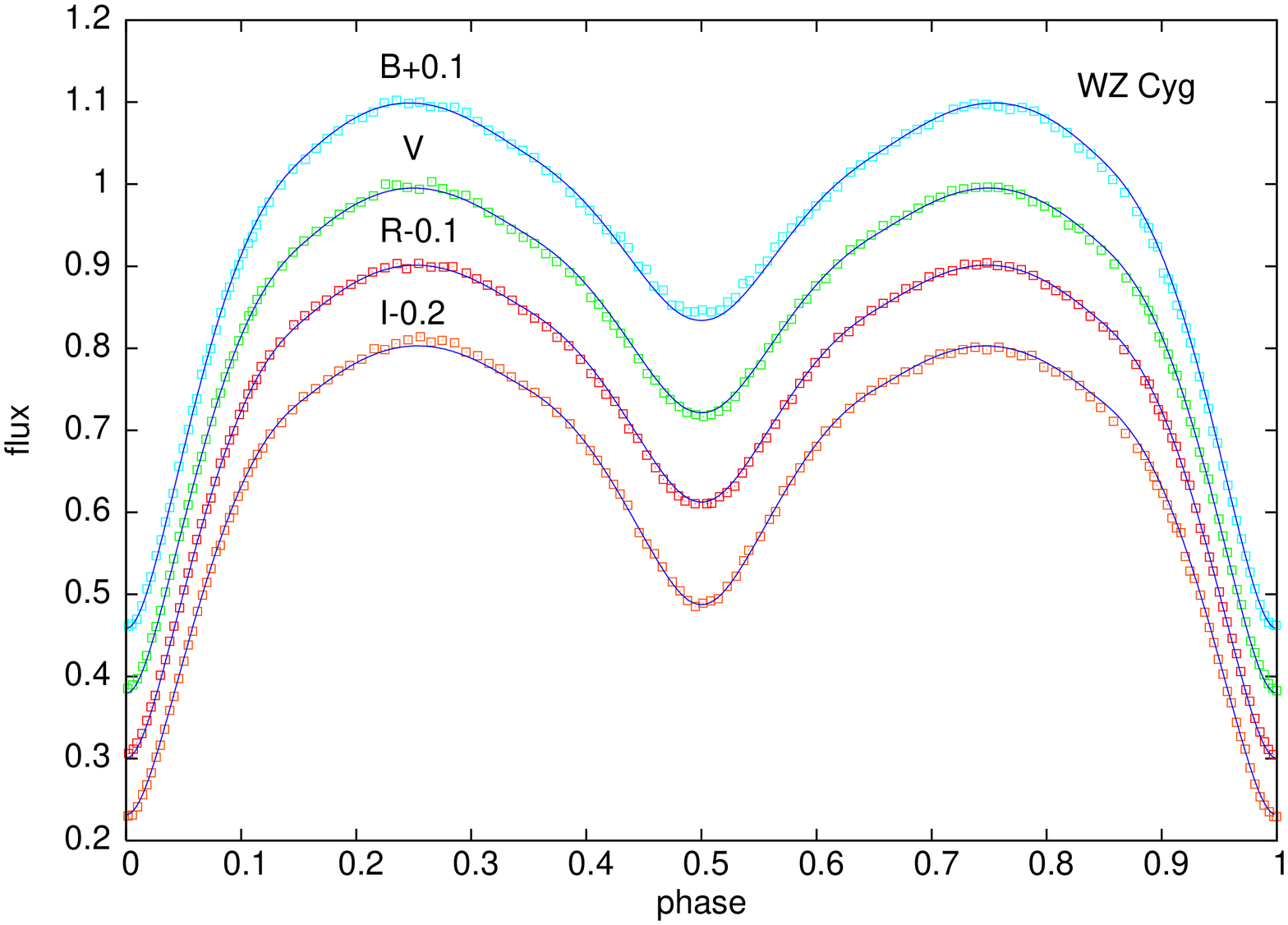}&
\includegraphics[width=1.8in,angle=-90]{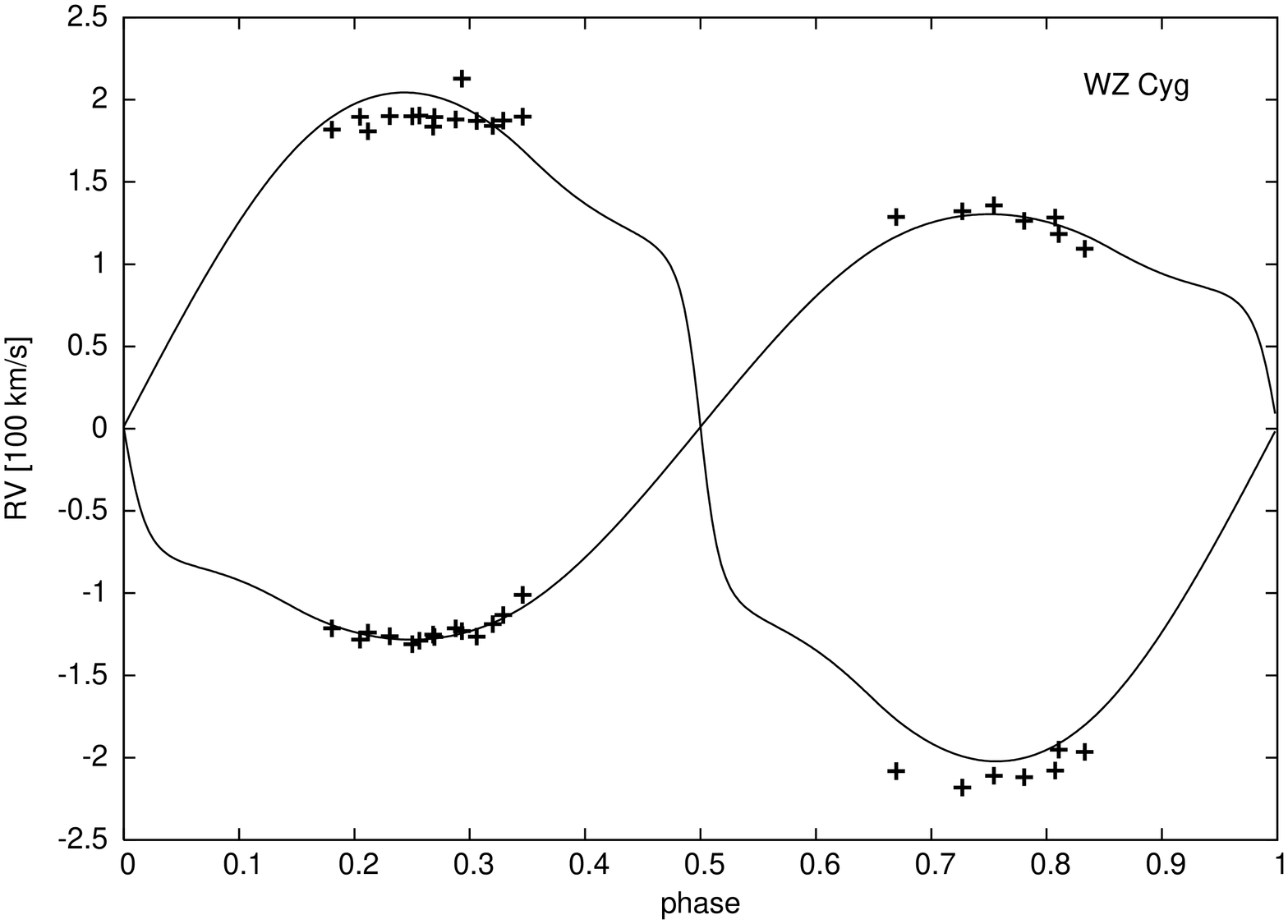}\\
\includegraphics[width=1.8in,angle=-90]{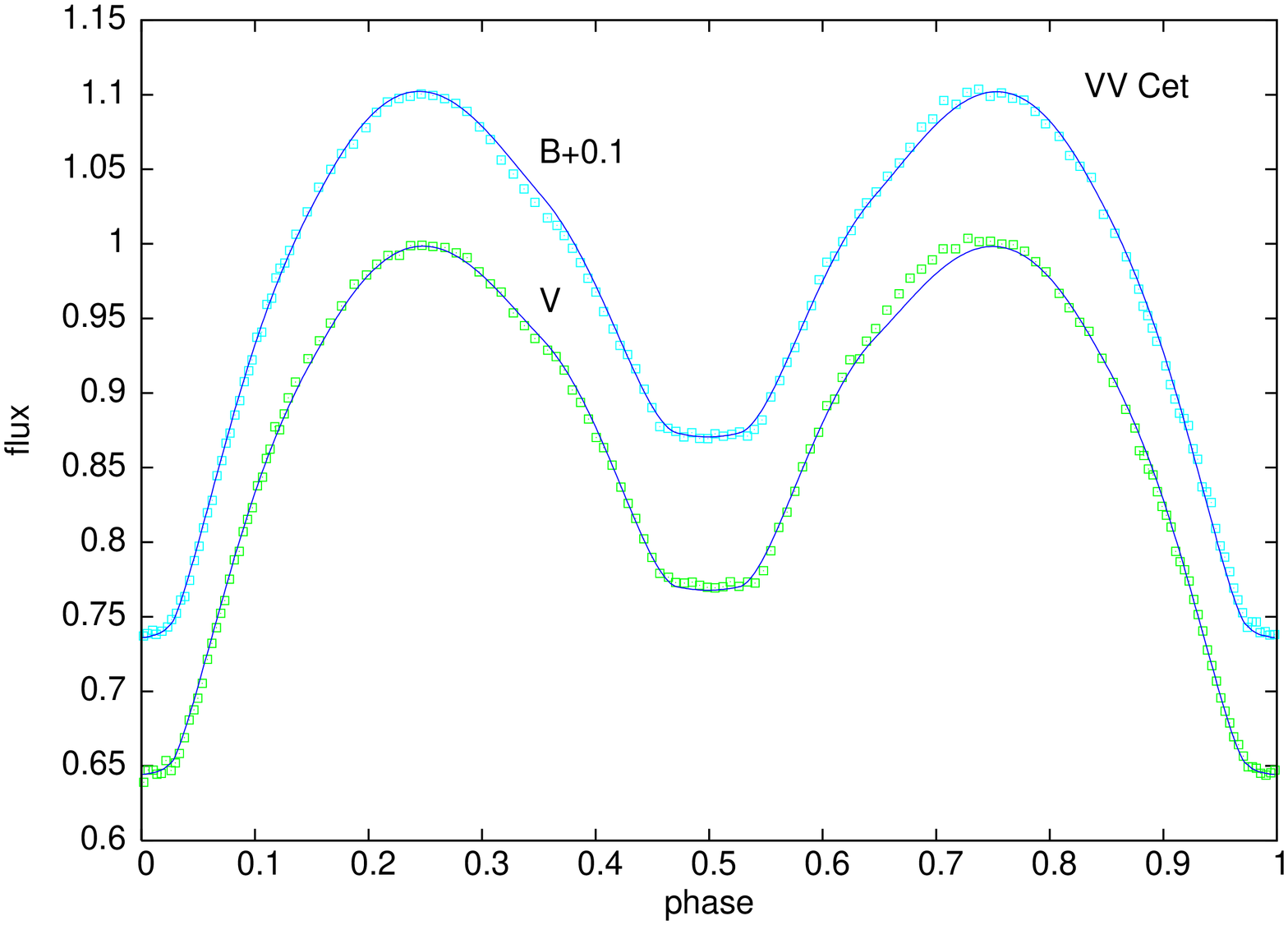}&
\includegraphics[width=1.8in,angle=-90]{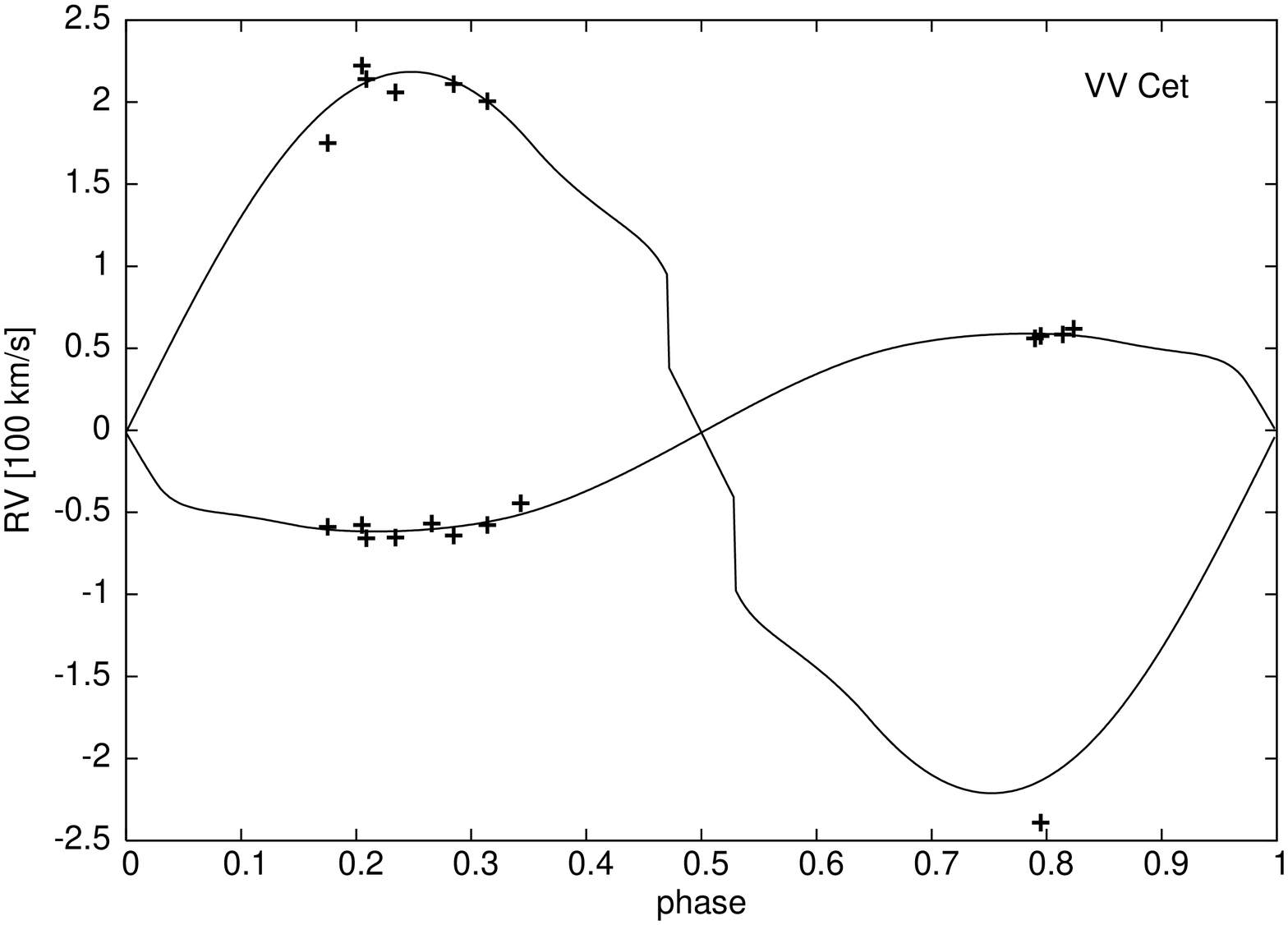}\\
\includegraphics[width=1.8in,angle=-90]{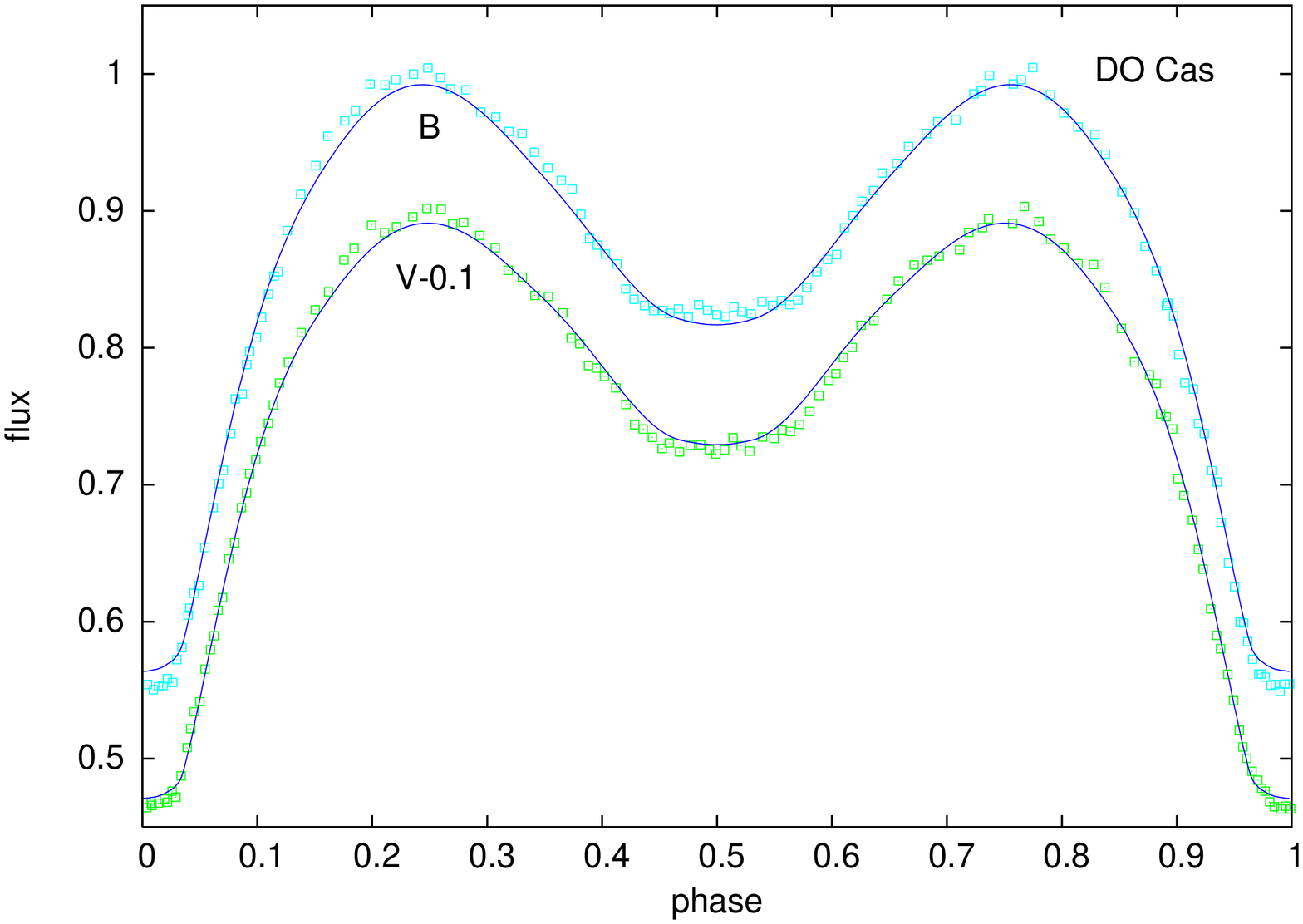}&
\includegraphics[width=1.8in,angle=-90]{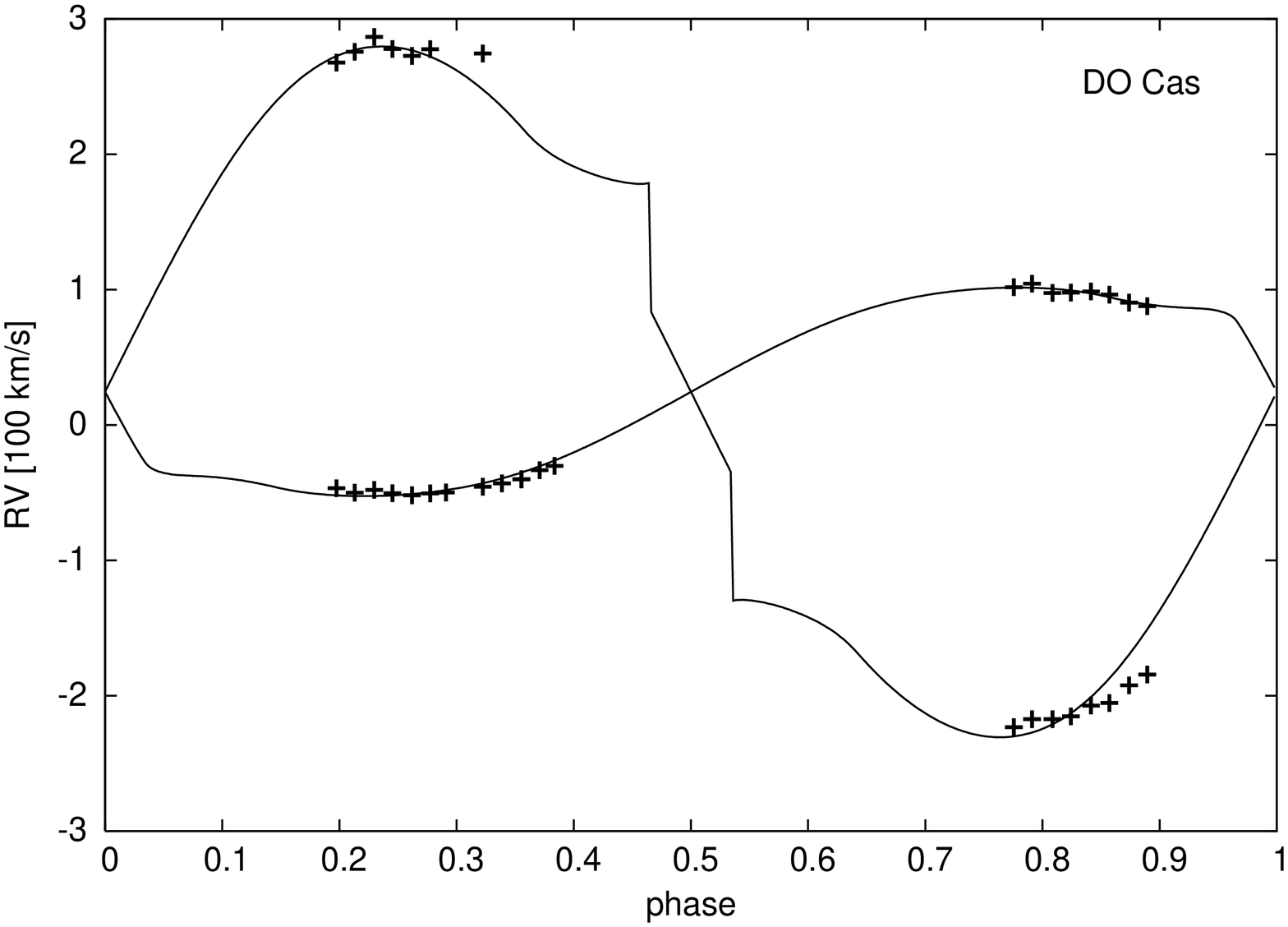}
\end{tabular}}
\vspace{0.5pc}
\FigCap{Comparison between the observed and theoretical photometric (left panels) 
and RV (right panels) curves of V1010~Oph, WZ~Cyg, VV~Cet and DO~Cas.}
\end{figure}

\begin{figure}[h]
\centerline{%
\begin{tabular}{c@{\hspace{1pc}}c}
\includegraphics[width=1.8in,angle=-90]{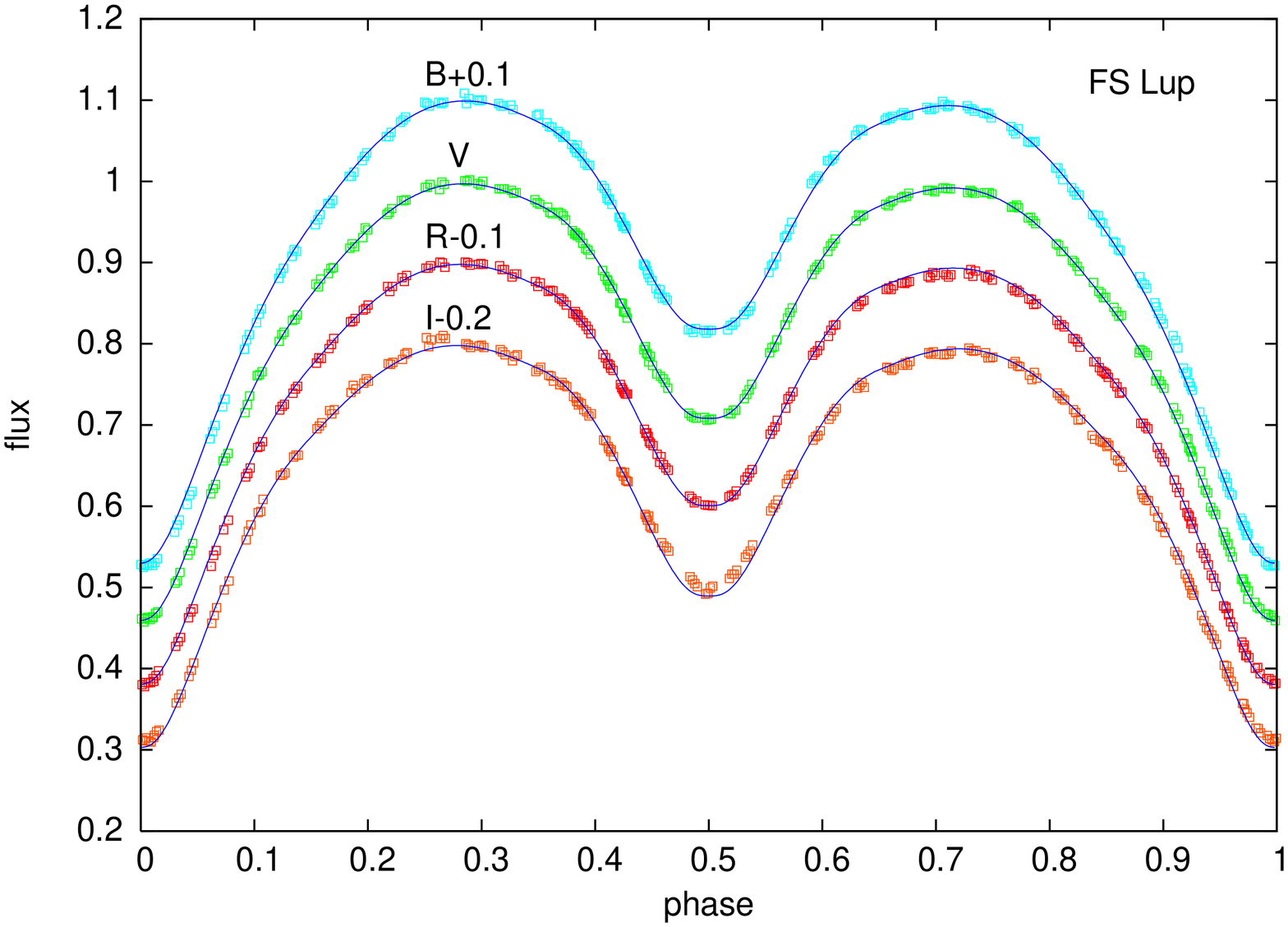}&
\includegraphics[width=1.8in,angle=-90]{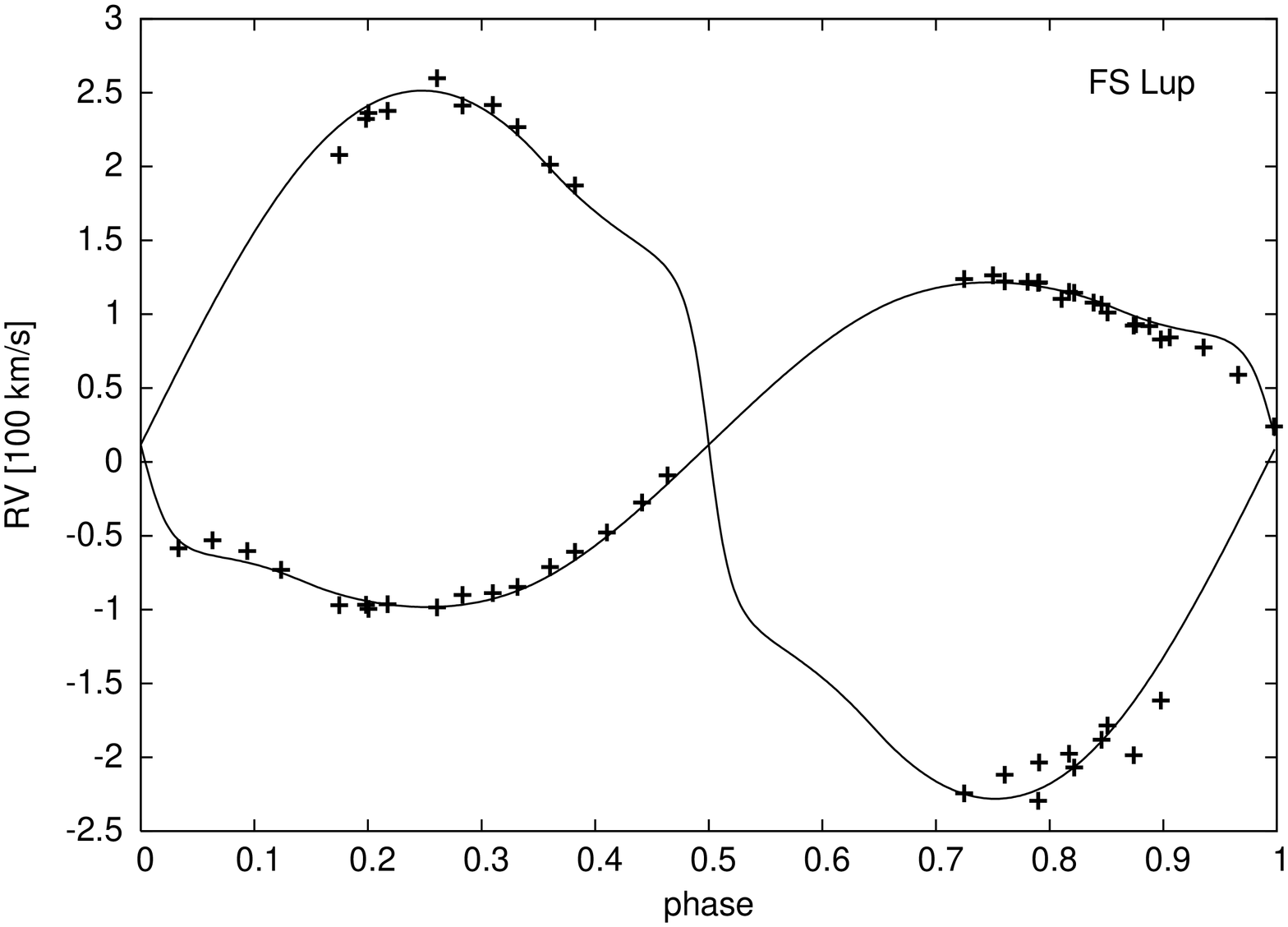}\\
\includegraphics[width=1.8in,angle=-90]{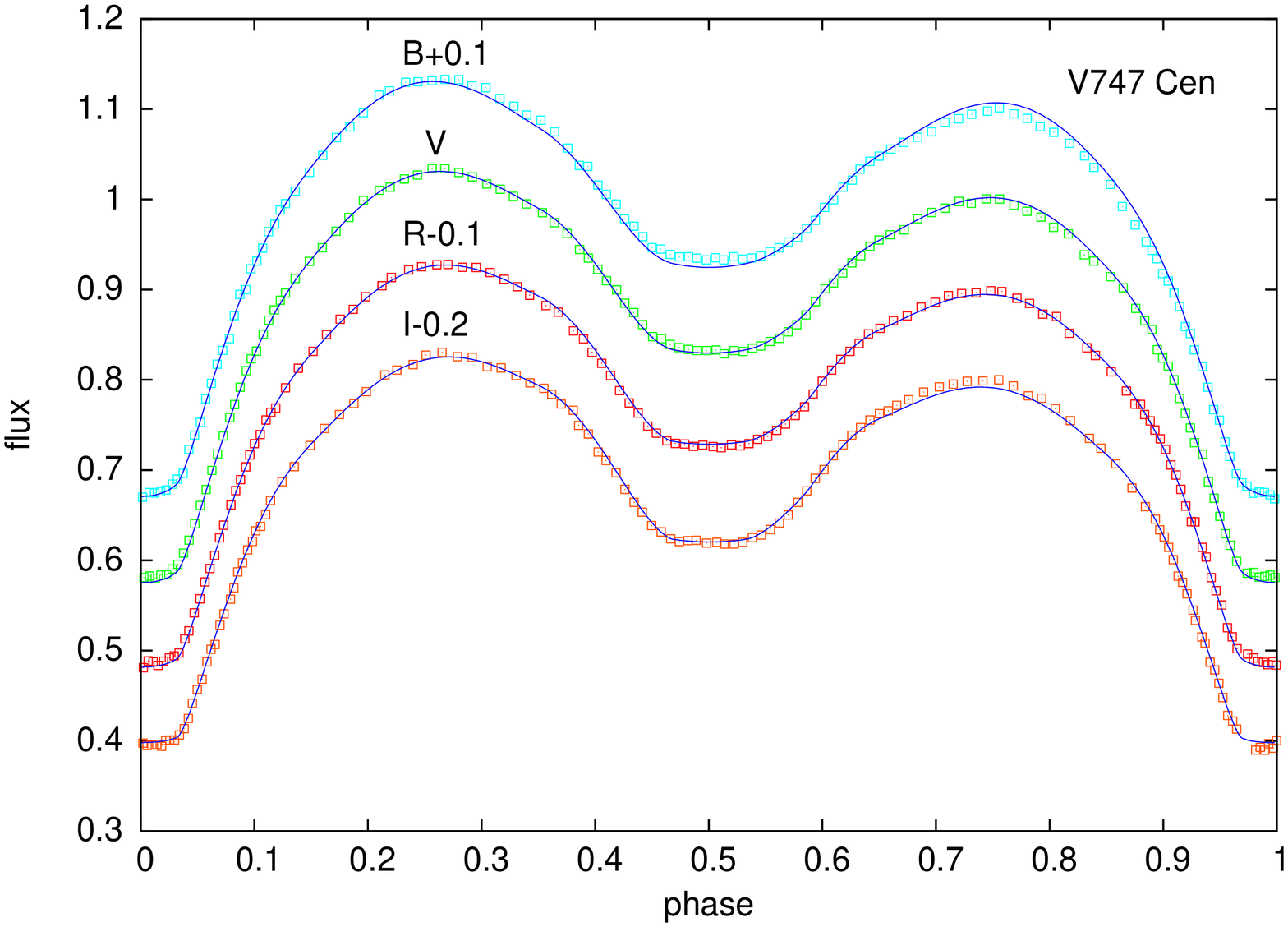}&
\includegraphics[width=1.8in,angle=-90]{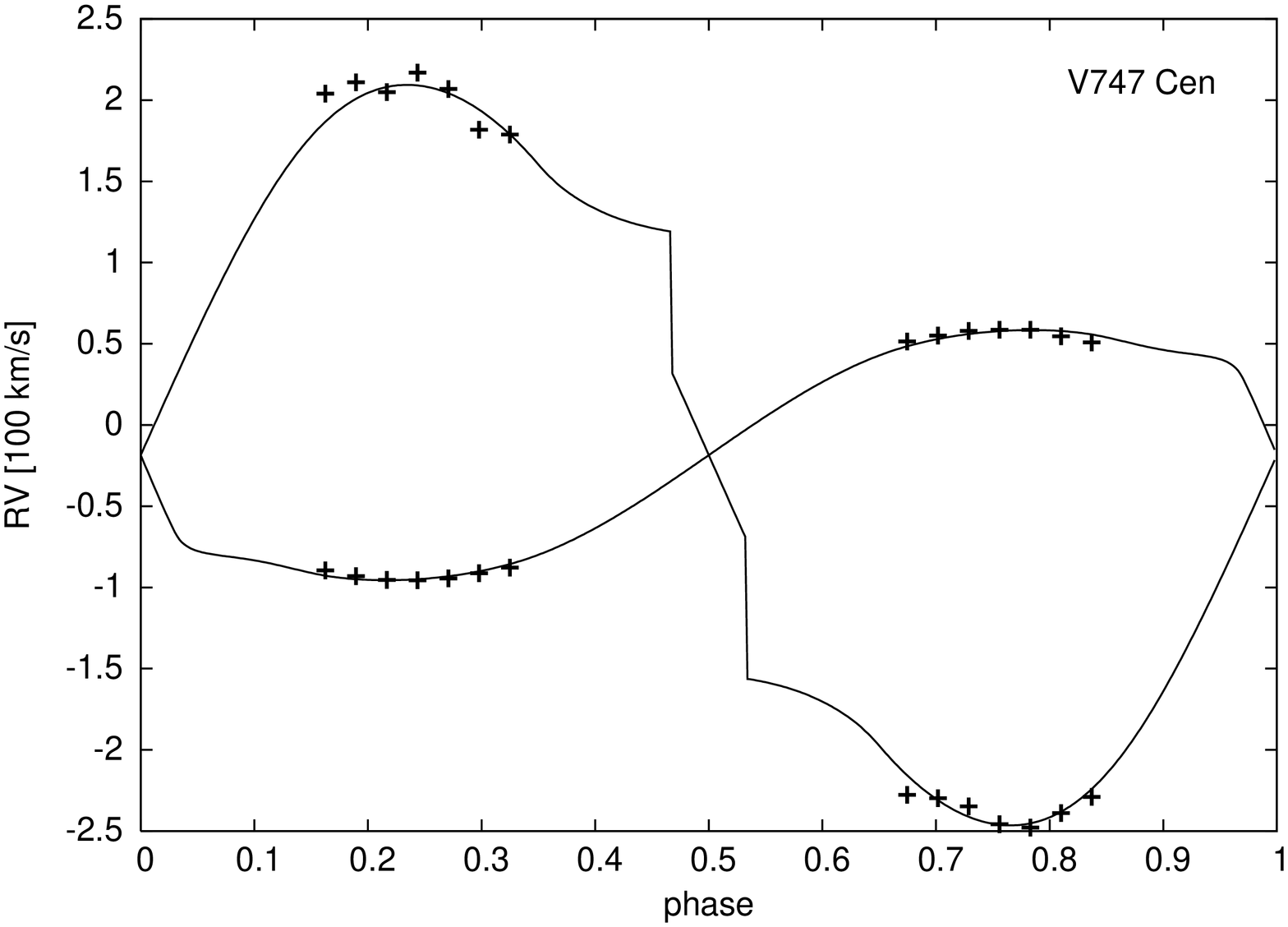}
\end{tabular}}
\vspace{0.5pc}
\FigCap{Comparison between the observed and synthetic photometric (left panels) 
and RV (right panels) curves of FS~Lup and V747~Cen.}
\end{figure}

\section{Near-contact binary stars}

\subsection{V1010 Oph}

For light-curve modelling, we used the
only available BV data obtained by
Leung (1974).  As an O'Connel effect is
visible in the light curves, we also
introduced a spot into the model.
During computations, we observed
good convergence of $\Omega_1$
to a critical value of the 
first Lagrangian point $L_1$, and
finally a semi-detached configuration
was obtained as the best solution.
This result naturally explains both
the presence of the hot spot on the
secondary star that is present in our
model and the period shortening seen
in the O--C diagram (Kreiner et al.,
2001), as the result of mass transfer
from the more massive primary to the
less massive secondary.  Although
BFs (Fig.~3) were
calculated from high signal-to-noise
spectra, they did not reveal any
features that could be interpreted as
a hot spot.  This may be a consequence of
exceptional characteristics 
of this hypothetical spot in RV
space, preventing it from being
easily detected; or of insufficient
spectral resolution; or of both 
these factors. 

\begin{figure}[h]
\centerline{%
\begin{tabular}{l@{\hspace{0.5pc}}r}
\includegraphics[width=1.75in,angle=-90]{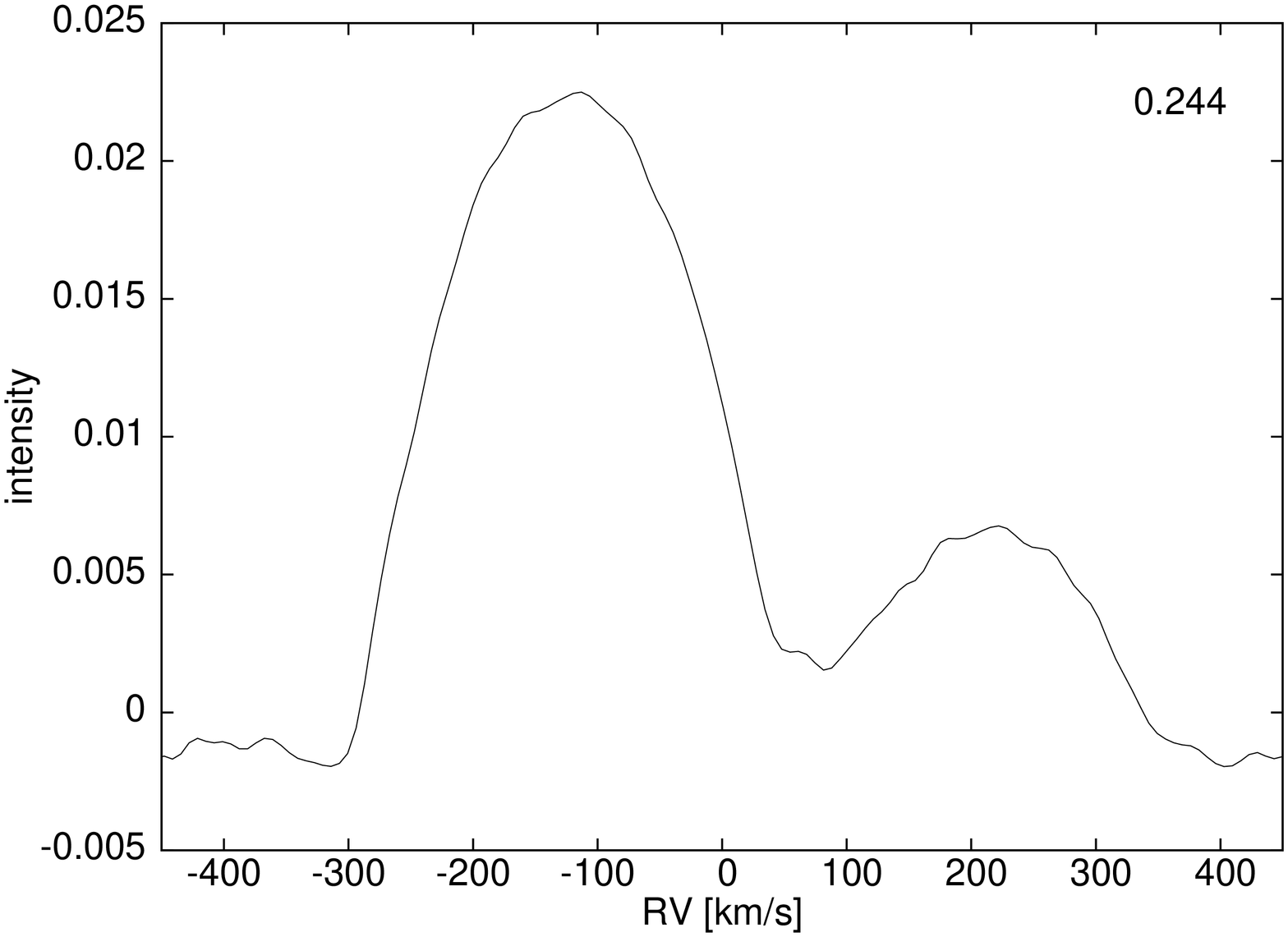}&
\includegraphics[width=1.75in,angle=-90]{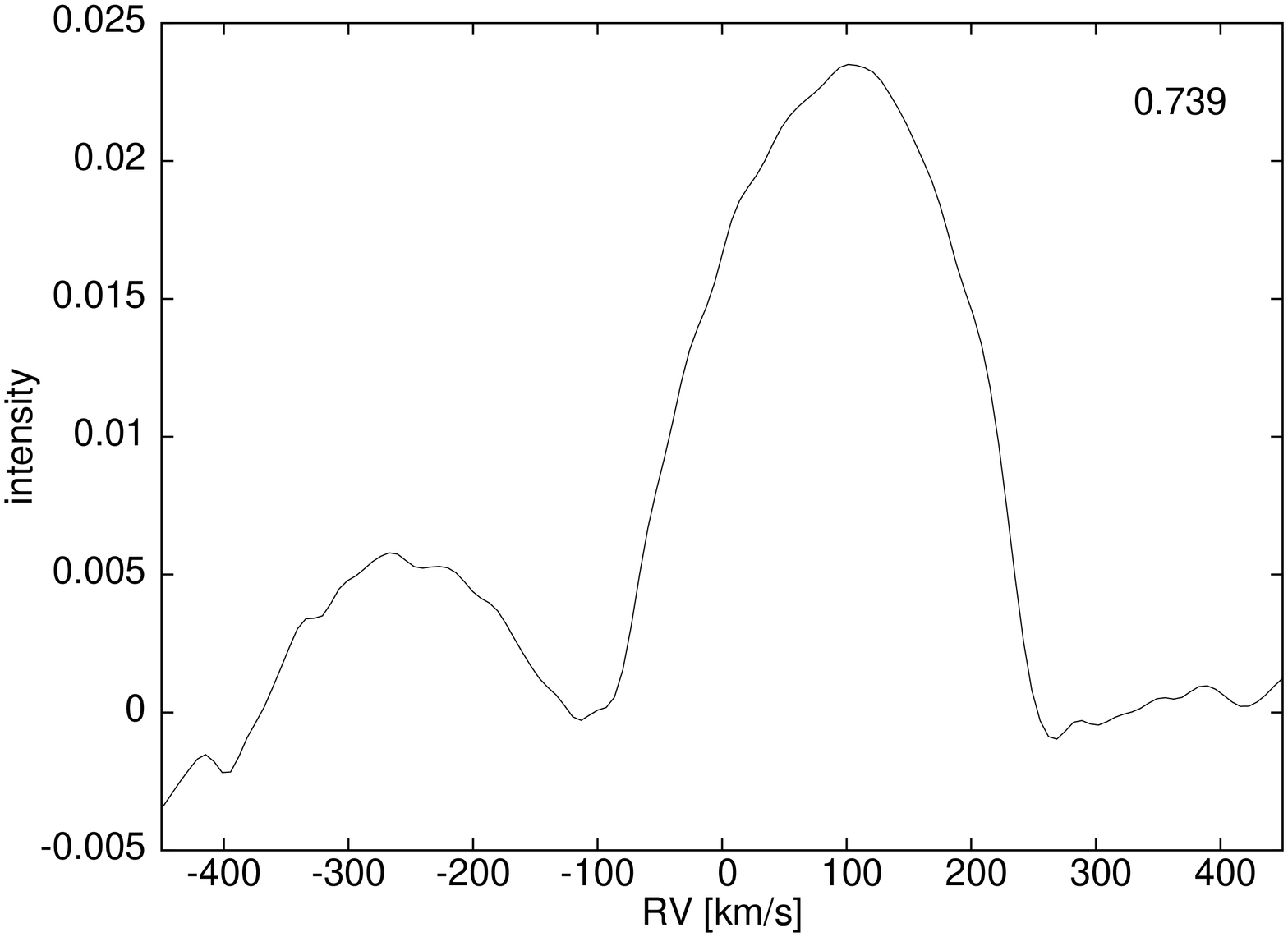}
\end{tabular}}
\vspace{0.5pc}
\FigCap{BFs of V1010~Oph obtained with F2V 
RV standard in both quadratures. The phases calculated for the times 
of mid-exposure are given in the upper-right corner.}
\end{figure}

\subsection{WZ Cyg}

We took a new set of light and
RV curves over a two-month
period.  Despite many attempts, we were
not able either to remove or to explain
the significant differences among the
values of $v_{\gamma}$ determined
separately for the RV
curves of each component. 
High-resolution, good-SN spectra 
might resolve the ambiguity.
Upon assuming a spectroscopic mass ratio
of $q_\mathrm{spec}=0.631(36)$, we obtained
a near-contact configuration
(Table~6, Fig.~1) for this system.
Because of the rather large uncertainty in
$q_\mathrm{spec}$, we made two additional
light-curve models with two extreme
possible values of $q_\mathrm{spec}$ (0.595
and 0.667), assumed constant in the 
computations. These models, however,
likewise yielded  
a near-contact configuration.

\subsection{VV Cet}

As the data obtained with the R filter turned
out to be affected by instrumental
problems, we used only the BV data
during light curve modelling.  As a
result, we obtained a semi-detached
configuration, with the primary
component filling its Roche lobe.
Only by placing a cold spot on the
surface of the primary component
was it possible to obtain a good
fit to the observed light curves.
Since the primary
component has a radiative atmosphere,
this result indicates the
possibility that the primary 
is of type Ap.
Unfortunately, our BF profiles are not of
a quality high enough for us to detect and
confirm the existence of such a spot on
the star.

\subsection{DO Cas}

As our three attempts to obtain new
high-quality light curves failed,
we used already-published BV data
(Gleim \& Winkler, 1969).  Our light
curve modelling used the spectroscopic
mass ratio determined for the first
time ($q_\mathrm{spec}=0.304$) and resulted
in a near-contact configuration.
 
\subsection{FS Lup}

We decided to improve the spectral
classfication of FS~Lup using our new
data: a comparison of the spectra
obtained at SAAO in the 3920--4750{\AA}
region with respective spectra of
RV standards indicated that 
the spectral type
of FS~Lup should be an early G.
Additionally, from the $B-V$
colour index, determined from our
new photometric observations at the
secondary minimum, we estimated the
spectral type of the primary component as
G2V (Table~5). The uncertainty deriving
from the  
interstellar-reddening correction
should not affect the effective
temperature of the primary component
(5860~K) by more than 150~K, 
as in the case of other systems with
determined spectral types.  

To account for the O'Connel effect, a
cold spot on the primary component was
placed in that model which best reproduced
the observed light curves.
BF profiles obtained
from medium-resolution spectra seem to
reveal a cold spot on the primary-star
hemisphere facing the secondary star,
at coordinates in accordance
with those obtained from light-curve
modelling. However, spectra
of higher resolution
are necessary to confirm
this finding.  In investigating 
configuration, we found FS~Lup
to be a near-contact system, possibly
semi-detached, with the primary star
almost or entirely filling its Roche
lobe (Table~6).

\subsection{V747 Cen}

The most interesting feature of
V747~Cen is one which is, 
because of the high quality
of our BFs, 
for the first time noticed
in close binary systems: there is               
a well-defined region of increased
intensity, visible in almost all
profiles of the secondary component
during second quadrature (Figs.~4,~5). 
We note that BFs of
the secondary component in first
quadrature are also higher in the
vicinity of the primary component
(as compared with the corresponding 
BF behaviour of
V1010~Oph in Fig.~3).  This brighter
region is underlined only in the plot
obtained at phase 0.190, but one
can easily locate it on the remaining
plots presented in the left column
of Fig.~4.  The region 
may be a result of a
hypothetical stream of matter, leaving
the primary star through the first
Lagrangian point, 
undergoing a slight deflection
through centrifugal force, 
and then directly 
striking the side of the secondary-star 
photosphere visible in first
quadrature.

Light-  and RV-curve modelling
supported these
conclusions: 
upon assuming that a hot spot
on the secondary component is 
responsible for the O'Connel effect, we
obtained a semi-detached configuration,
with the primary component filling its
Roche lobe (Table~6, Fig.~2).

\begin{figure}
\centerline{%
\begin{tabular}{c@{\hspace{4pc}}c@{\vspace{-0.7pc}}}
\includegraphics[width=1.10in,angle=-90]{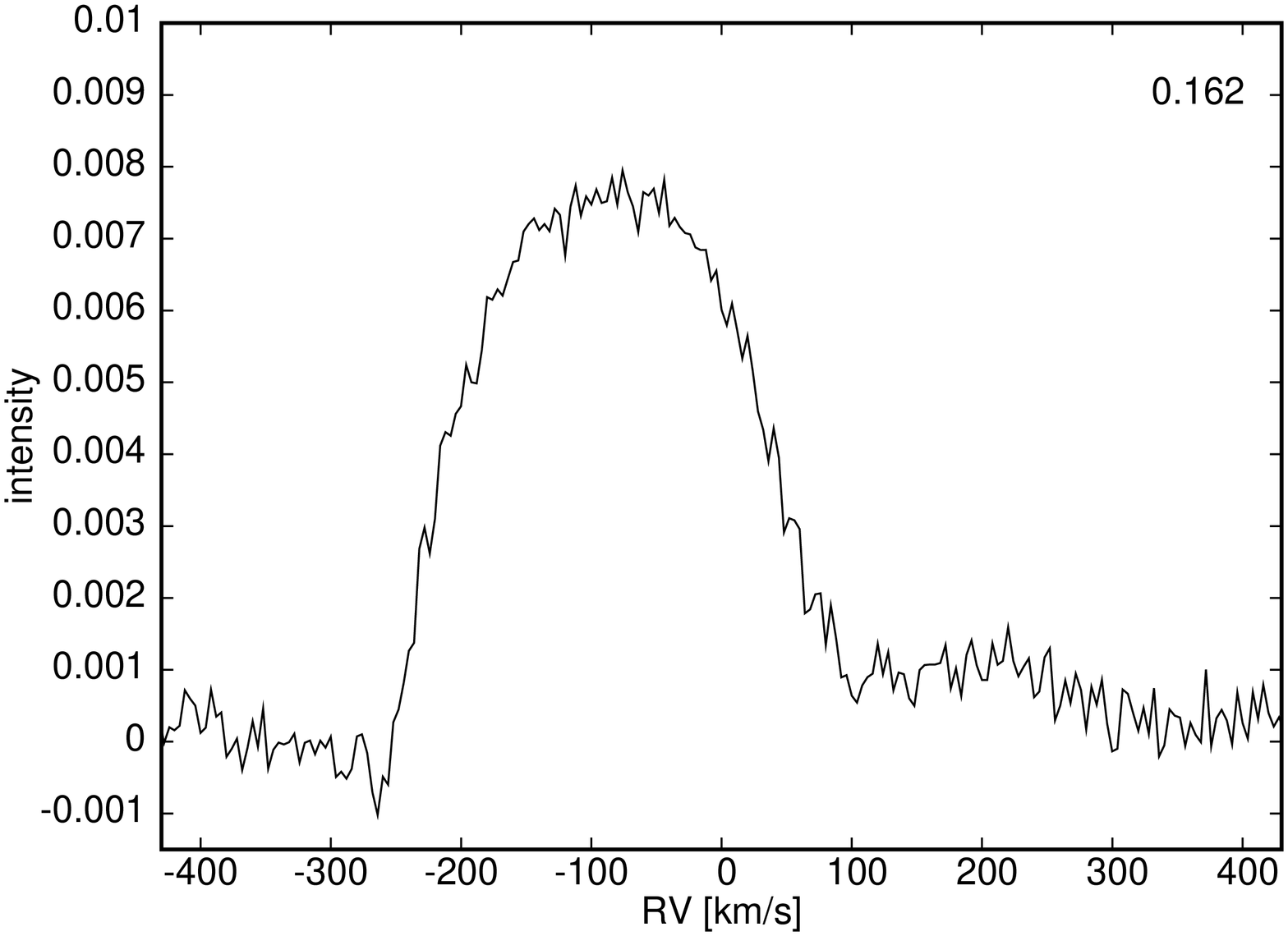}&
\includegraphics[width=1.10in,angle=-90]{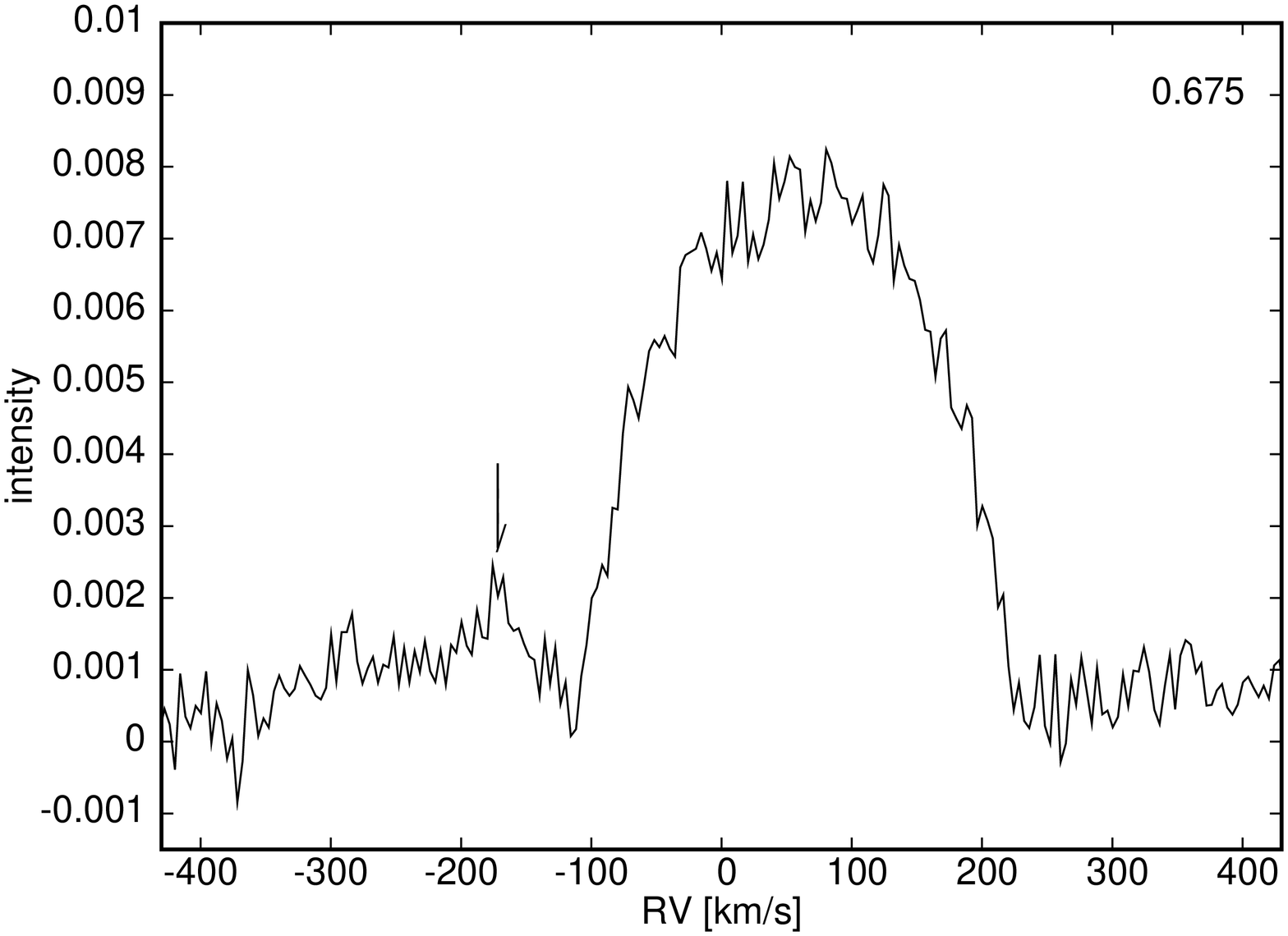}\\
\includegraphics[width=1.10in,angle=-90]{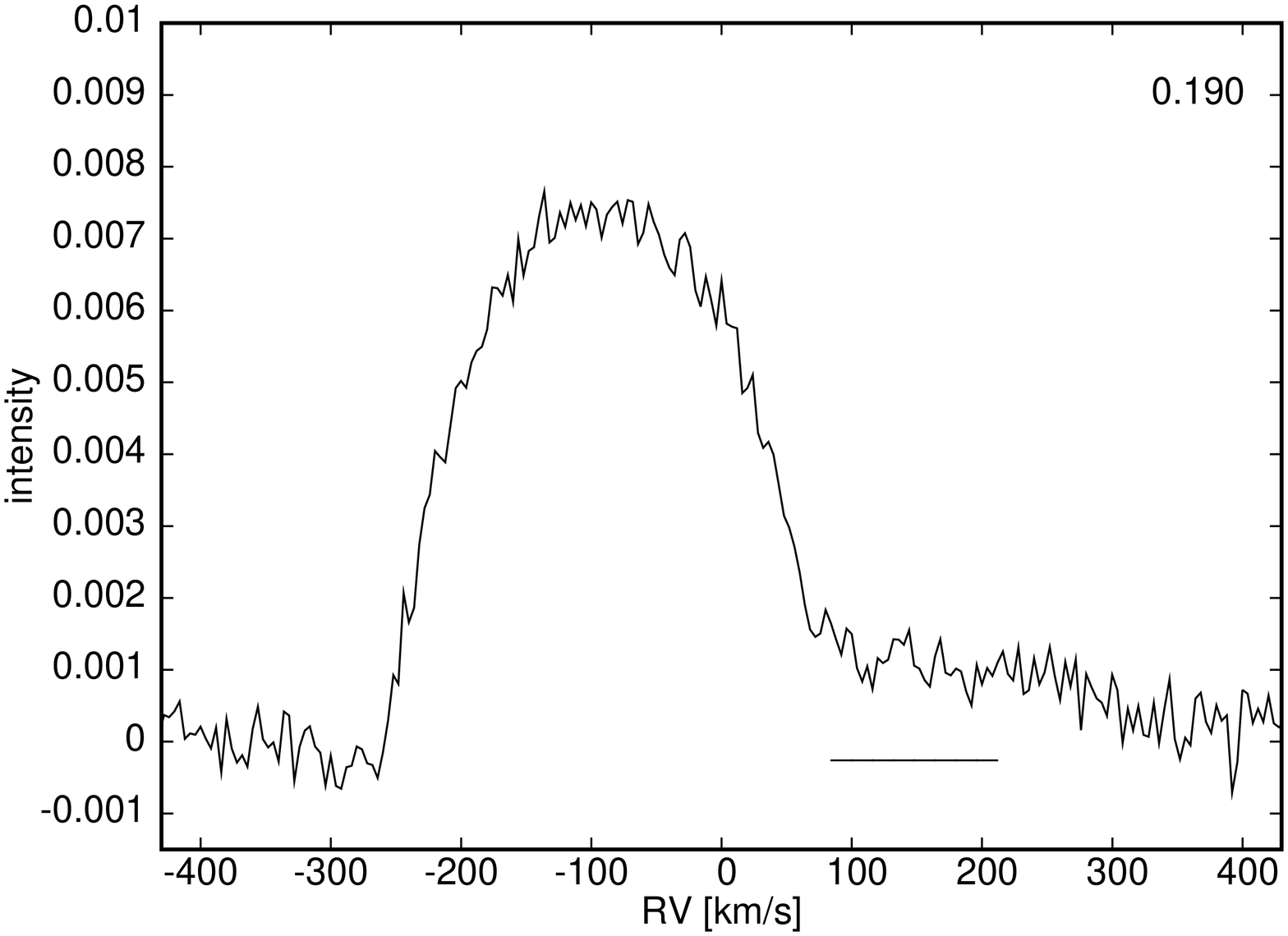}&
\includegraphics[width=1.10in,angle=-90]{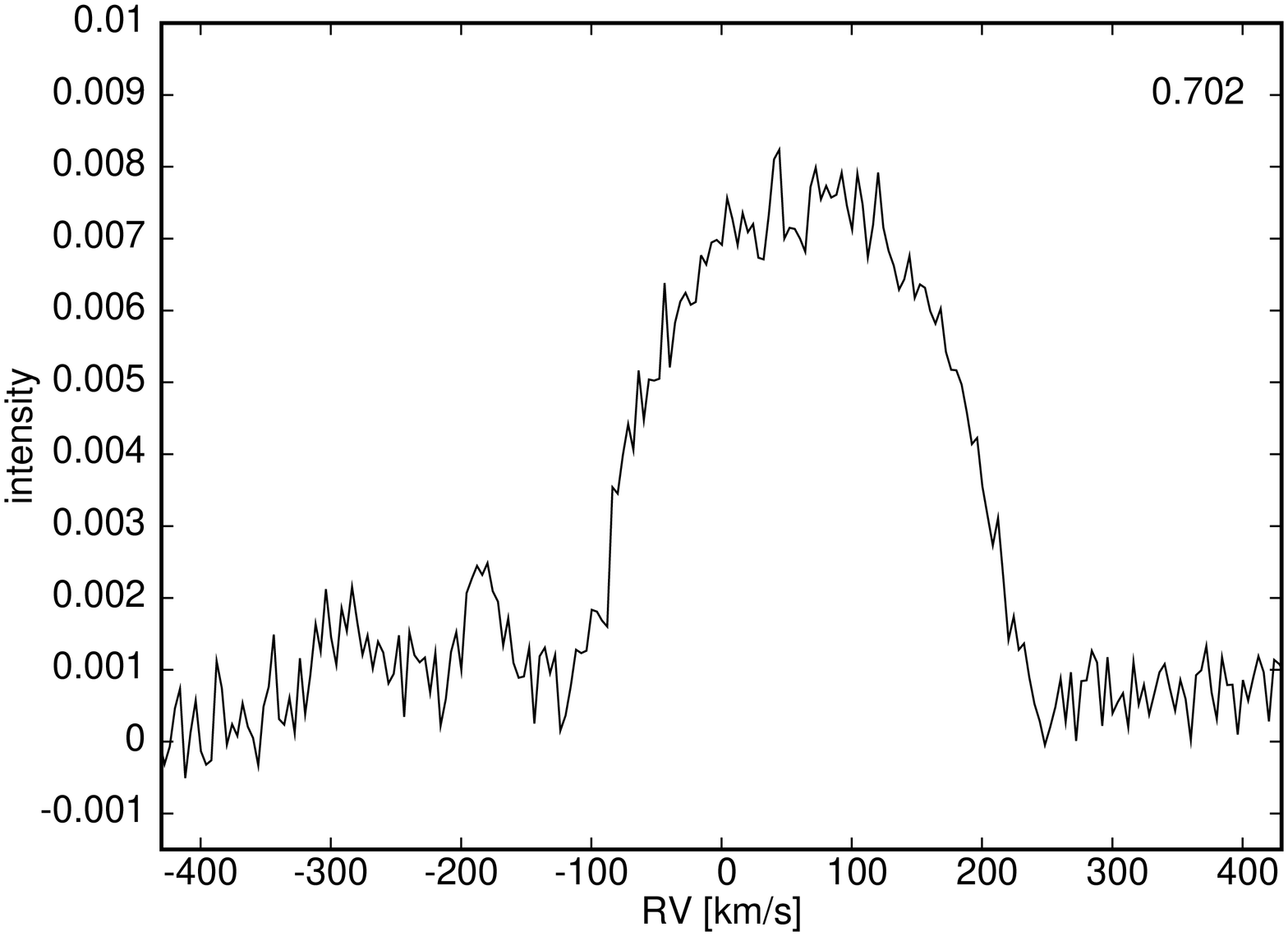}\\
\includegraphics[width=1.10in,angle=-90]{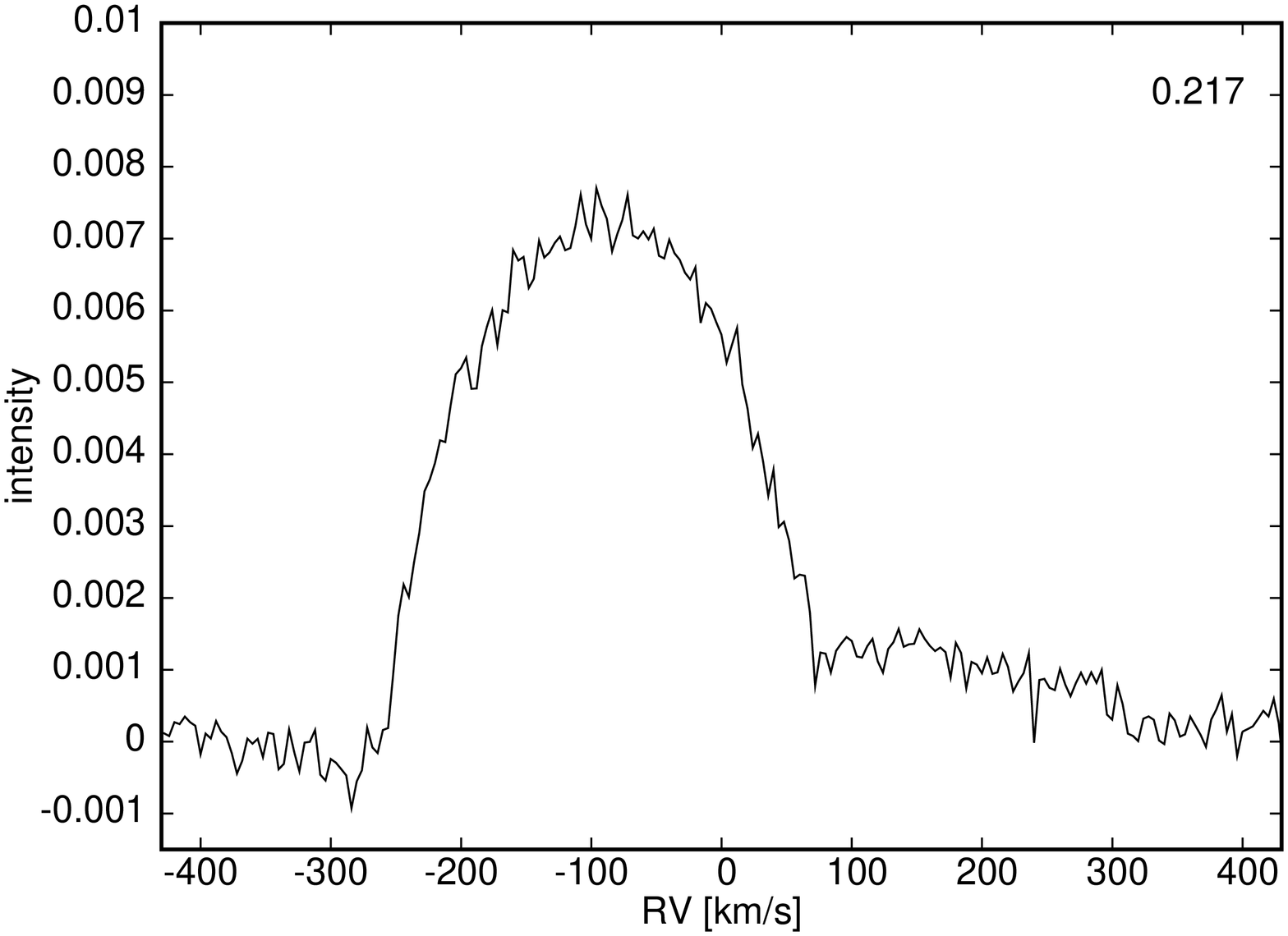}&
\includegraphics[width=1.10in,angle=-90]{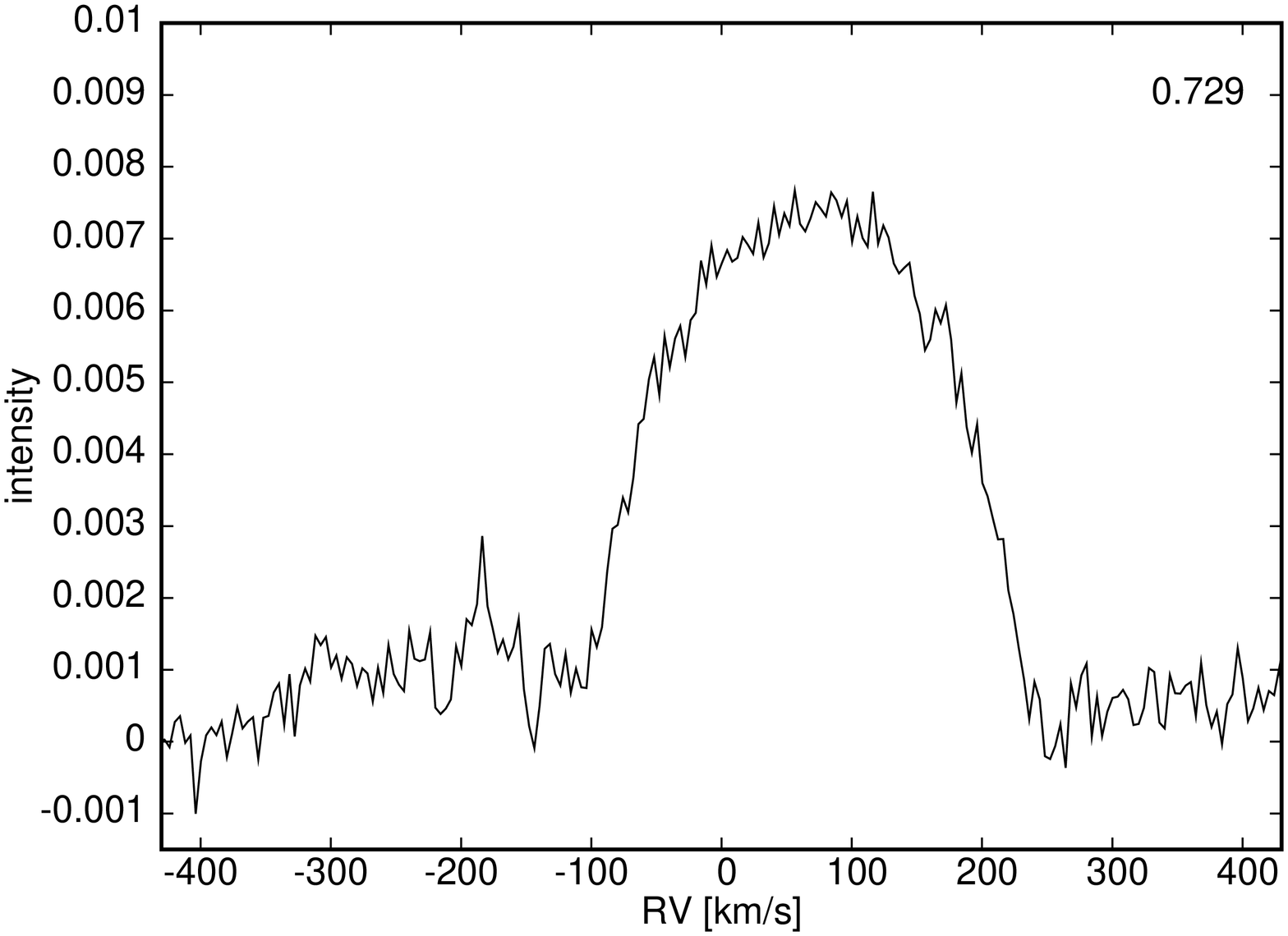}\\
\includegraphics[width=1.10in,angle=-90]{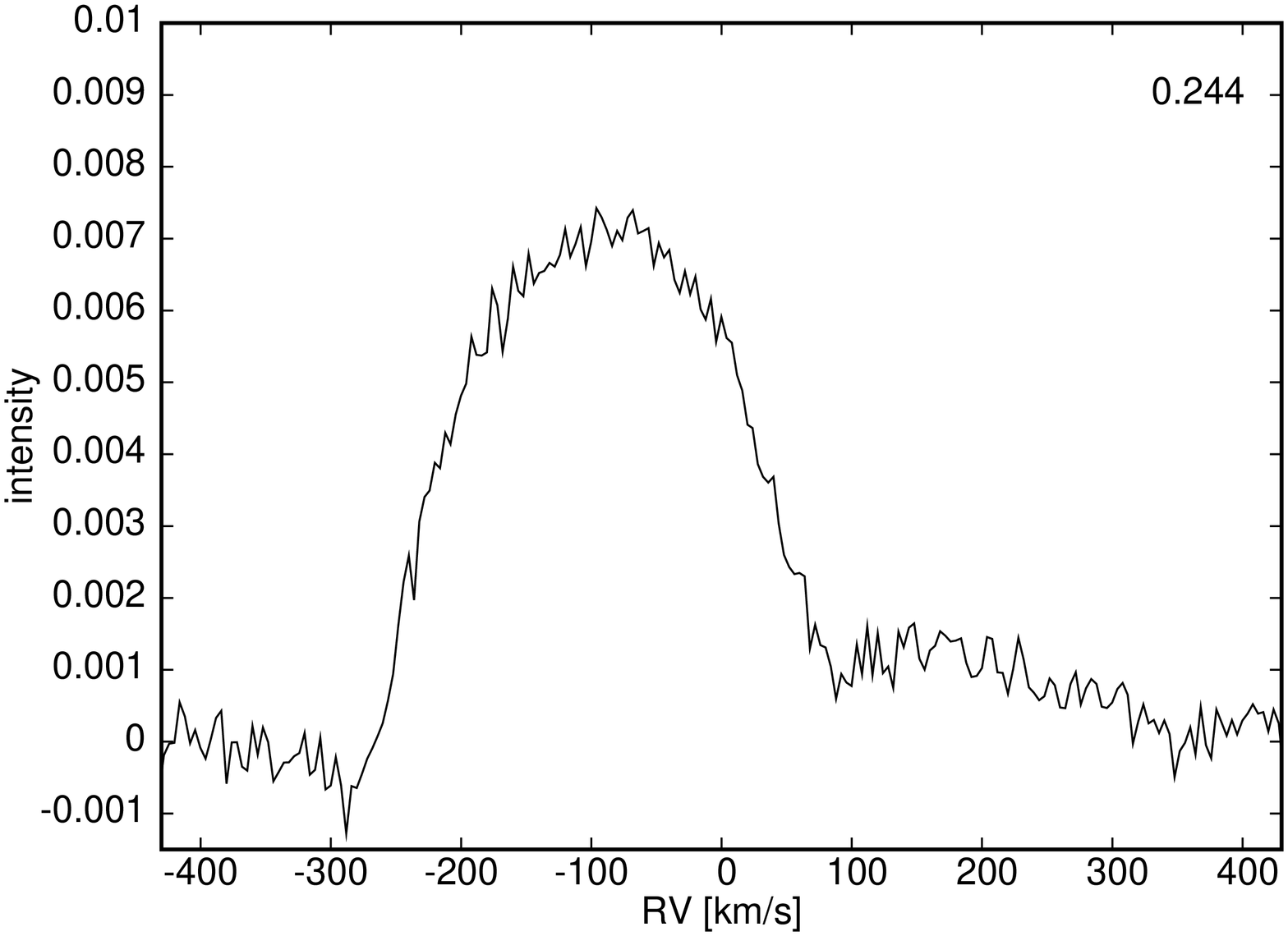}&
\includegraphics[width=1.10in,angle=-90]{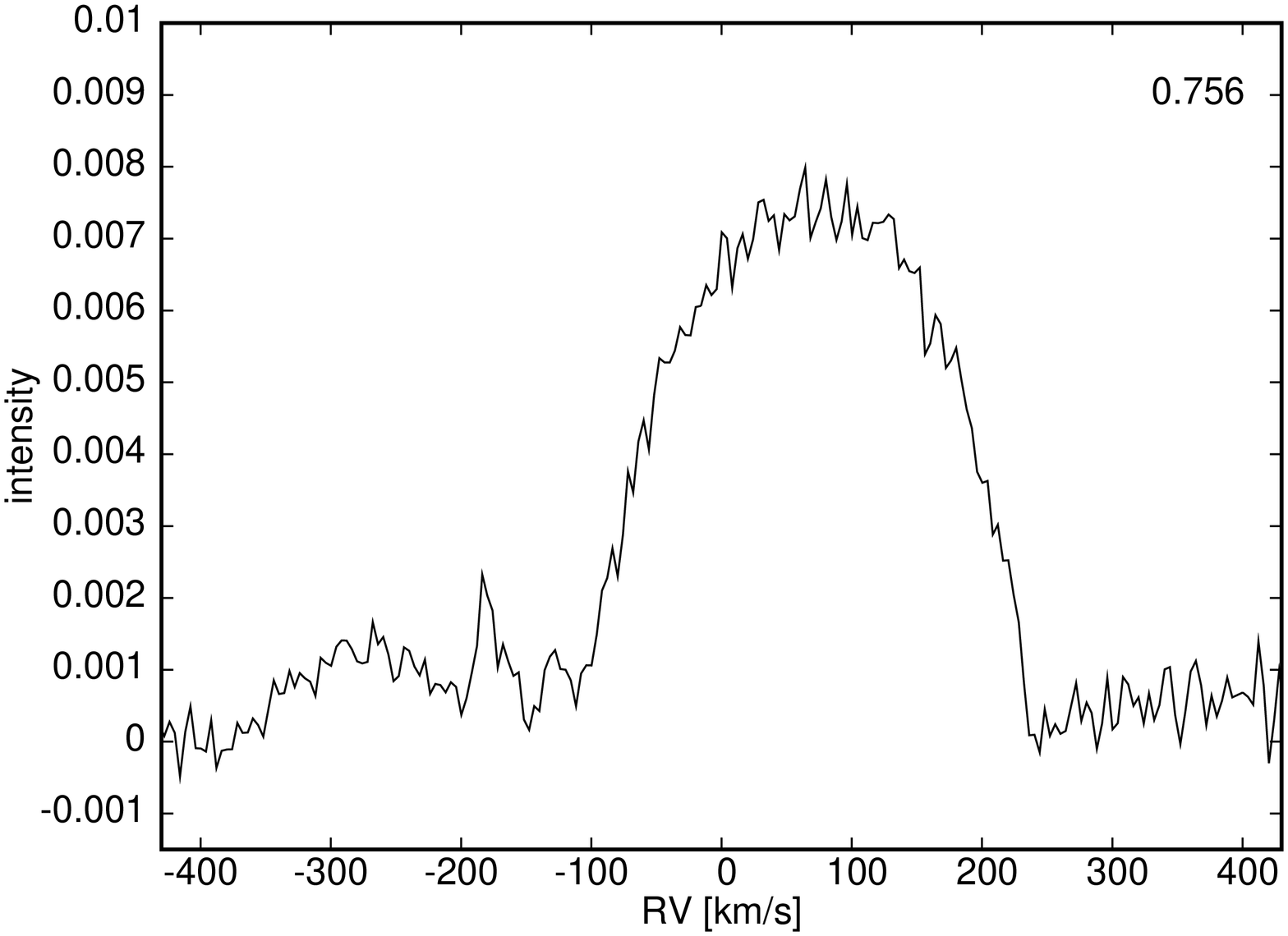}\\
\includegraphics[width=1.10in,angle=-90]{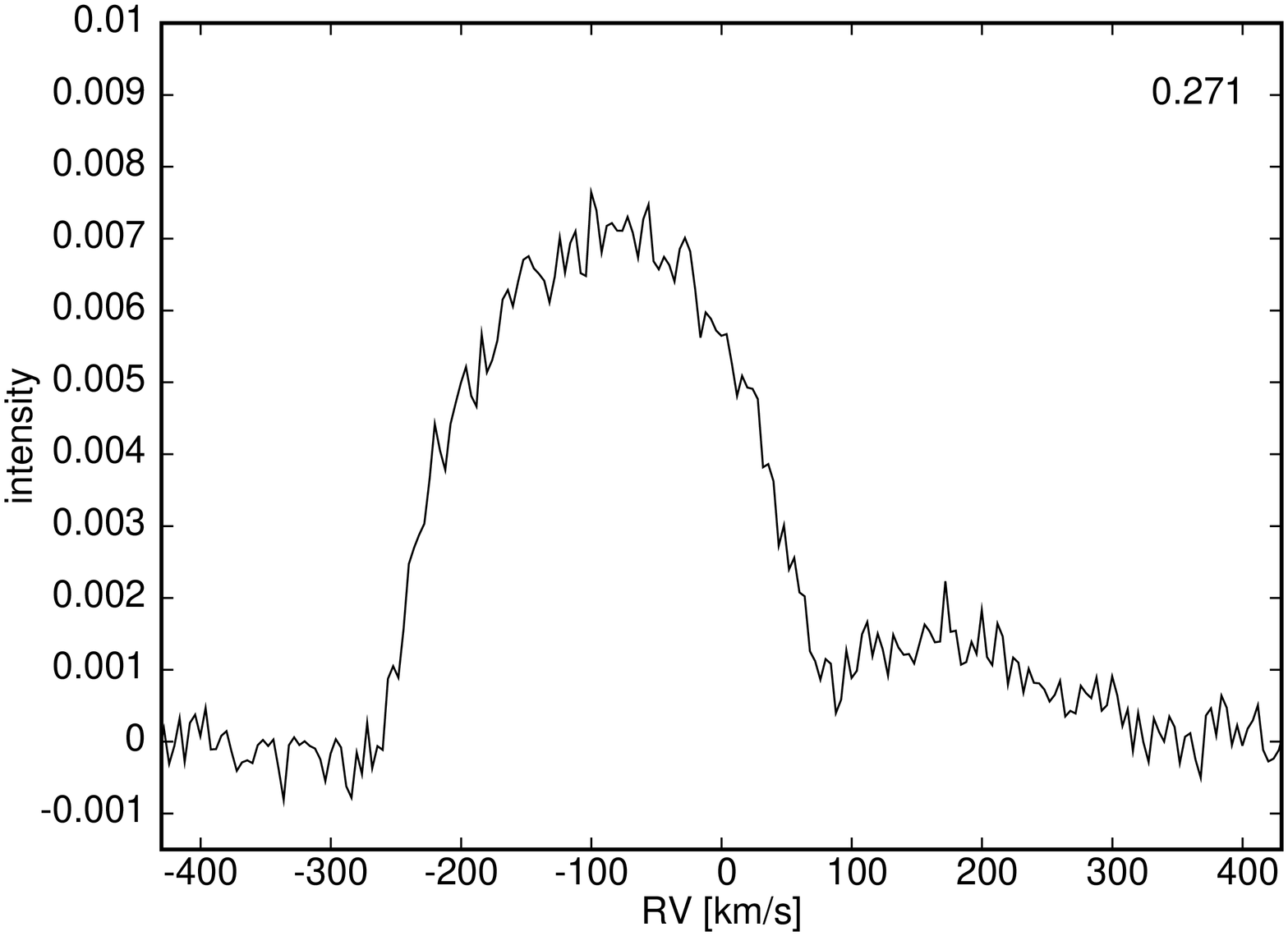}&
\includegraphics[width=1.10in,angle=-90]{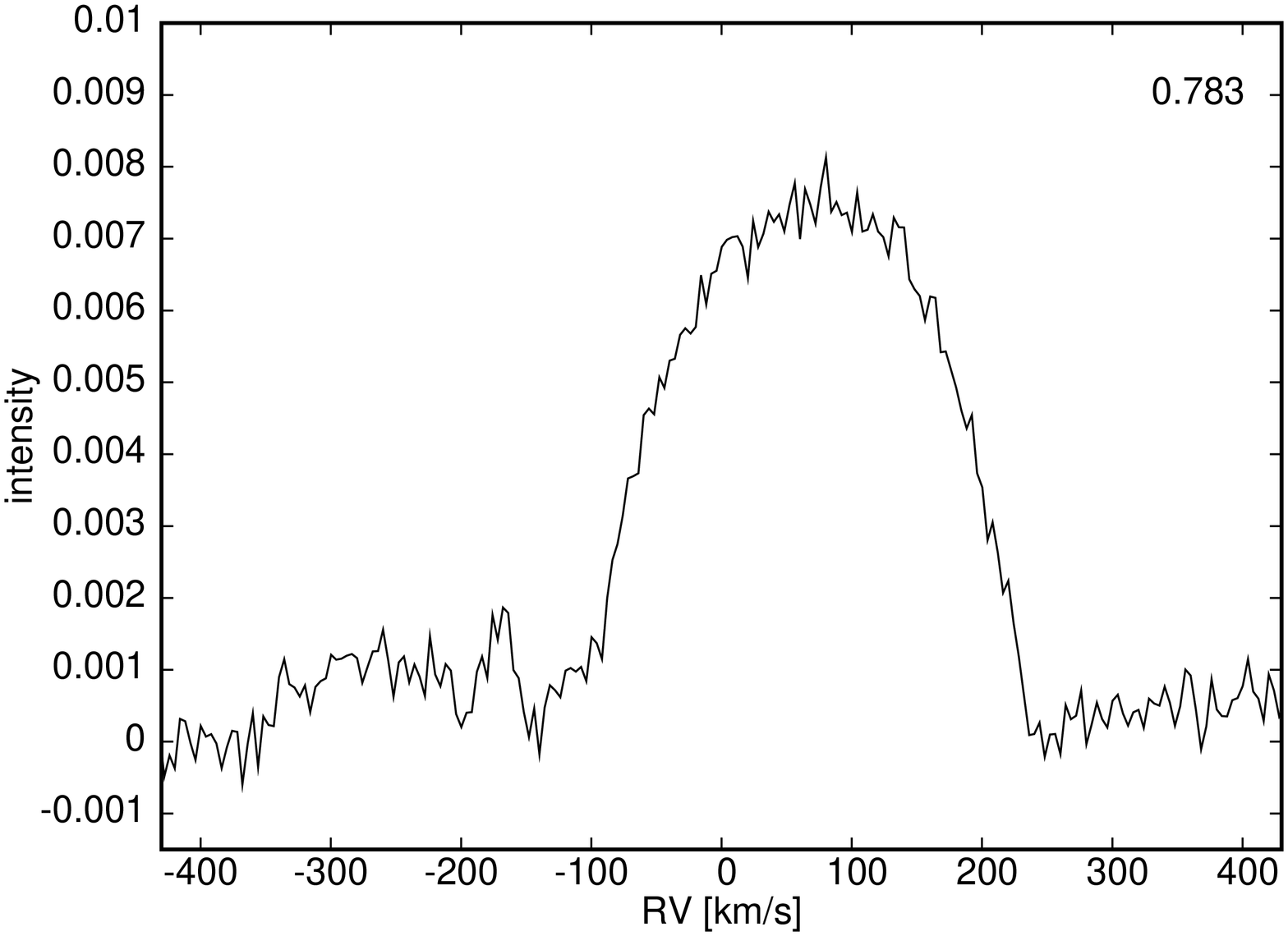}\\
\includegraphics[width=1.10in,angle=-90]{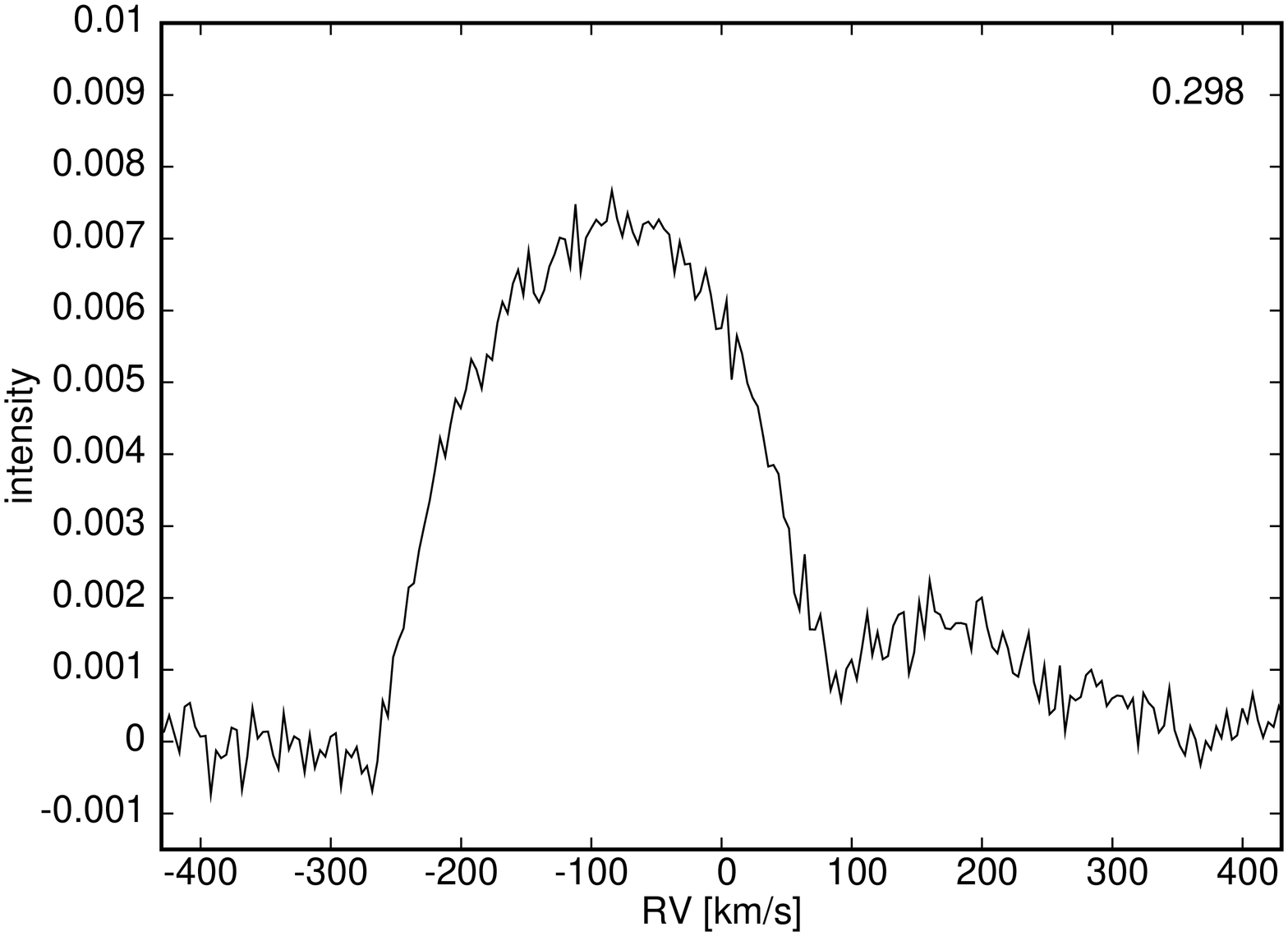}&
\includegraphics[width=1.10in,angle=-90]{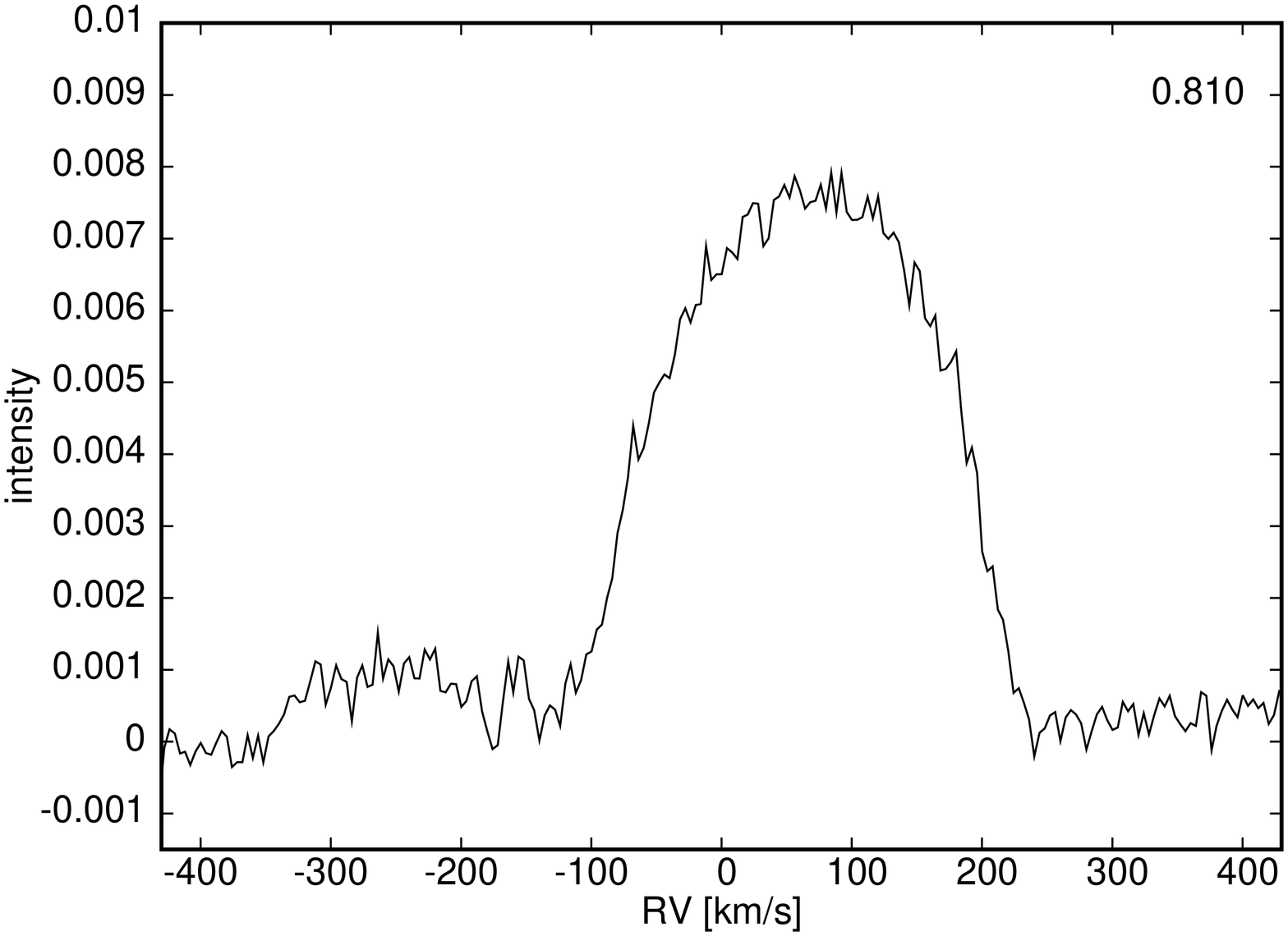}\\
\includegraphics[width=1.10in,angle=-90]{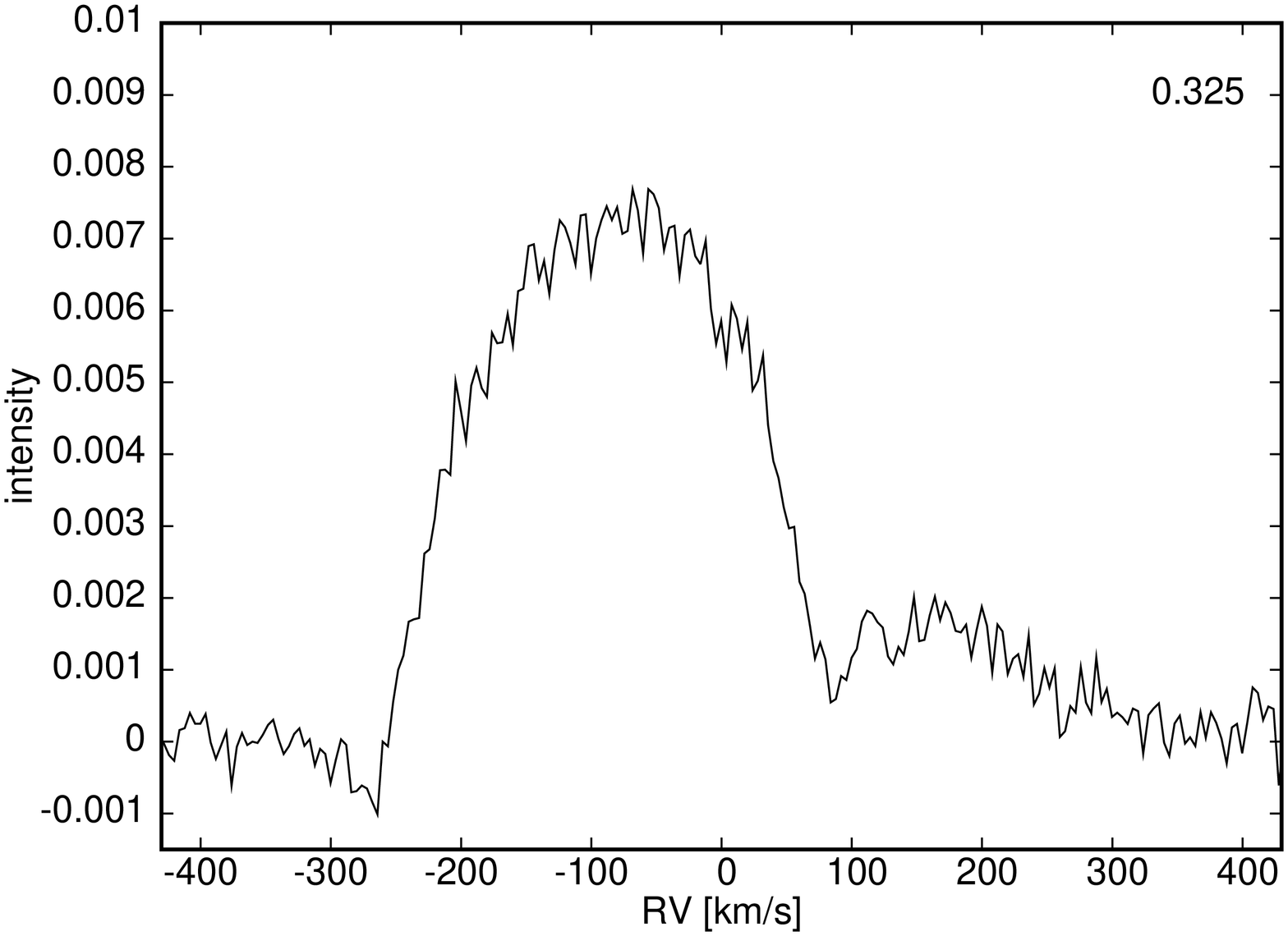}&
\includegraphics[width=1.10in,angle=-90]{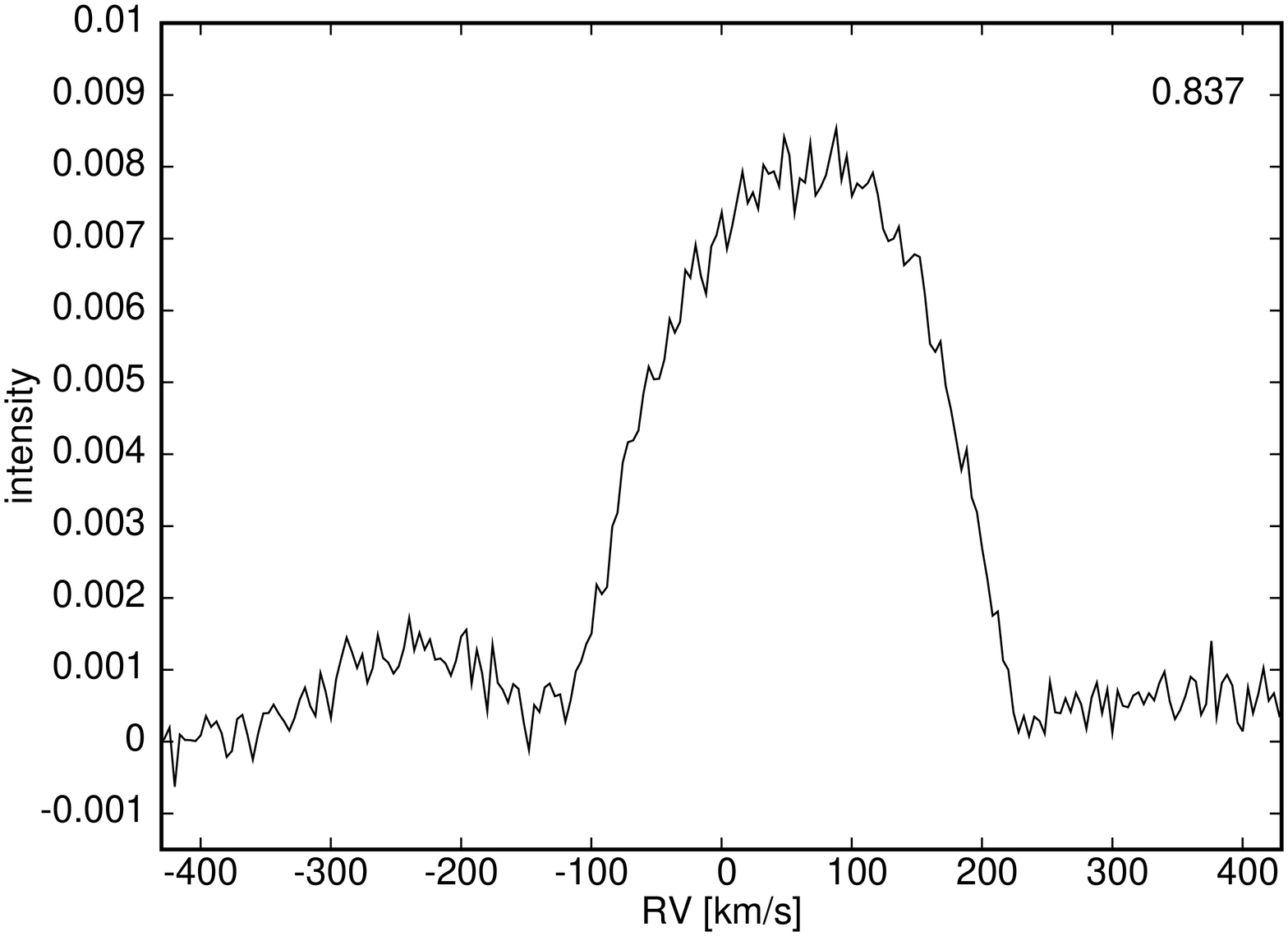}
\end{tabular}}
\vspace{1pc}
\FigCap{BFs of V747~Cen 
obtained with G5V RV standard. 
The brighter region on the secondary
component in the vicinity of the
primary component is underlined
at phase 0.190 (see Sec.~4.6).
The well defined brighter region
visible in second quadrature
(right panels) is indicated by an
arrow in the first plot.  Only one BF
profile of the secondary component,
obtained at phase 0.837, seems to
be free from accretion effects.}
\end{figure}

\begin{figure}
\centering
\includegraphics[]{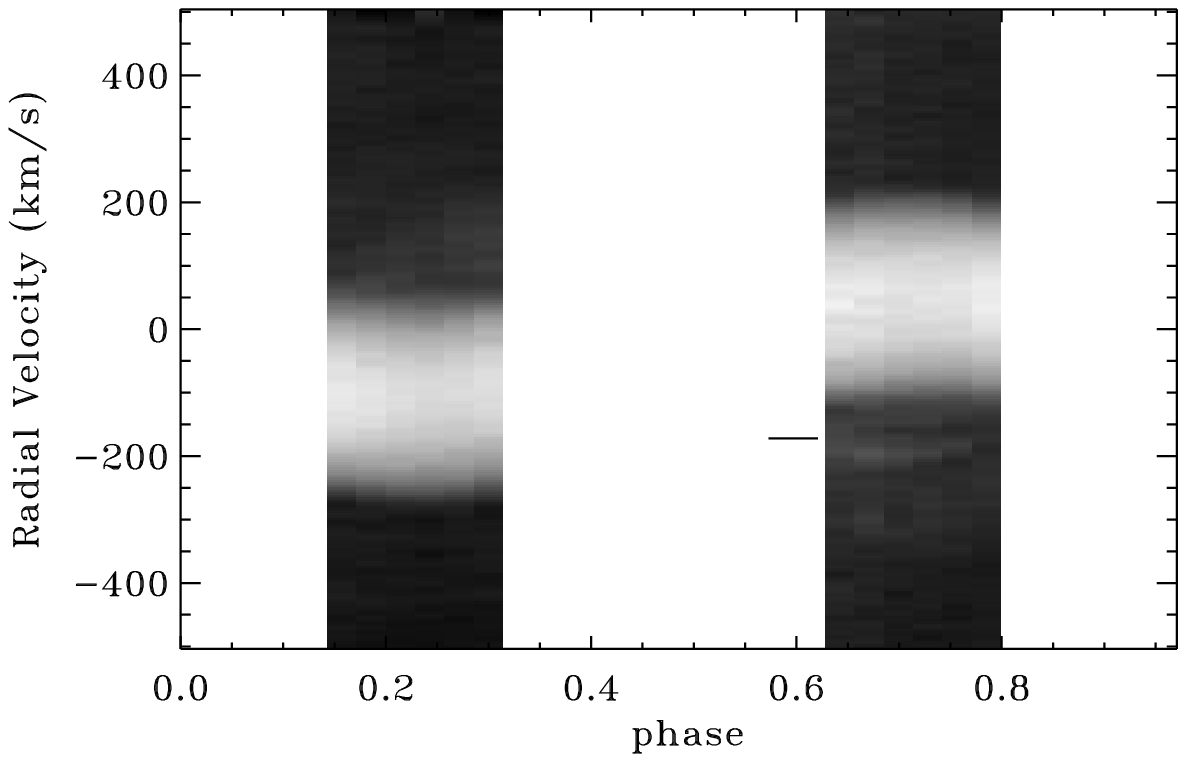}
\FigCap{BFs of
V747~Cen (as in Fig.~4) in 
phase-velocity space, rebinned
in phase with a constant     
increment of 0.03.
The bright region on the secondary
component visible in second
quadrature is marked.  As expected, the
profile of the radiative A5V primary
component does not reveal dark spots.}
\end{figure}

\MakeTable{lllllll}{12.5cm}{The results obtained from light- and RV-curve modelling 
of contact systems with large temperature differencies between components (CLdT).
\newline $^*$ - parameters assumed during light curve modelling (see Sections~3 and 5 for explanation)}
{\hline\hline
parameter          &  CX Vir       & FT Lup      & BV Eri     & FO Hya     & CN And       & BX And        \\ \hline
$i[^o]$            &  83.33(17)    & 89.26(51)   & 80.31(11)  & 79.00(34)  & 69.416(92)   & 75.862(33)    \\
$T_{eff}^{prim}[K]$&  6450$^*$     & 6700$^*$    & 6700$^*$   & 7000$^*$   & 6450$^*$     & 6650$^*$      \\
$T_{eff}^{sec}[K]$ &  4694(26)     & 4651(4)     & 5387(20)   & 4667(27)   & 4726(34)     & 4758(5)       \\
$\Omega$           &  2.5146(8)    & 2.7231(3)   & 2.3737(7)  & 2.2714(25) & 2.6403(12)   & 2.7760(5)     \\
$q_{spec}$         & 0.343(3)$^*$  & 0.440(3)$^*$&0.274(2)$^*$&0.238(6)$^*$& 0.385(5)$^*$ & 0.455(8)$^*$  \\
$v_{\gamma}[km/s]$ & -11.65(32)    & -2.30(30)   & -39.62(35) & 54.42(30)  & -23.47(63)   & -24.00(60)    \\
$a[R_{\odot}]$     &  4.584(19)    & 3.537(13)   & 3.365(12)  & 3.184(21)  & 3.163(22)    & 4.424(36)     \\ \hline
$f[\%]$            & 21.2          & 13.4        & 20.0       & 36.0       & 3.0          & 4.5           \\ \hline
$L_{1U}$           &  ---          &  ---        & 8.798(40)  & ---        &  ---         &   ---         \\ 
$L_{1B}$           & 9.350(51)     & 10.597(23)  & 8.920(40)  & 10.746(81) &  9.952(18)   & 11.0925(51)   \\
$L_{1V}$           & 9.242(48)     & 10.315(24)  & 8.748(37)  & 10.886(78) &  9.830(16)   & 10.8293(52)   \\
$L_{1R}$           & 9.039(49)     & 10.164(24)  &  ---       & 11.091(74) &  9.710(15)   & 10.6043(56)   \\
$L_{1I}$           & 8.956(47)     &  9.909(22)  &  ---       & 11.093(96) &  9.531(13)   & 10.2546(58)   \\
$L_{2U}$           &  ---          &  ---        & 0.996(23)  & ---        &  ---         &   ---         \\
$L_{2B}$           & 0.648(26)     &  0.7151(46) & 0.946(22)  & 0.361(13)  &  0.781(36)   &  0.9154(52)   \\
$L_{2V}$           & 0.789(30)     &  0.9002(56) & 1.093(23)  & 0.497(16)  &  0.946(40)   &  1.1300(58)   \\
$L_{2R}$           & 0.923(32)     &  1.0850(61) &  ---       & 0.630(18)  &  1.114(43)   &  1.3325(61)   \\
$L_{2I}$           & 1.185(35)     &  1.3902(69) &  ---       & 0.759(19)  &  1.395(44)   &  1.6874(60)   \\
$l_{3U}$           &  ---          &  ---        & 0.1407(30) & ---        &  ---         &   ---         \\ 
$l_{3B}$           & 0.1060(35)    &  0.0481(16) & 0.1409(30) & 0.0894(63) &  ---         &   ---         \\
$l_{3V}$           & 0.1048(34)    &  0.0463(18) & 0.1398(29) & 0.0672(60) &  ---         &   ---         \\
$l_{3R}$           & 0.1073(37)    &  0.0456(18) & ---        & 0.0374(58) &  ---         &   ---         \\
$l_{3I}$           & 0.0957(37)    &  0.0386(16) & ---        & 0.0304(74) &  ---         &   ---         \\ \hline
$\phi[^o]$         &  90.0$^*$     &  90.0$^*$   & 90.0$^*$   & 90.0$^*$   &  90.0$^*$    &  90.0$^*$     \\
$\lambda[^o]$      &  7.61(29)     &  19.66(43)  & 11.83(35)  & 64.9(3.8)  &  11.49(36)   &  3.00(0.45)   \\
$r[^o]$            & 104.1(2.1)    &  65.3(1.3)  & 107.9(1.8) & 102.9(4.4) &  113.9(1.9)  &  19.53(0.91)  \\
$T_{spot}/T_{eff}^{sec}$&1.2051(59)&  1.2053(33) & 1.141(4)   & 1.1015(55) &  1.2138(72)  &  1.177(15)    \\ \hline\hline
}

\begin{figure}
\centerline{%
\begin{tabular}{c@{\hspace{1pc}}c}
\includegraphics[width=1.8in,angle=-90]{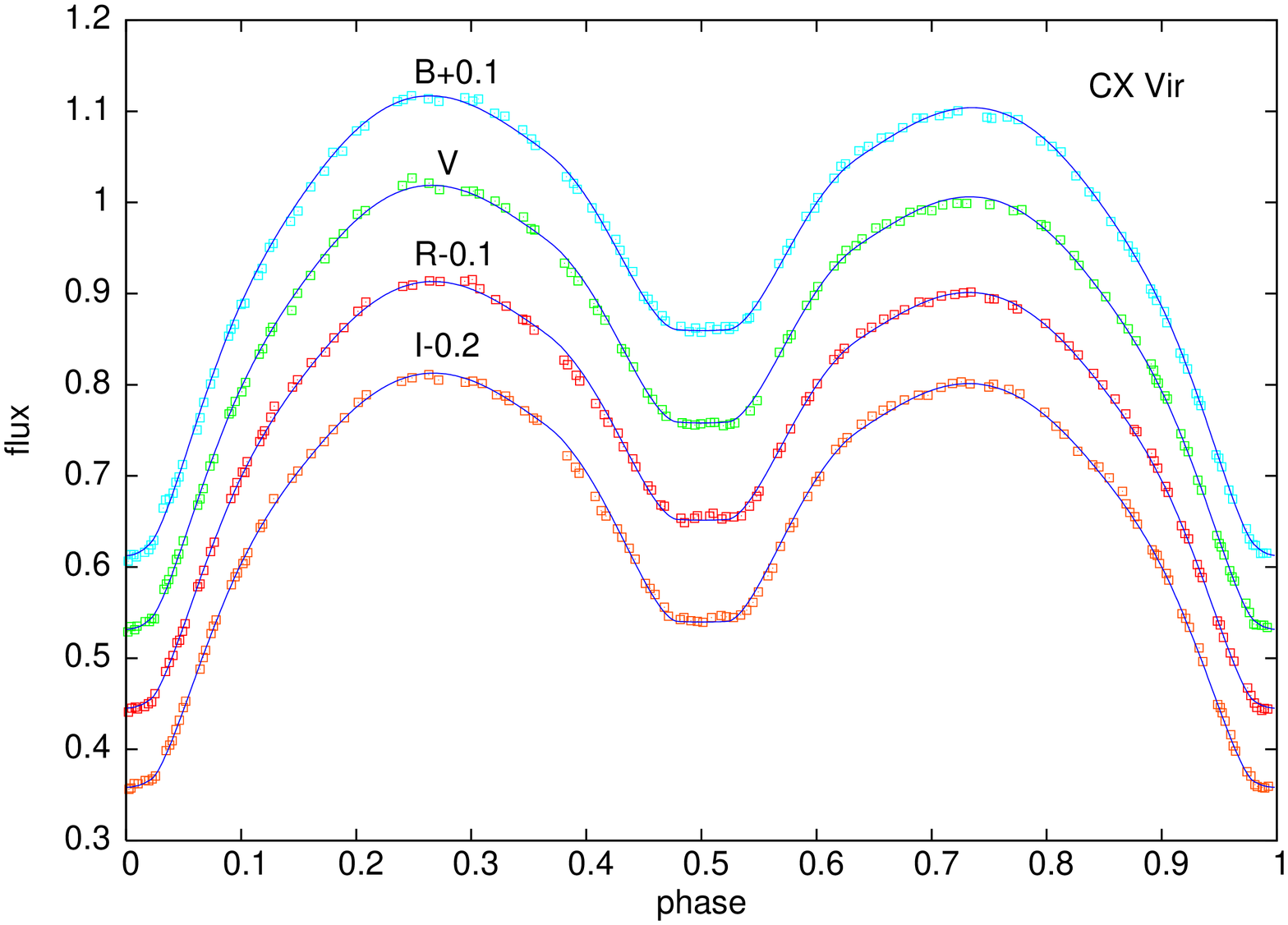}&
\includegraphics[width=1.8in,angle=-90]{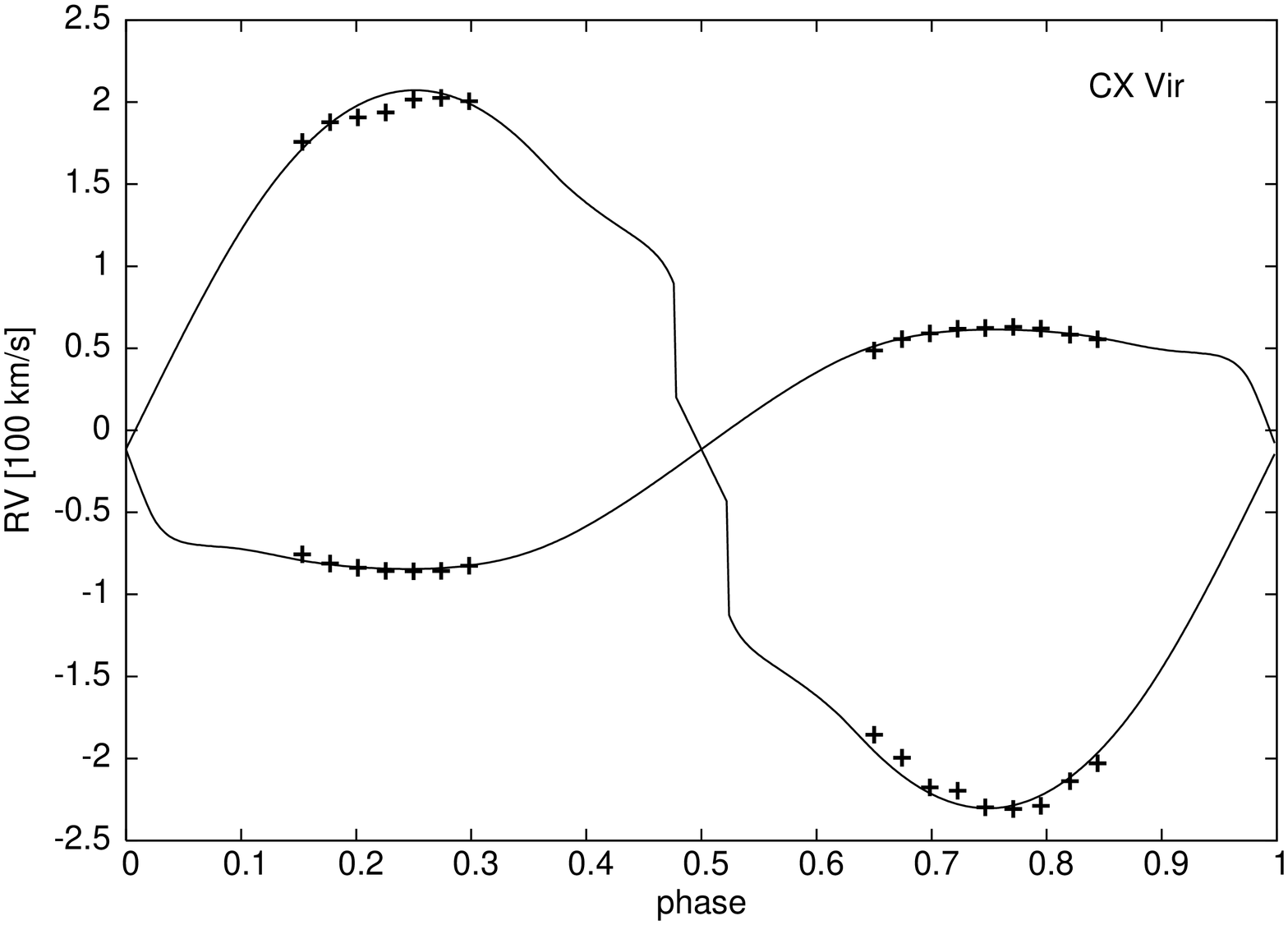} \\
\includegraphics[width=1.8in,angle=-90]{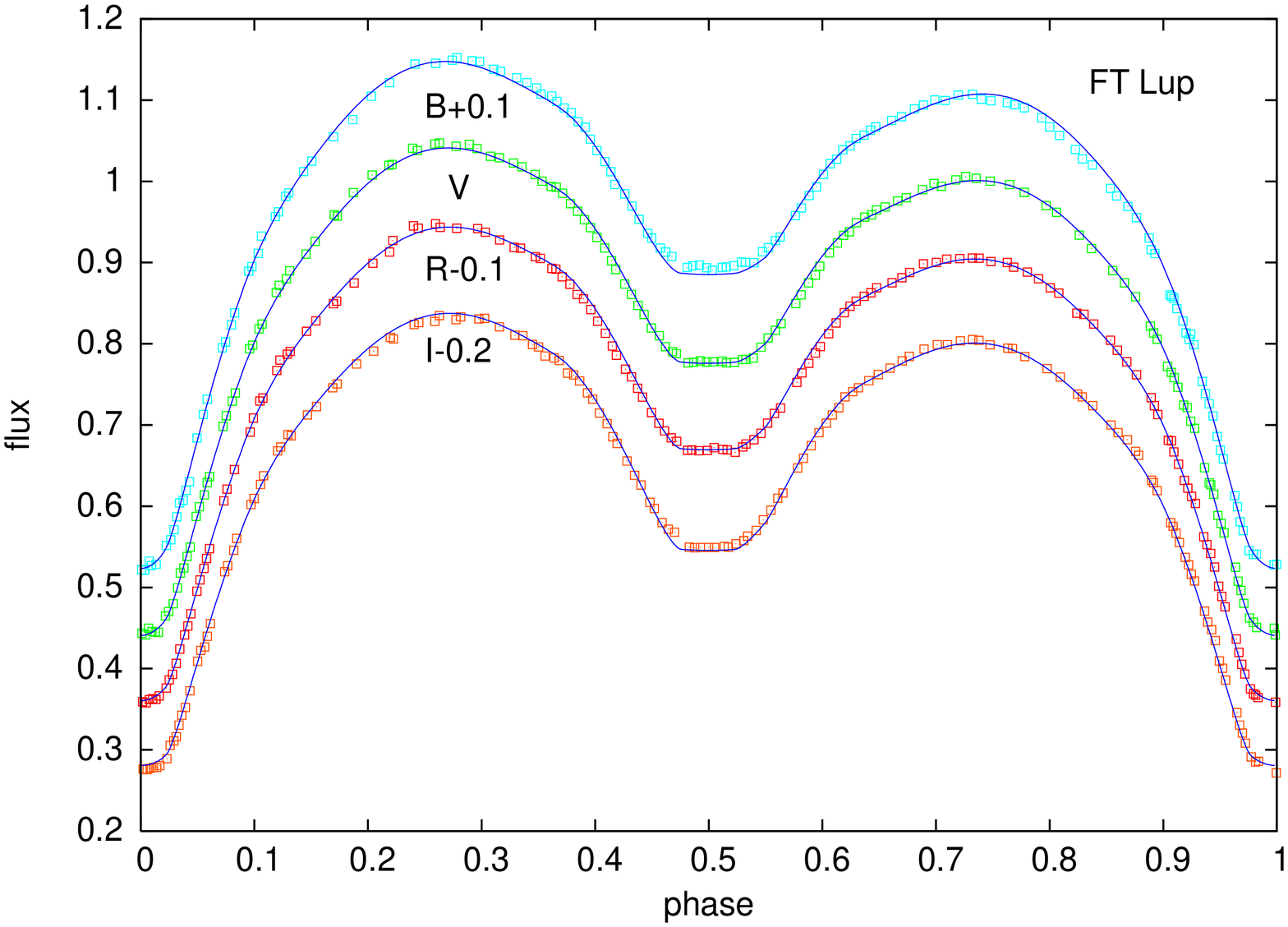}&
\includegraphics[width=1.8in,angle=-90]{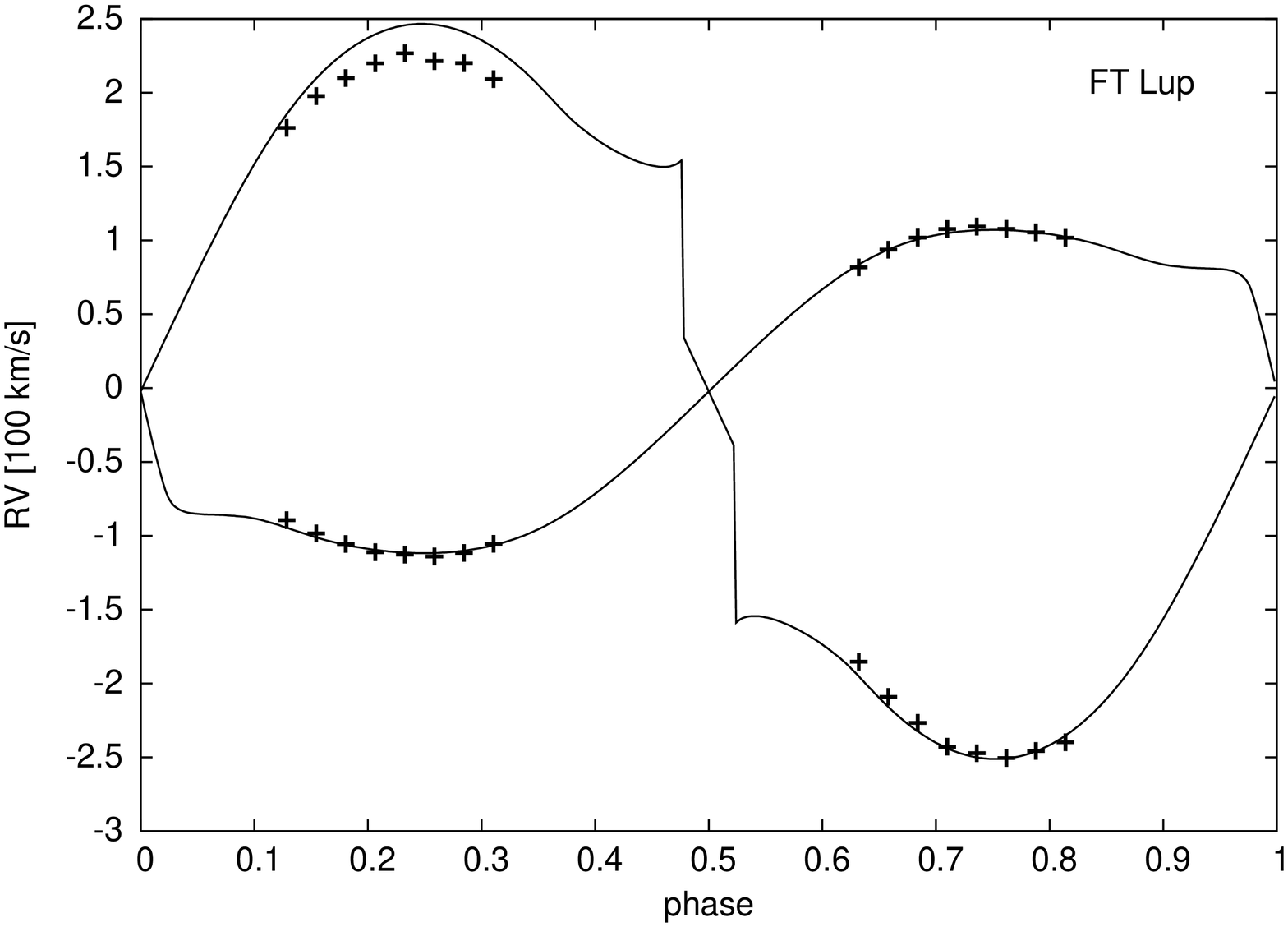}\\
\includegraphics[width=1.8in,angle=-90]{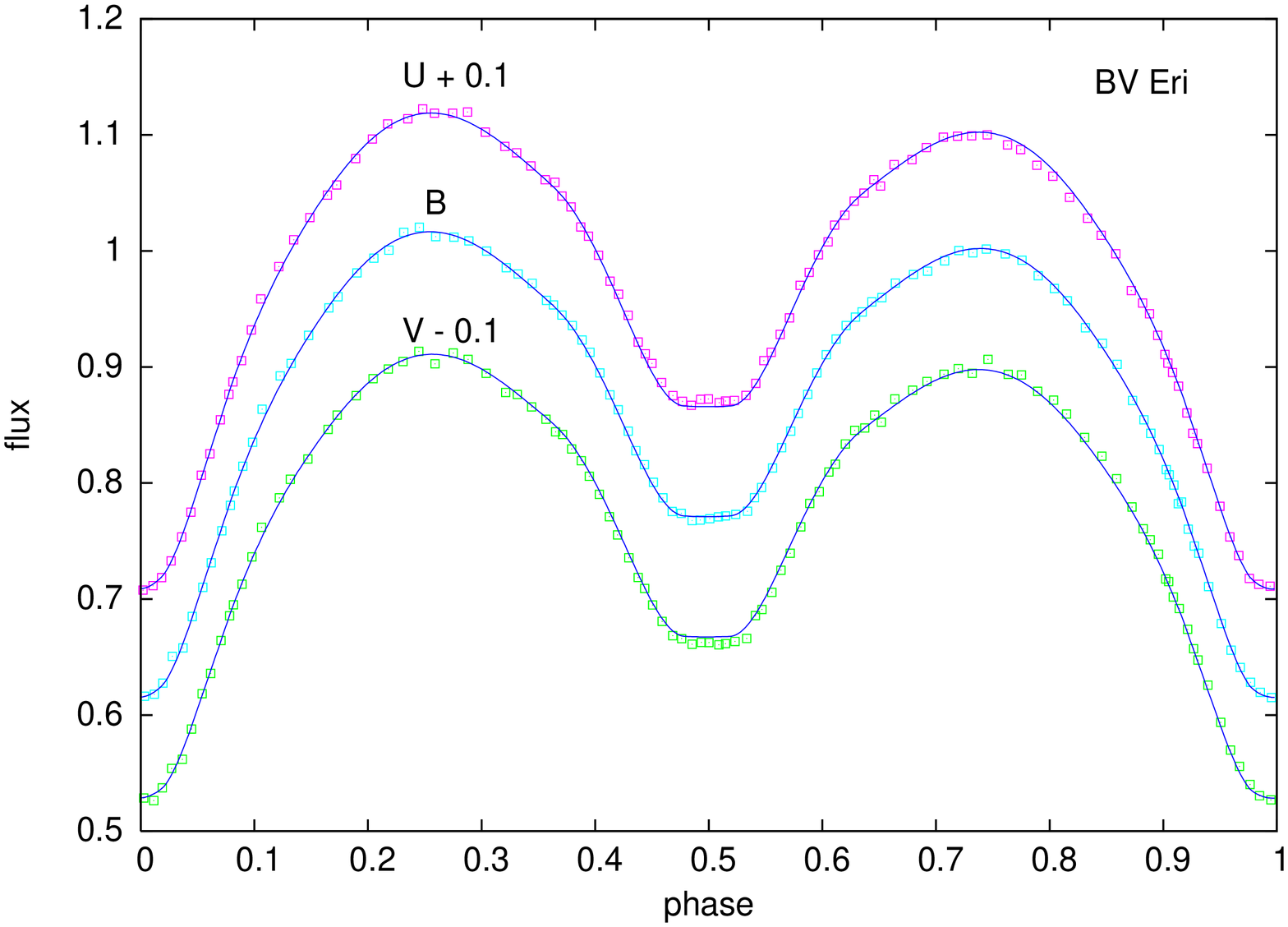}&
\includegraphics[width=1.8in,angle=-90]{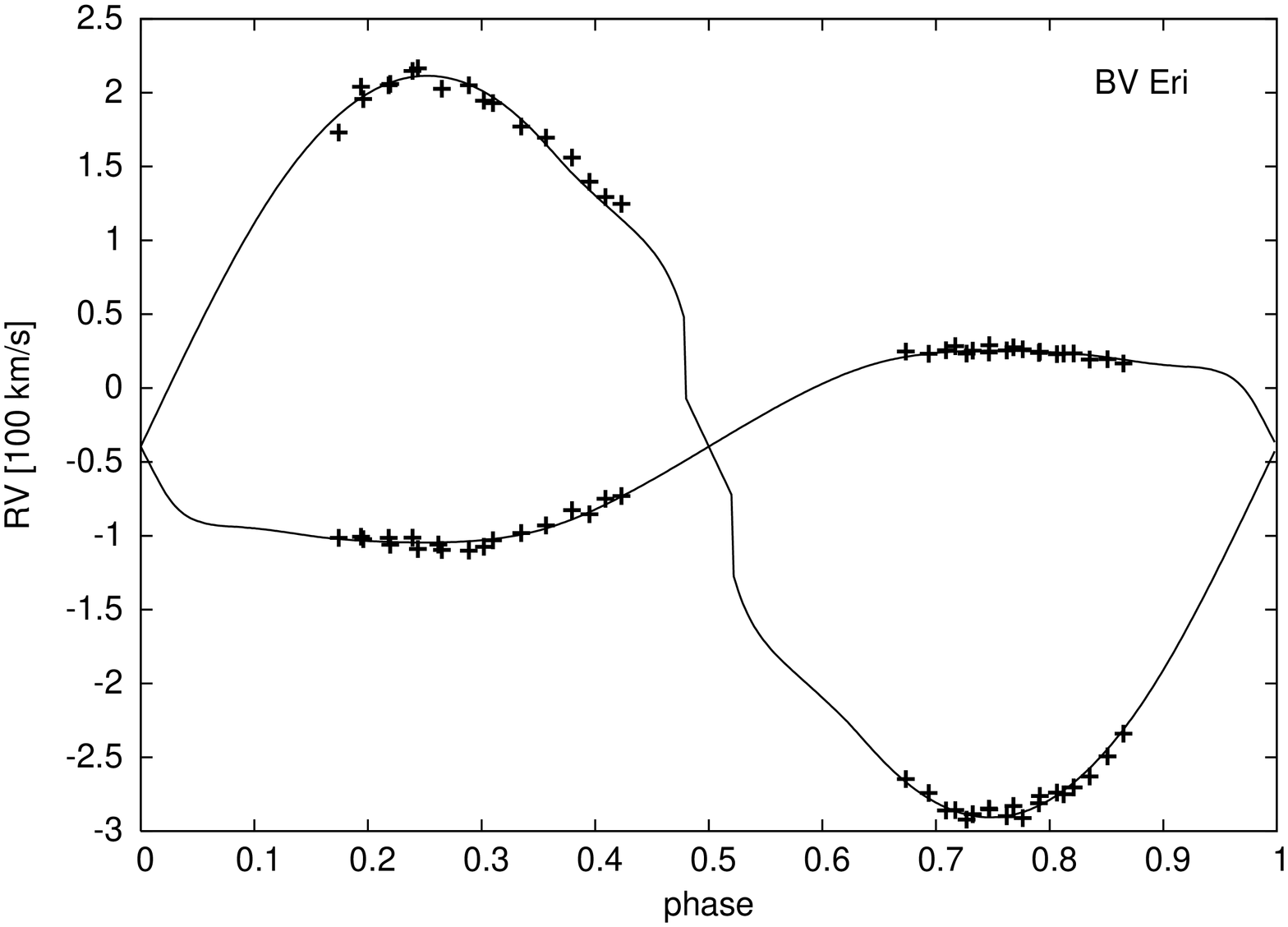}\\
\includegraphics[width=1.8in,angle=-90]{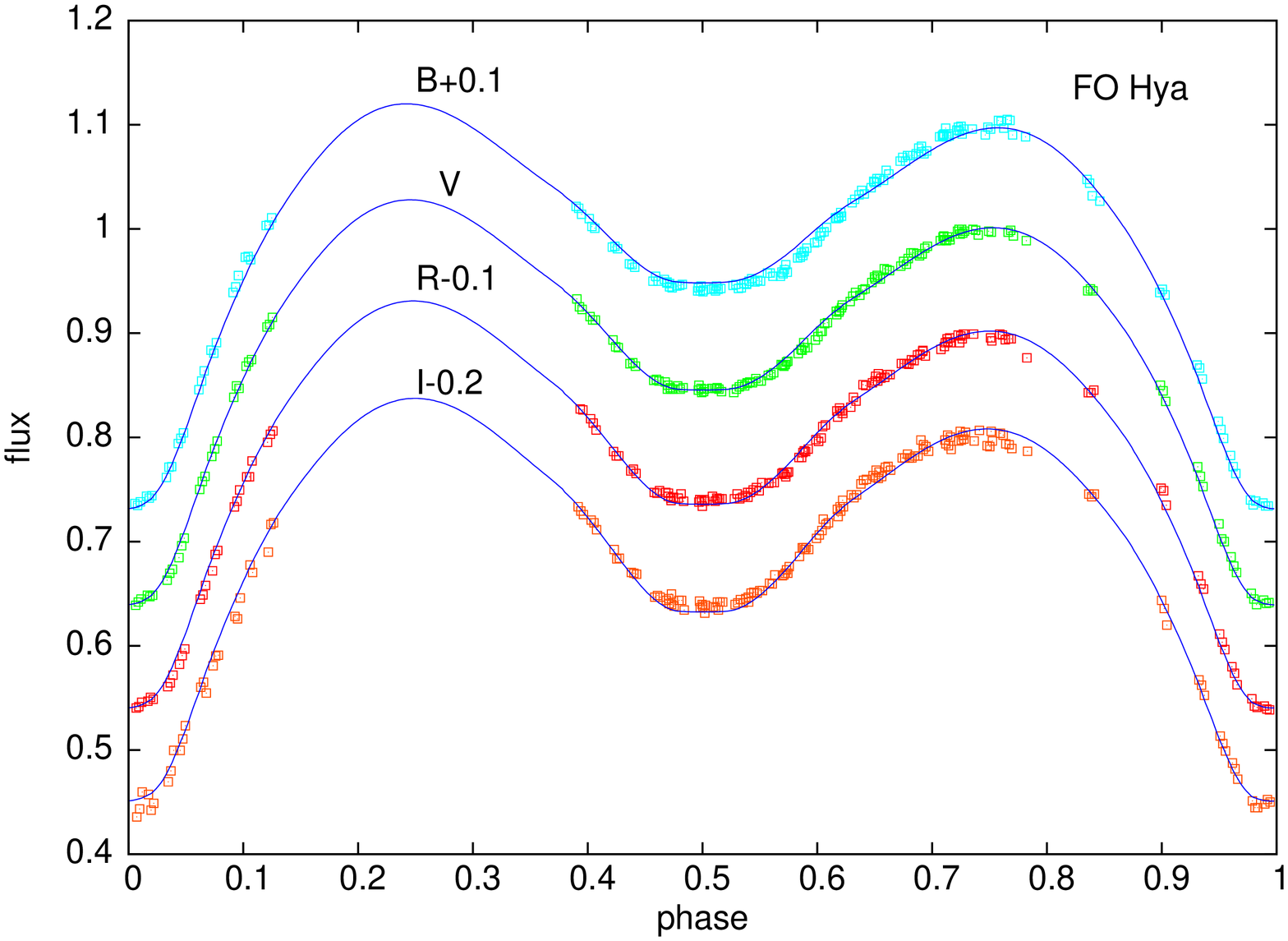}&
\includegraphics[width=1.8in,angle=-90]{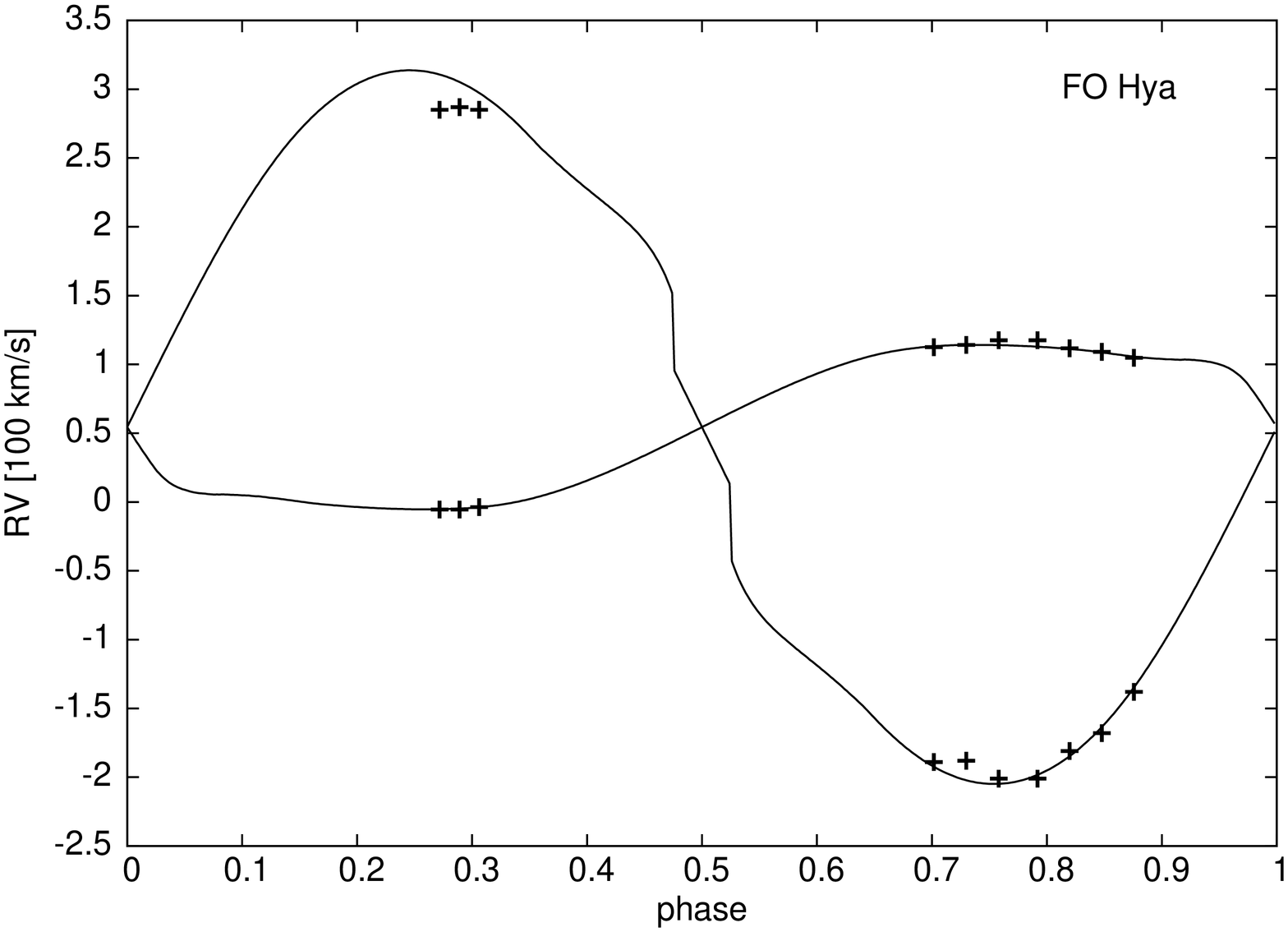}
\end{tabular}}
\vspace{0.5pc}
\FigCap{Comparison between the observed and synthetic photometric (left panels) 
and RV (right panels) curves of CX~Vir, FT~Lup, BV~Eri and FO~Hya.}
\end{figure}

\begin{figure}[h]
\centerline{%
\begin{tabular}{c@{\hspace{1pc}}c}
\includegraphics[width=1.8in,angle=-90]{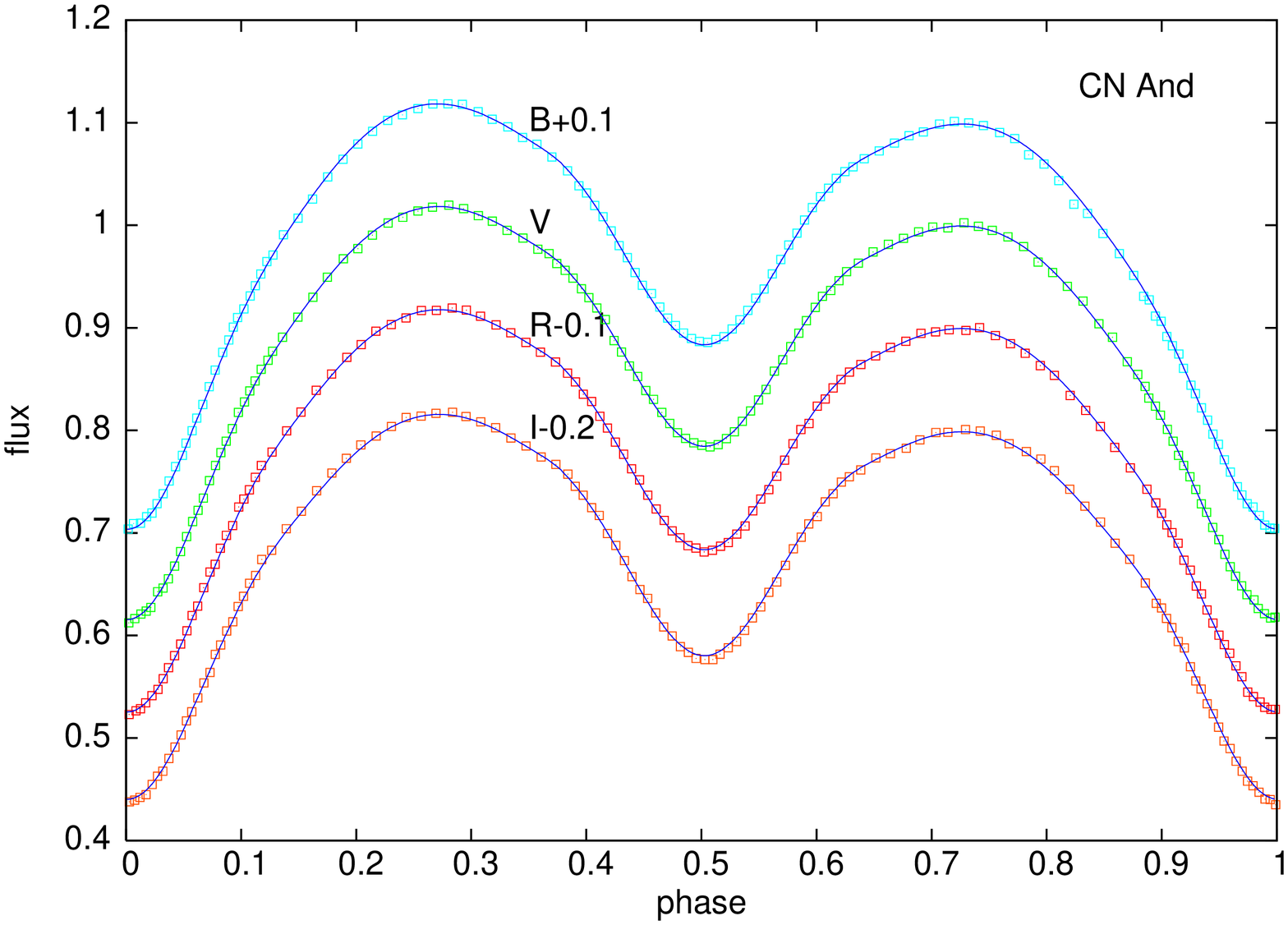}&
\includegraphics[width=1.8in,angle=-90]{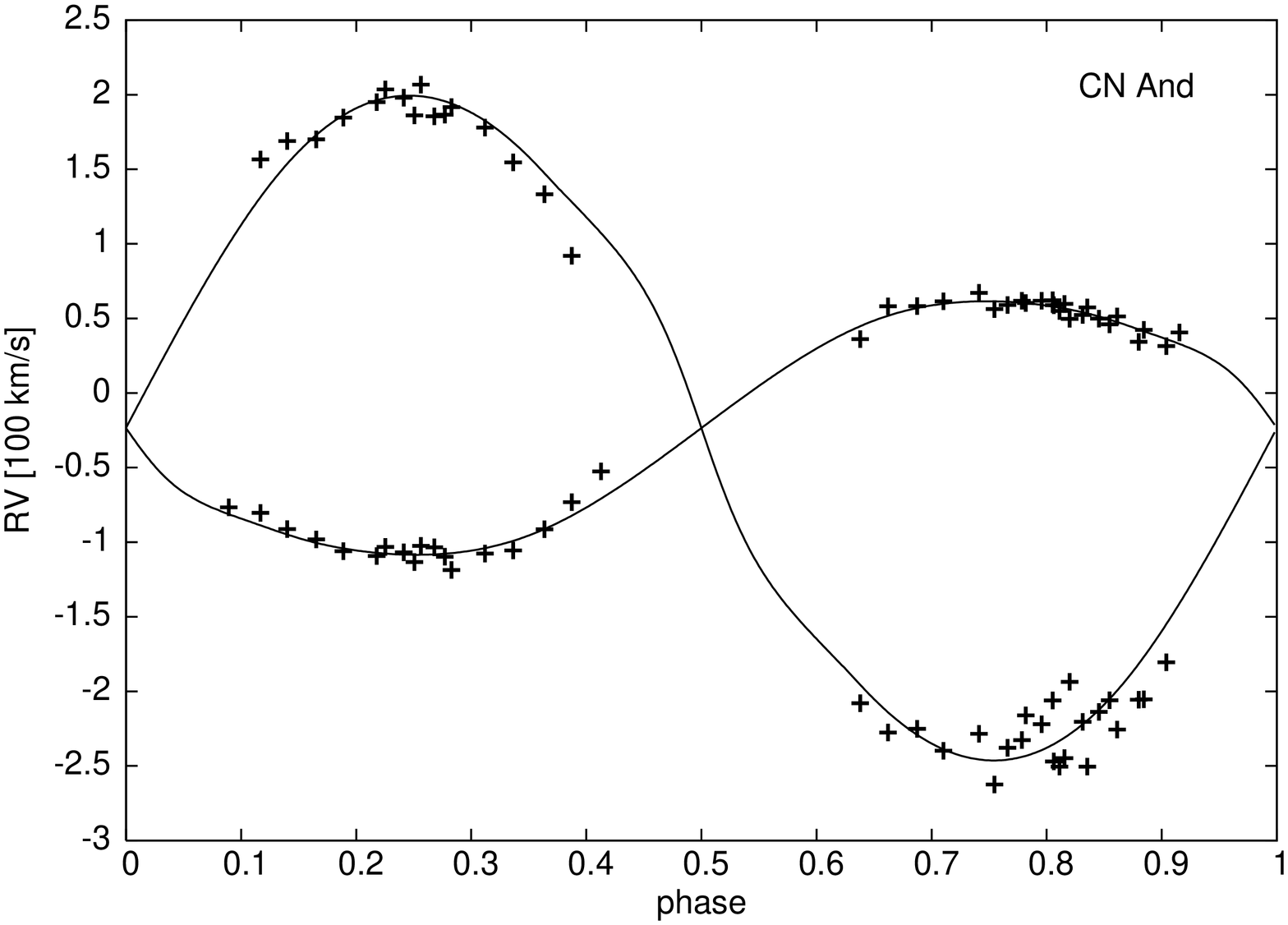}\\
\includegraphics[width=1.8in,angle=-90]{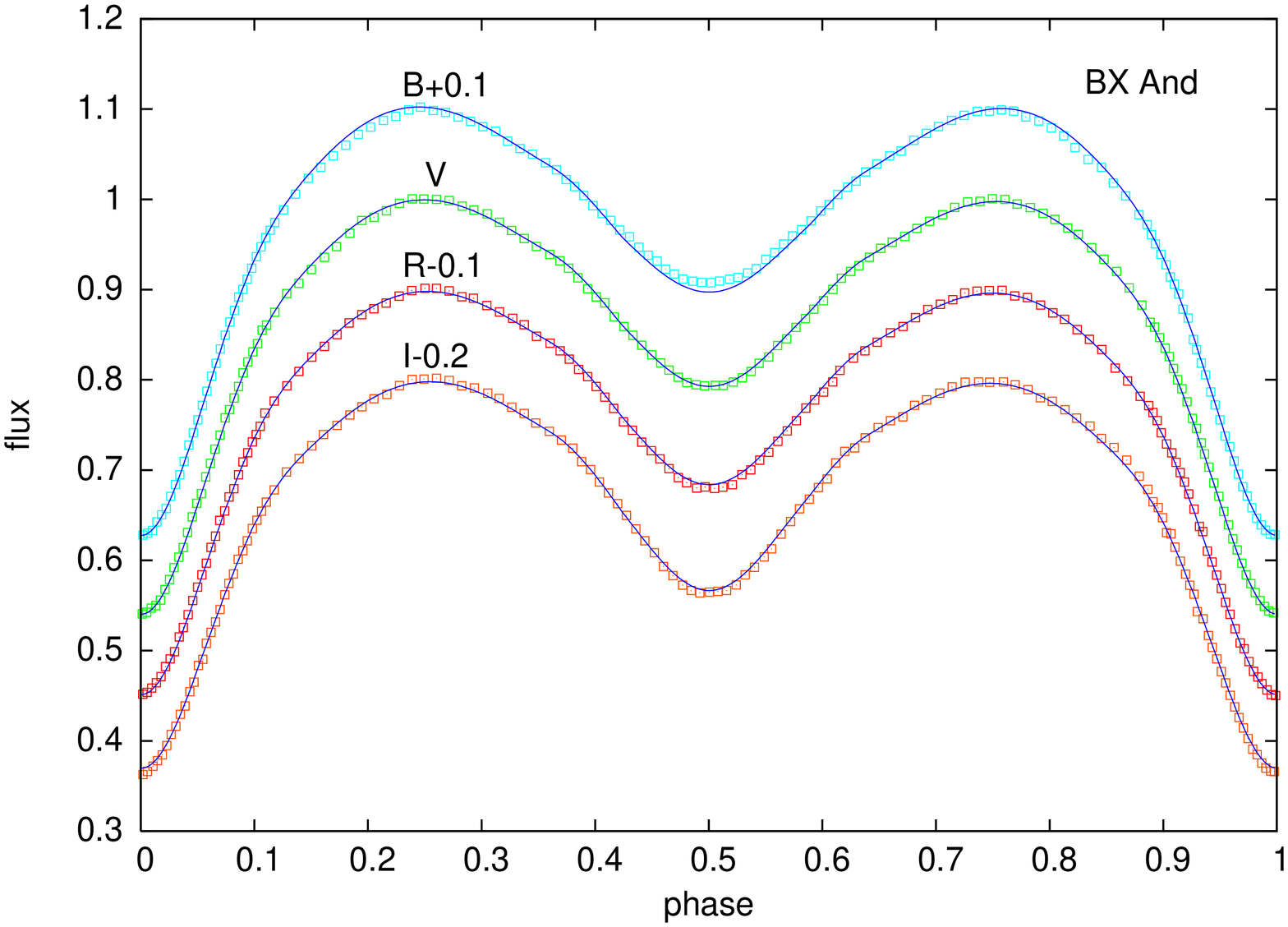}&
\includegraphics[width=1.8in,angle=-90]{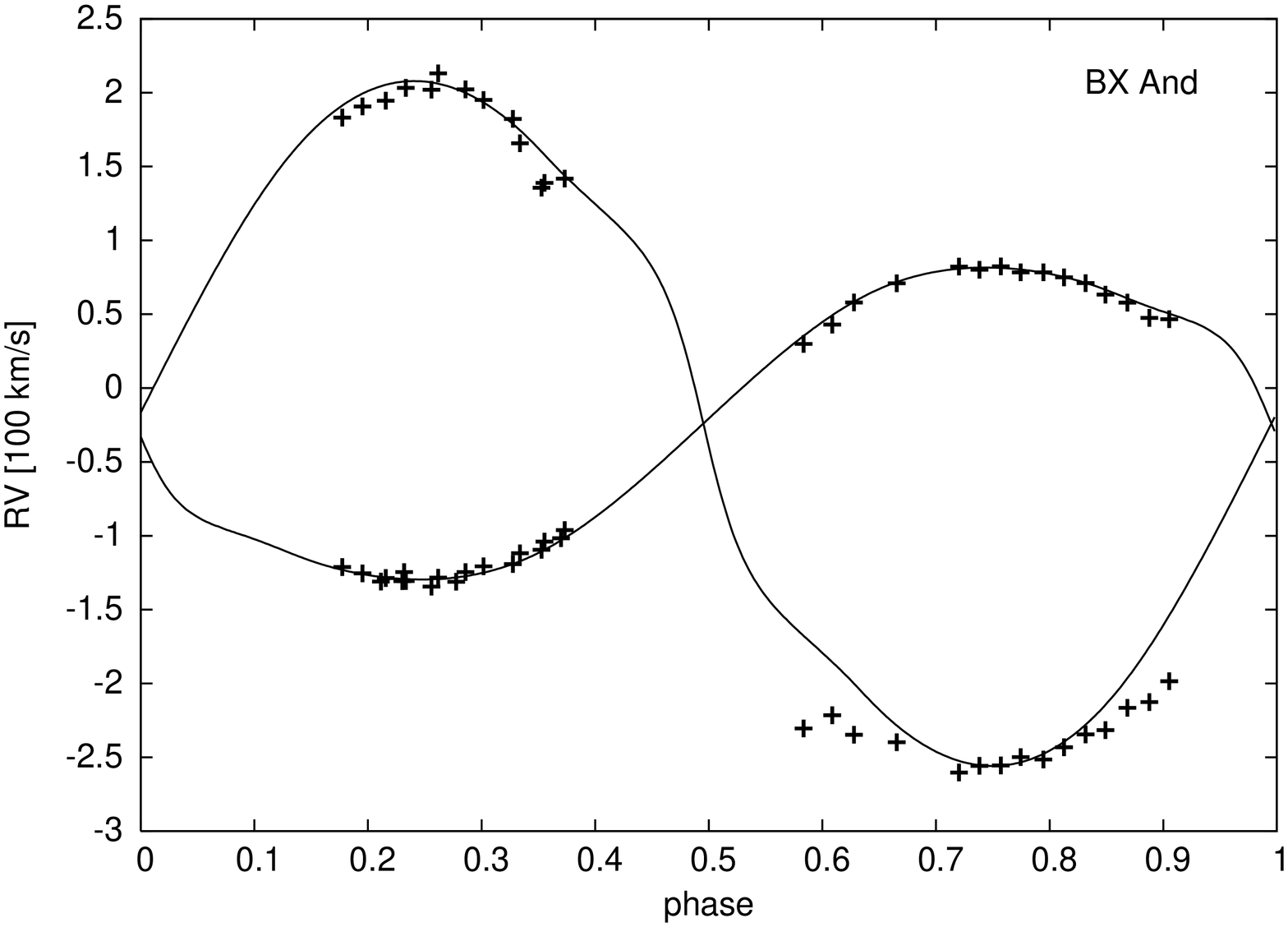}
\end{tabular}}
\vspace{0.5pc}
\FigCap{Comparison between the observed and synthetic photometric
and RV curves of CN~And and BX~And.}
\end{figure}

\section {Contact systems with large $\Delta T$}

\subsection{CX Vir}

In the high-quality BFs presented
in Fig.~8, we notice even stronger
differences between 
secondary-component 
velocity profiles obtained in
first and second quadrature than
for the semi-detached system 
V747~Cen (Sec.~4.6).  At phases
between 0.6 and 0.7, the triangular 
bright region (indicated by an
arrow on the BF obtained at phase
0.747 in Fig.~8) dominates the
signal of the secondary component.  The
latter begins to be distinguished from
the brighter area in the BF obtained in
phase 0.747 (Fig.~8, doubly underlined
area on the same plot).
The bright region becomes narrower
and fainter with phase, but 
remains visible even at phase 0.844. 
BFs of the secondary component
obtained in first quadrature
are more intense (brighter) in the
vicinity of the primary component
(underlined area at phase 0.202),
exactly as with V747~Cen.
This similarly suggests an ongoing
mass transfer from the primary to the
secondary.
In Fig.~9, a few cold spots
(i.e., the darker coherent trails) 
on the primrary are visible as well, 
outside the well defined bright
region on the secondary star (indicated
by longer marks). 

As the bright region on
the secondary component introduces the
largest effects in the BFs, we assumed
one circular hot spot on the equator
of the secondary star during 
light-curve modelling.  We expected to obtain
a semi-detached configuration as in the
case of V747~Cen, but in fact we 
obtained a contact configuration, with
a large temperature difference between
components and a significant value of
third light $l_3$, comprising about 10\% of the
total flux (Table~7).  Hilditch \& King
(1988) did not account for a third
light in their analysis, as it is not
obvious from their data whether the
secondary minimum is total or partial. 
This apparently small detail always, 
however, has a large impact on the final 
light-curve solution.

\begin{figure}
\centerline{%
\begin{tabular}{c@{\hspace{4pc}}c@{\vspace{-0.7pc}}}
\includegraphics[width=1.10in,angle=-90]{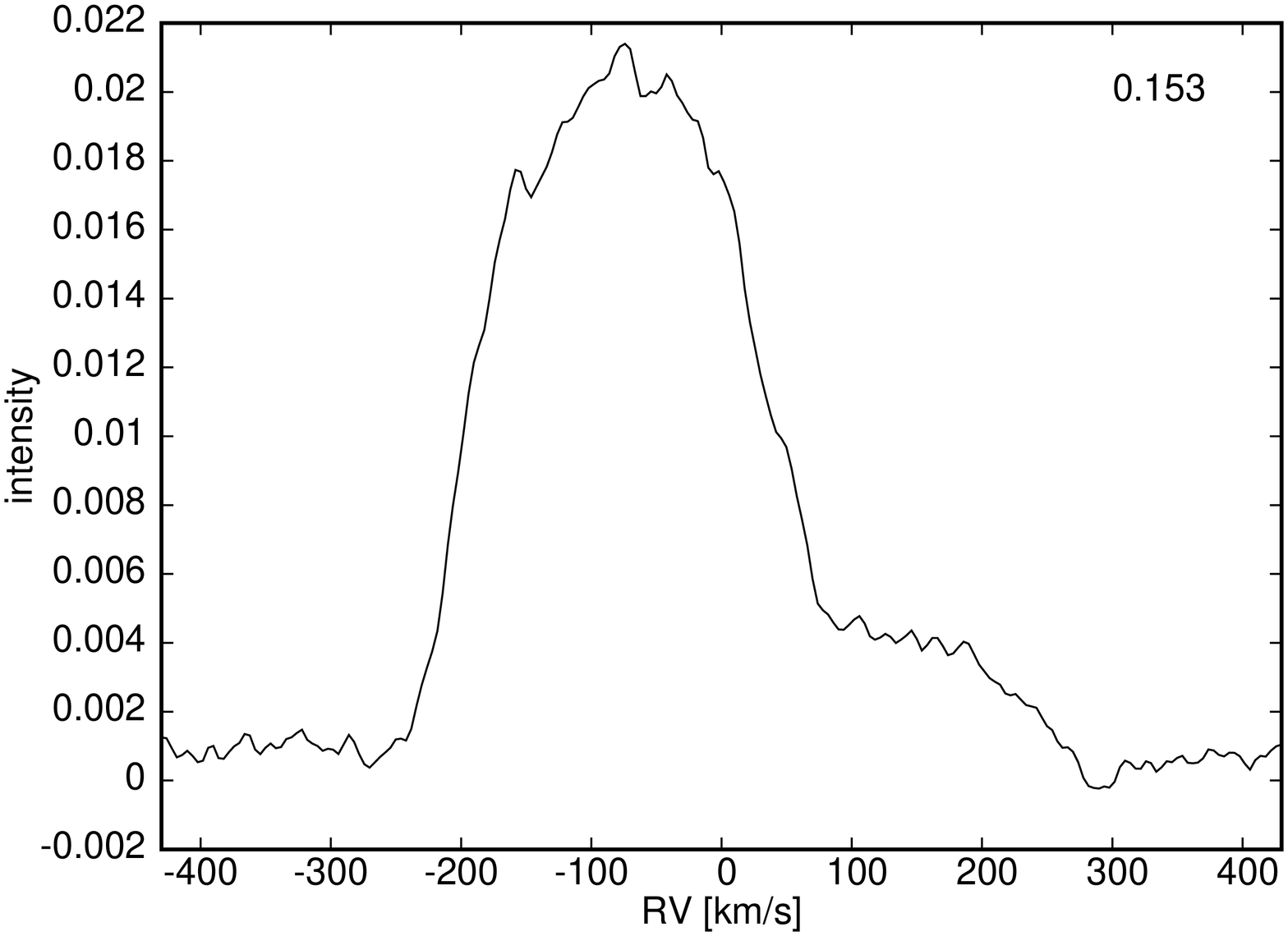}&
\includegraphics[width=1.10in,angle=-90]{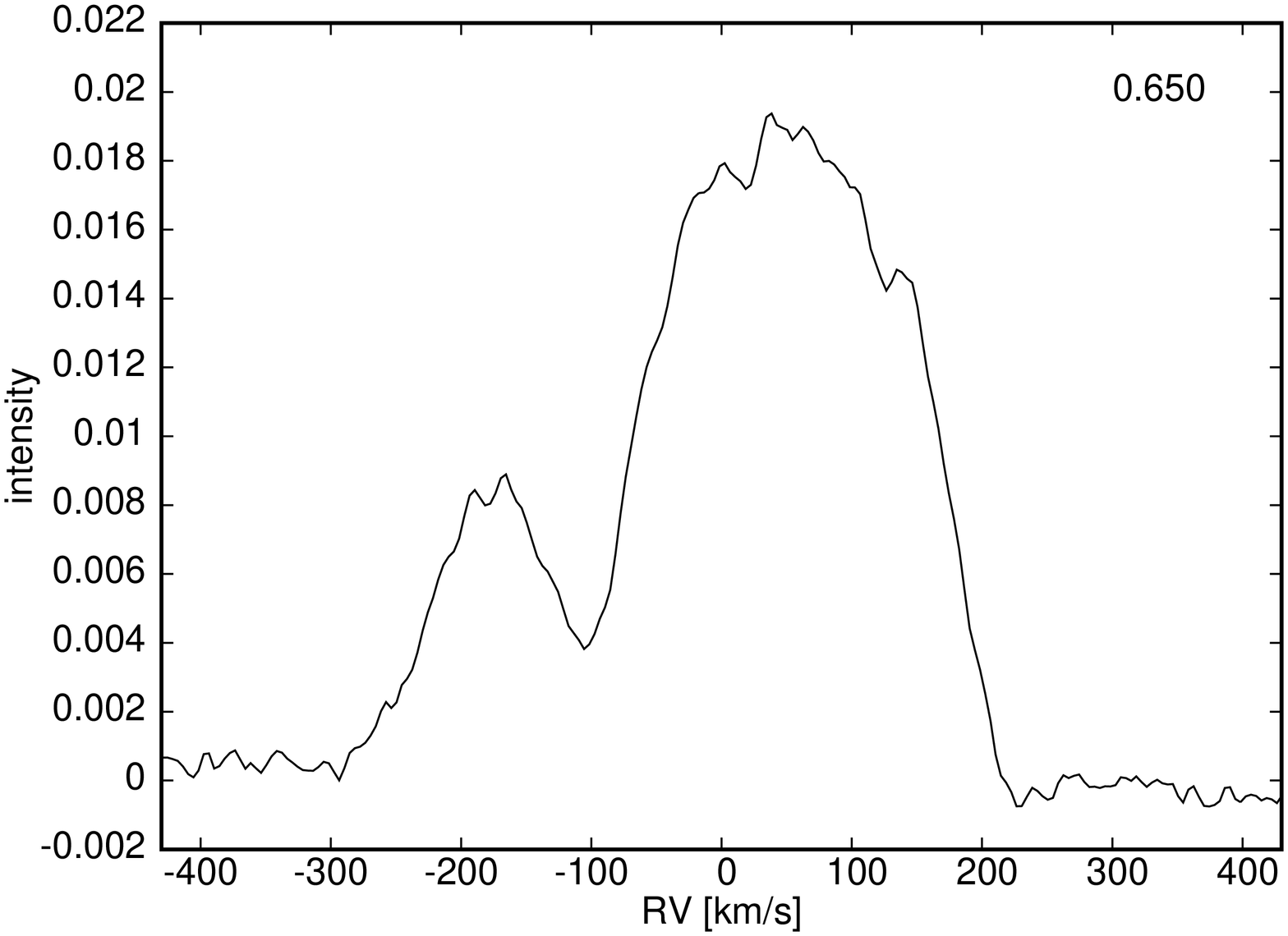}\\
\includegraphics[width=1.10in,angle=-90]{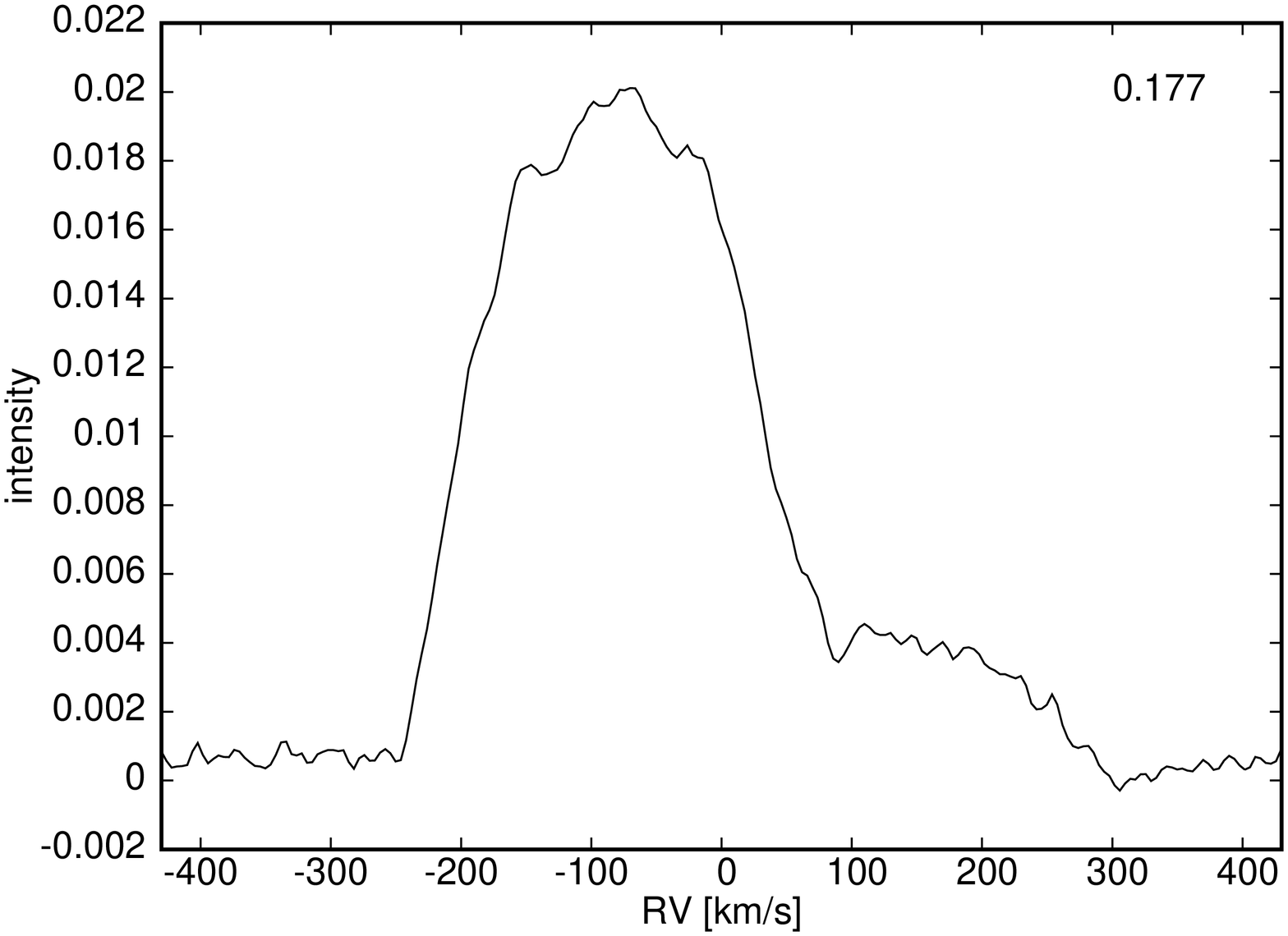}&
\includegraphics[width=1.10in,angle=-90]{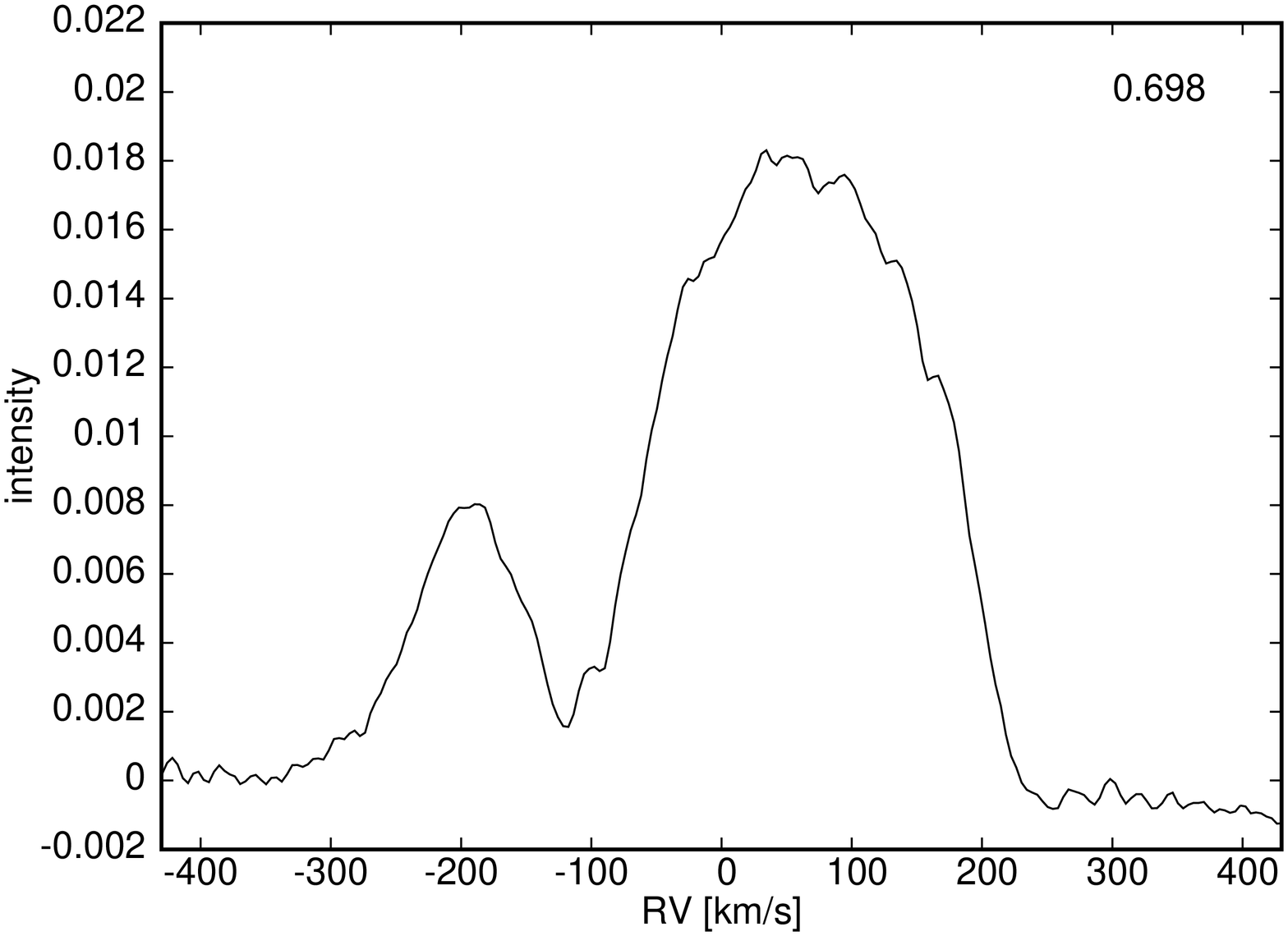}\\
\includegraphics[width=1.10in,angle=-90]{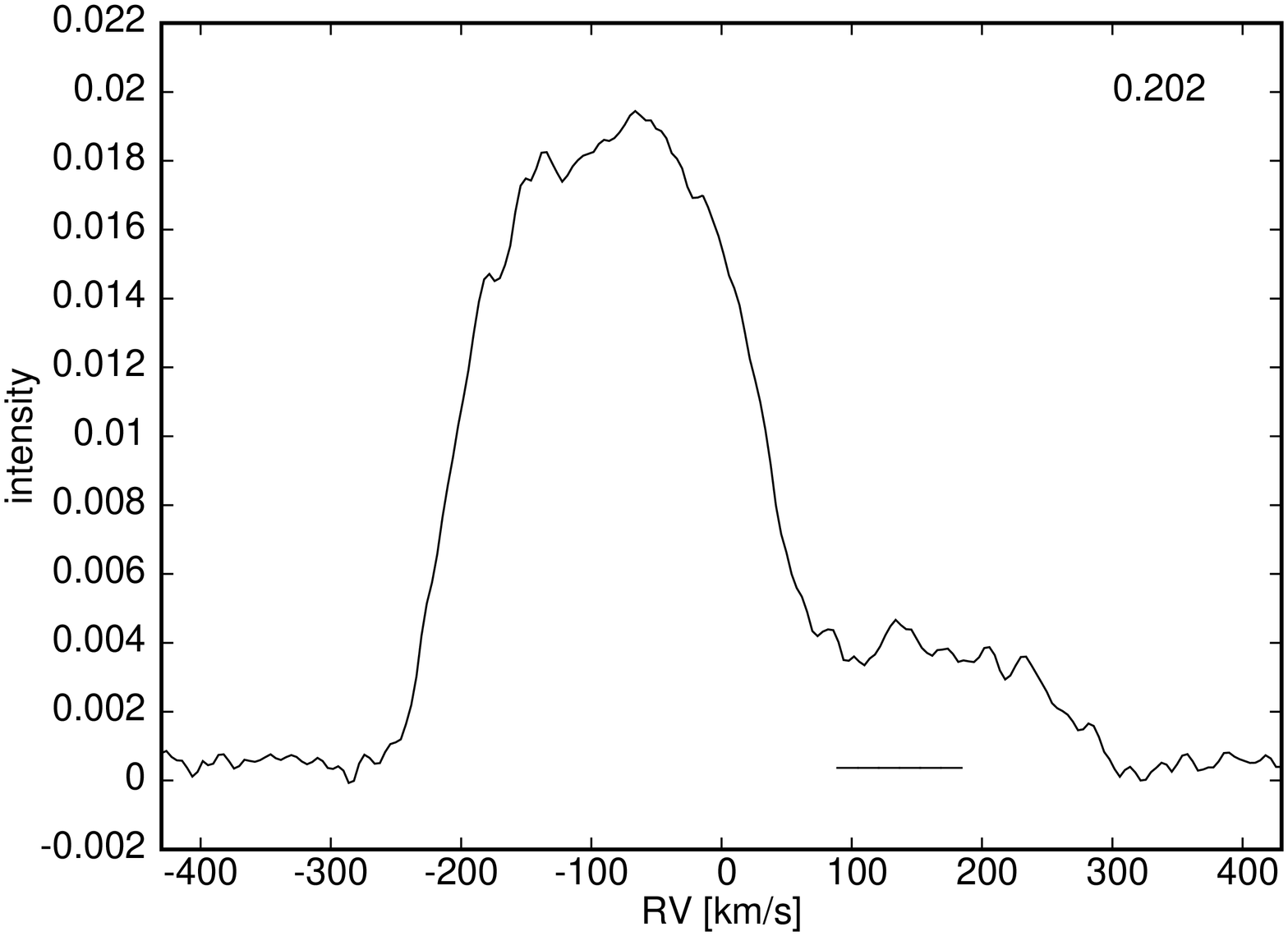}&
\includegraphics[width=1.10in,angle=-90]{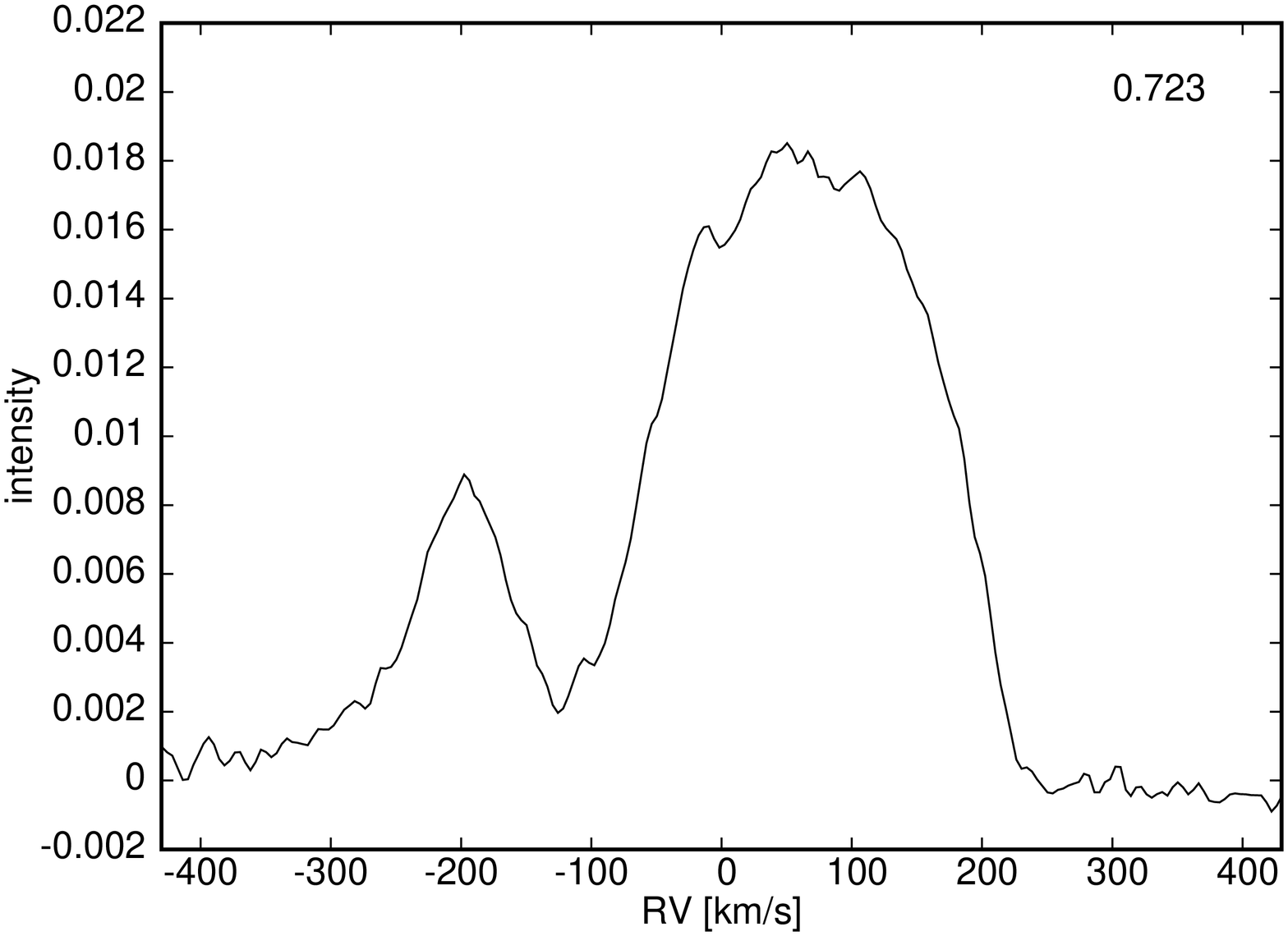}\\
\includegraphics[width=1.10in,angle=-90]{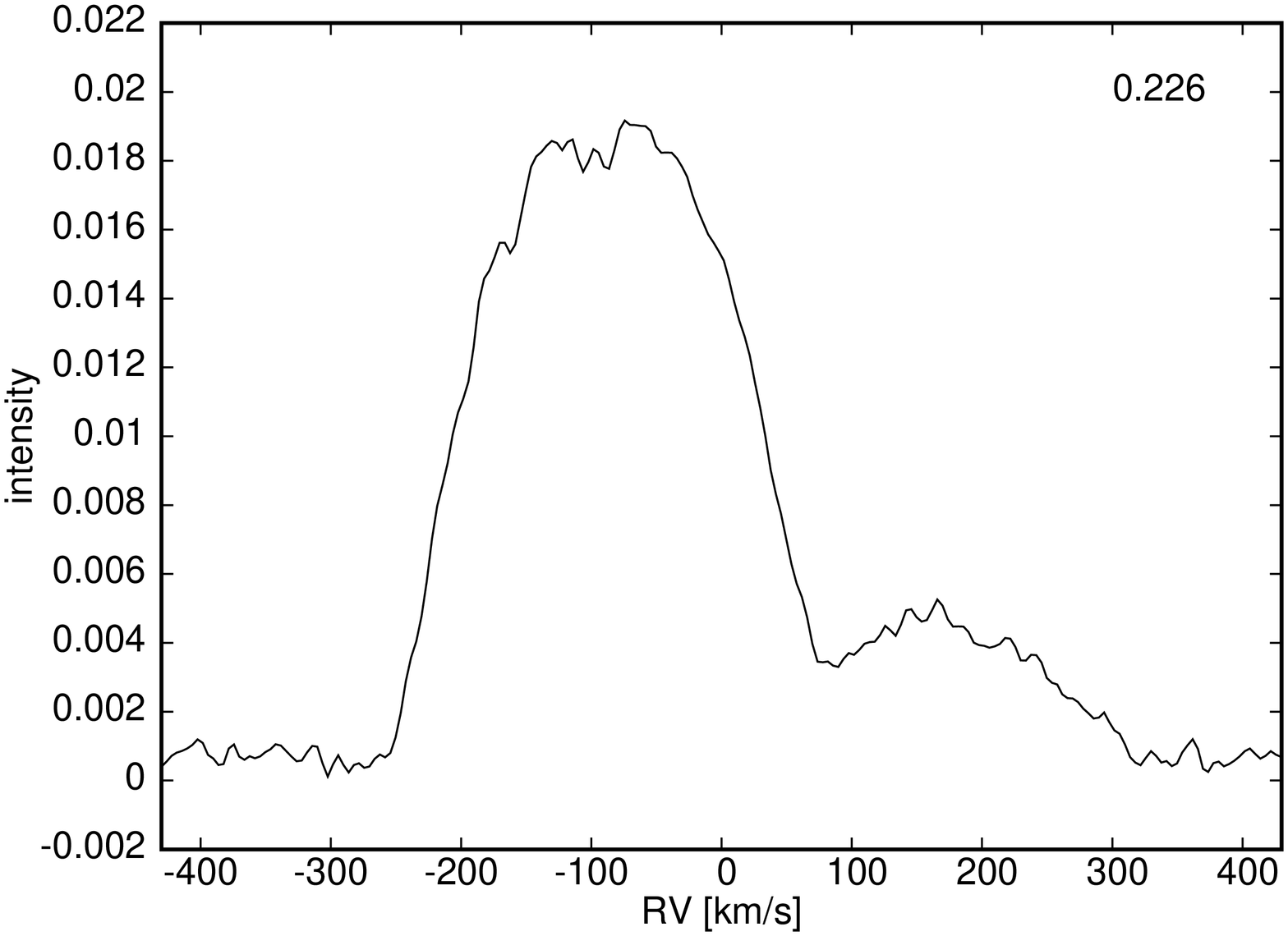}&
\includegraphics[width=1.10in,angle=-90]{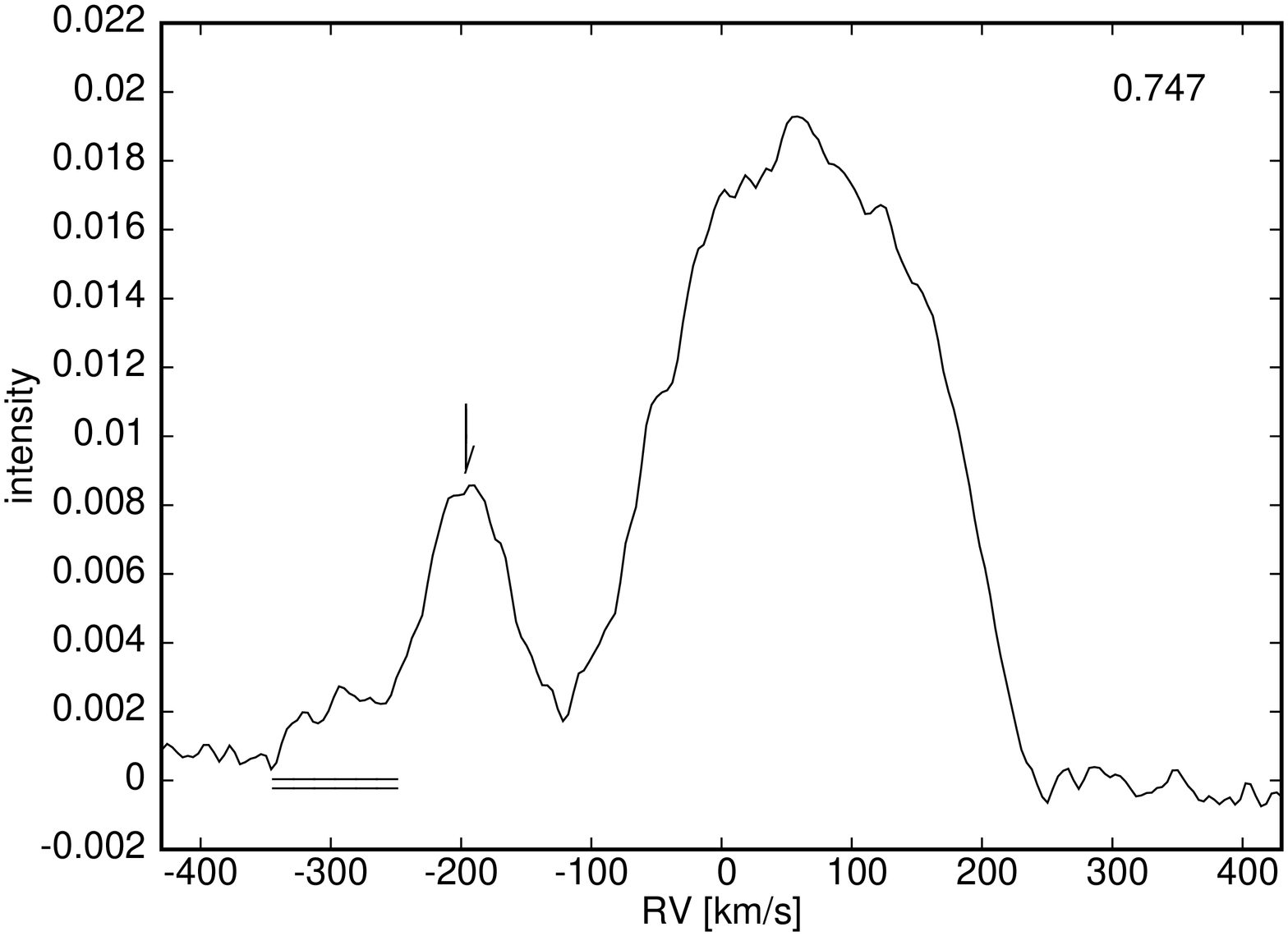}\\
\includegraphics[width=1.10in,angle=-90]{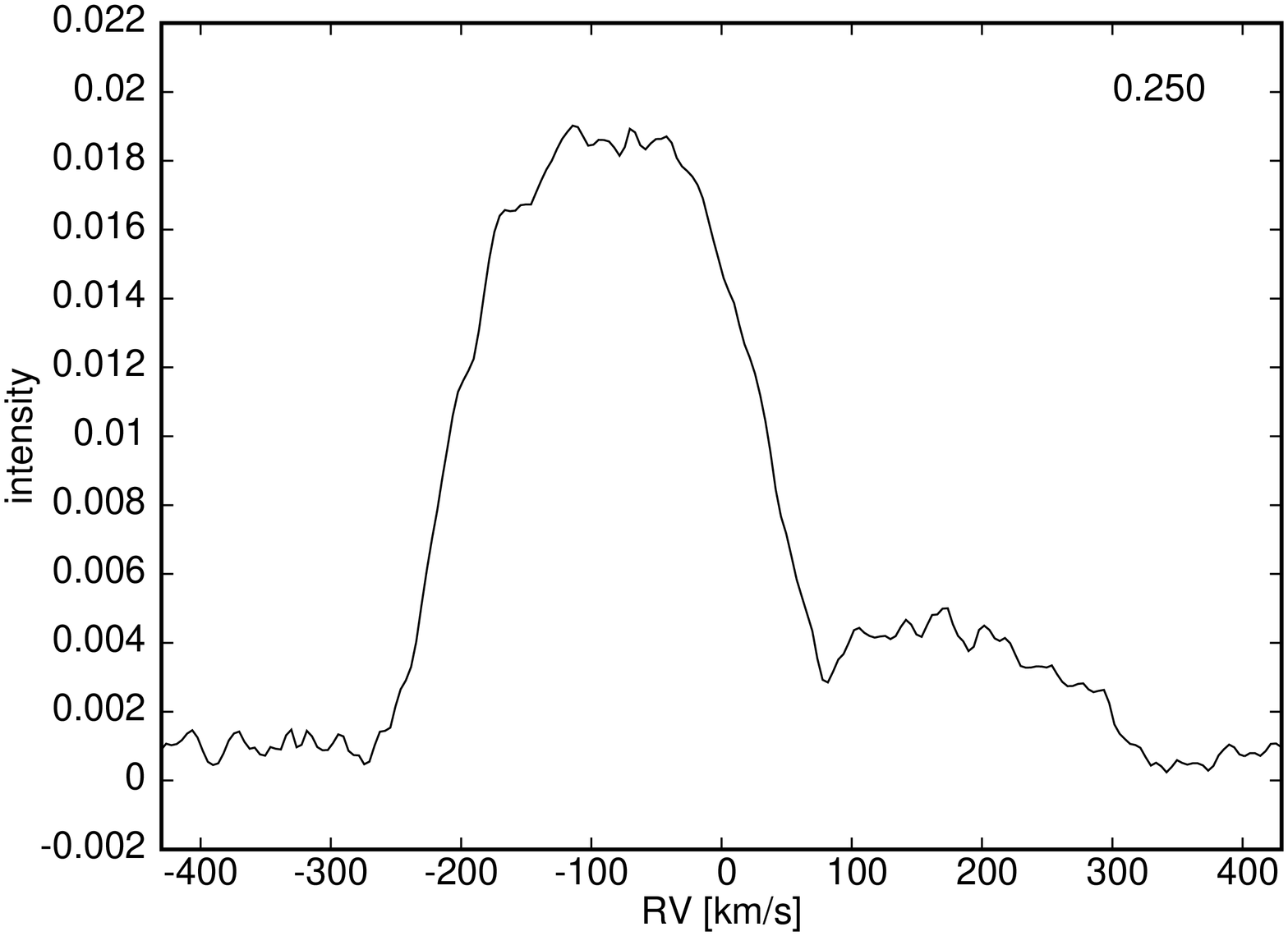}&
\includegraphics[width=1.10in,angle=-90]{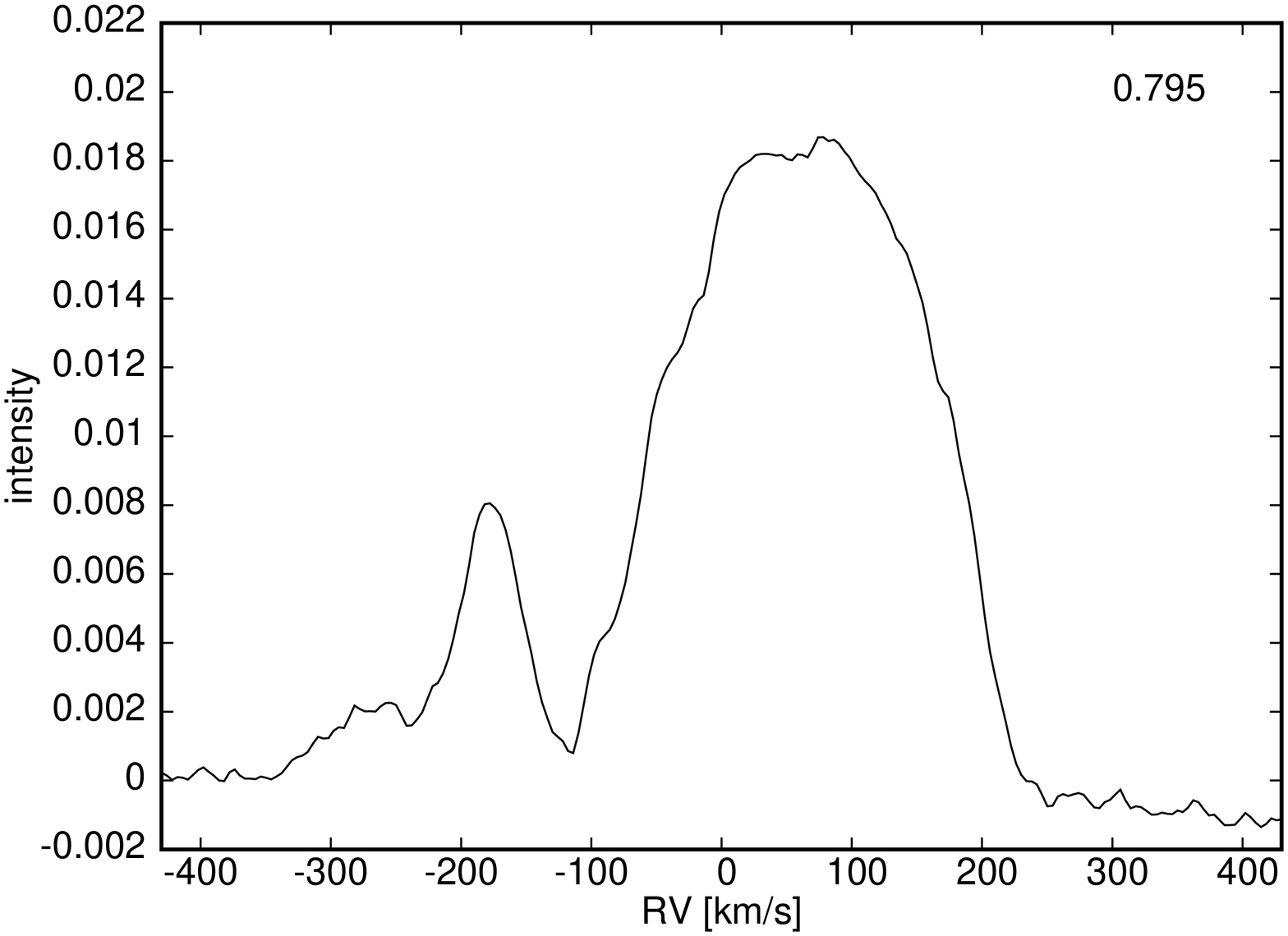}\\
\includegraphics[width=1.10in,angle=-90]{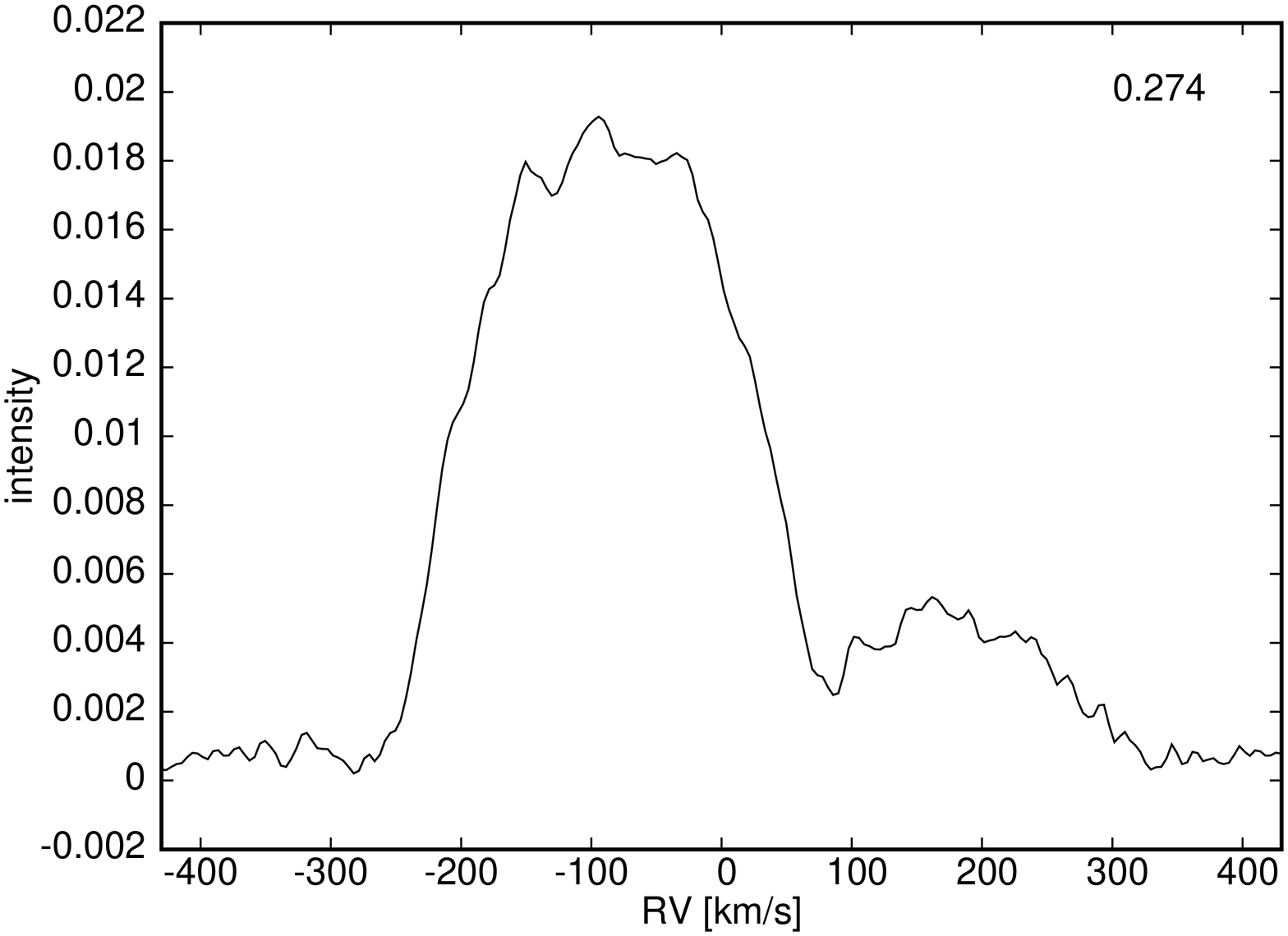}&
\includegraphics[width=1.10in,angle=-90]{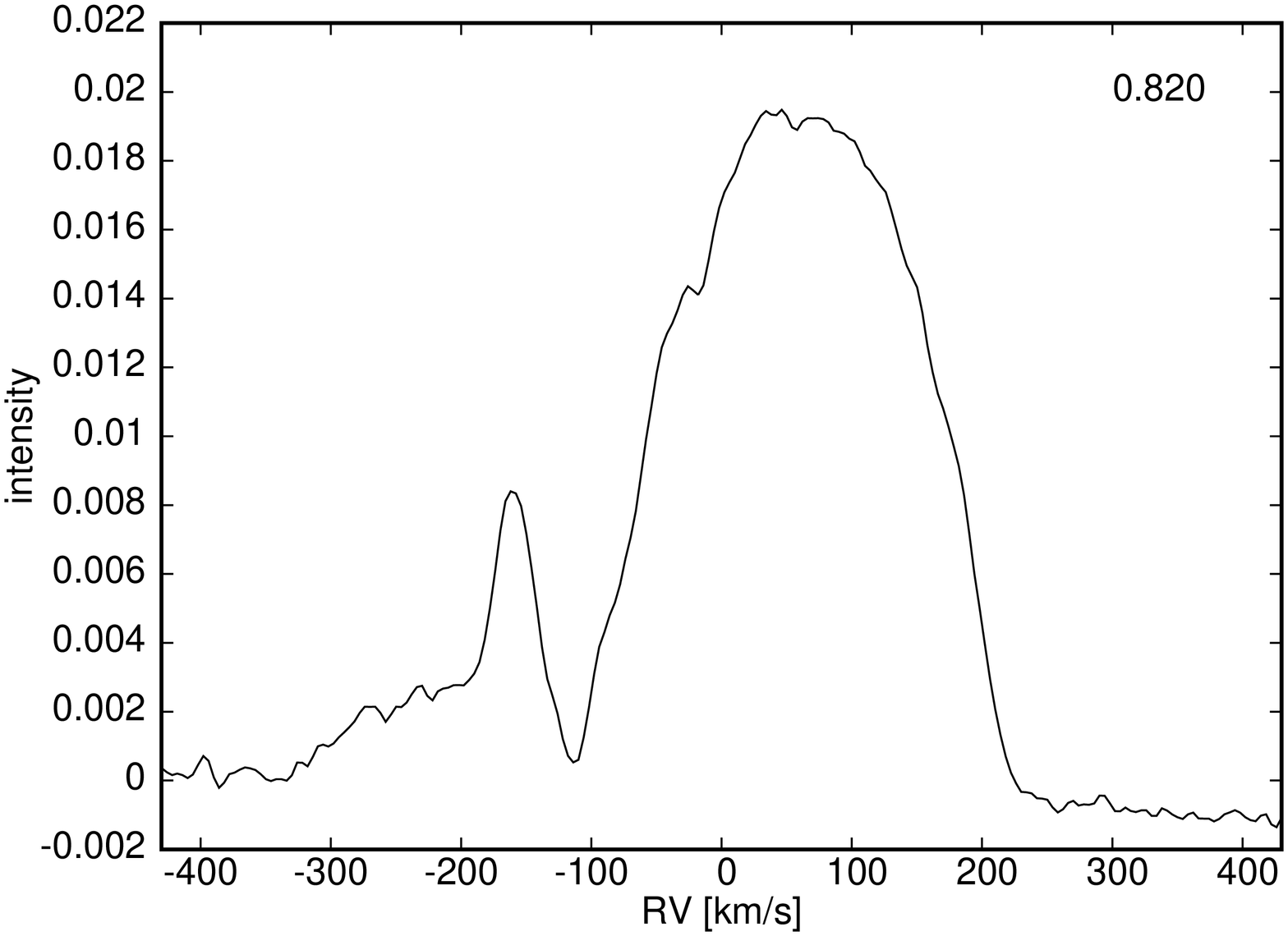}\\
\includegraphics[width=1.10in,angle=-90]{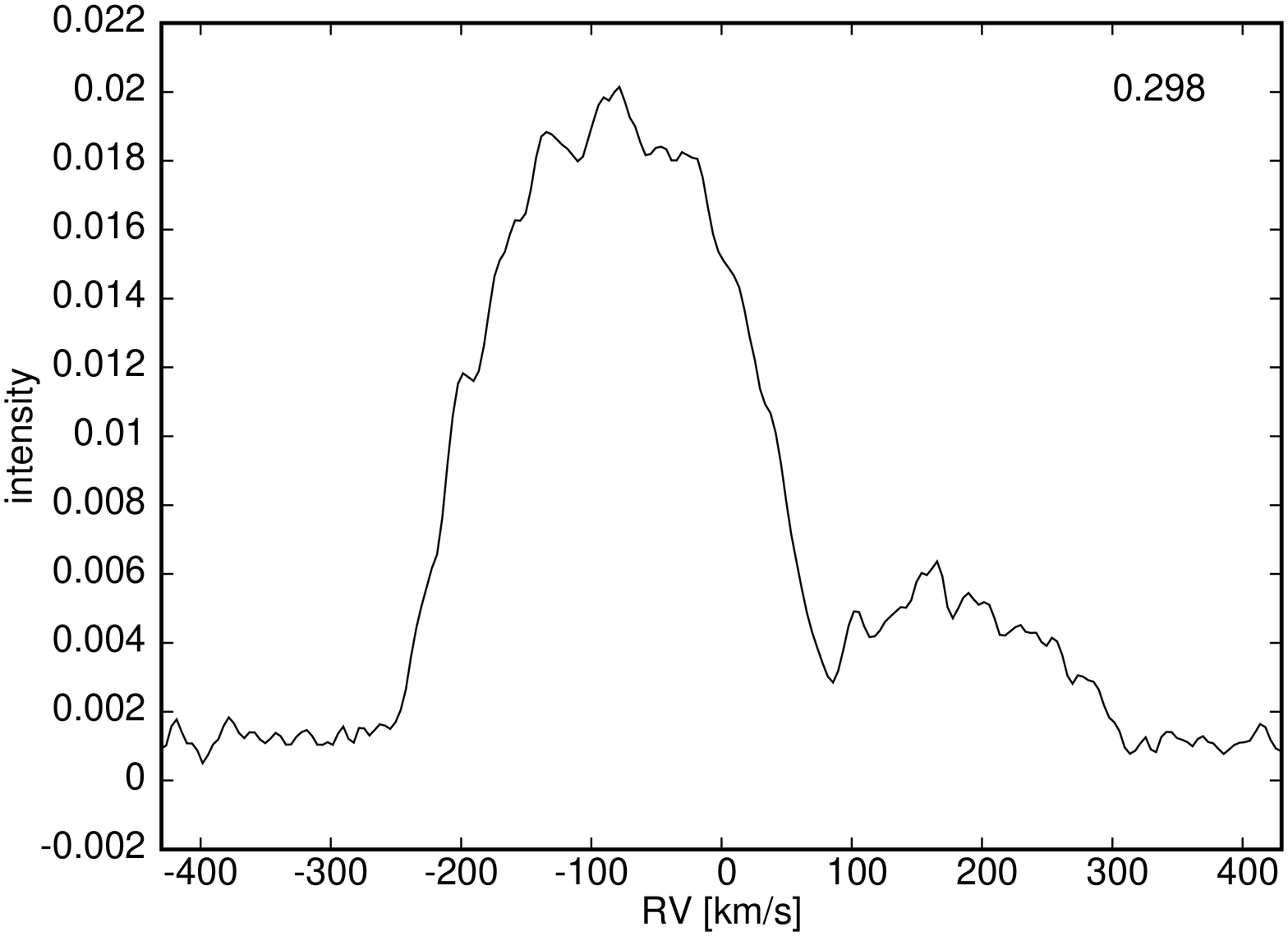}&
\includegraphics[width=1.10in,angle=-90]{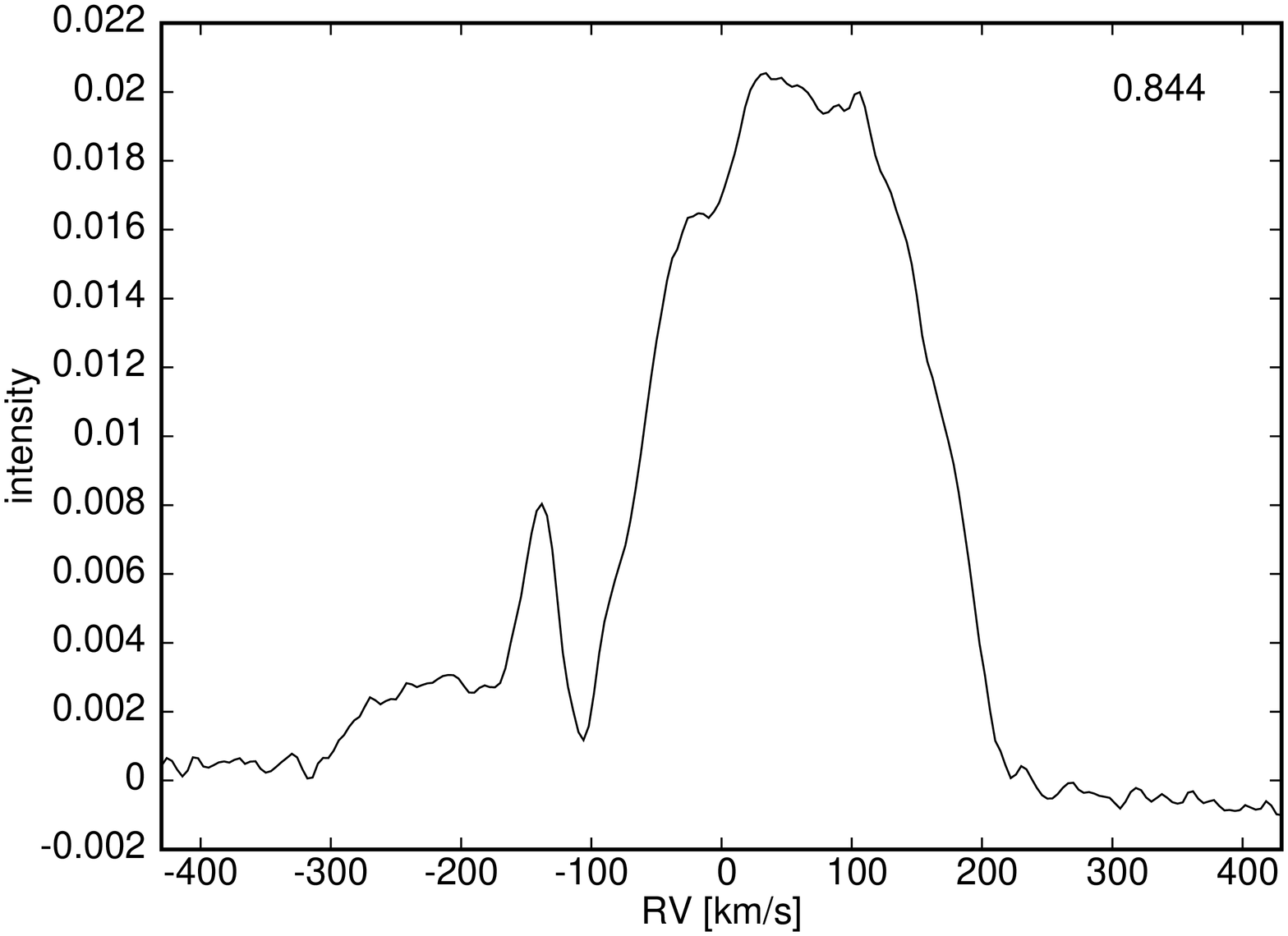}
\end{tabular}}
\vspace{1pc}
\FigCap{BFs of CX~Vir
obtained with F5V RV standard.  The brighter
region on the secondary component in
the vicinity of the primary component
is underlined at phase 0.202 (see
Section~5.1).  The well defined very
bright region visible at second
quadrature (right panels) is indicated
by an arrow on the BF obtained at
phase 0.747, at which point
it begins to be
distinguishable from the apparently
pure profile of the secondary component
(doubly underlined on the
same plot).  The
outstanding accuracy of the BF
profiles is noteworthy here.}
\end{figure}

\begin{figure}
\centering
\includegraphics[]{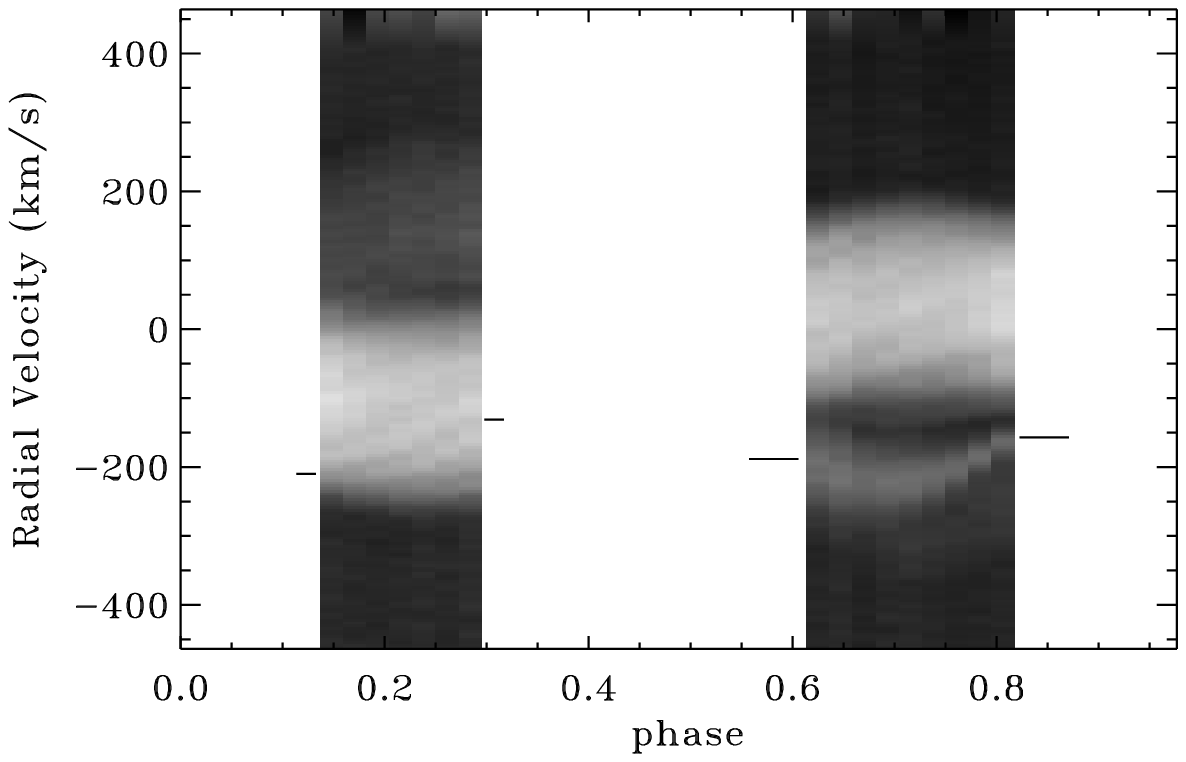}
\FigCap{BFs of CX~Vir
in phase-velocity representation,
rebinned in phase with a 
constant increment of 0.0235. 
In addition to the very bright
region on the secondary component seen
in second quadrature (as indicated
by longer marks), cold spots on the
primary component are visible. 
(One of these, near first
quadrature, is indicated by shorter
marks).}
\end{figure}

\subsection{FT Lup}

As in the case of V747~Cen and CX~Vir,
the BF profiles of the secondary
component obtained in second
quadrature reveal a bright region.
In first quadrature, the BF
profiles of the secondary are
brighter in the vicinity of the primary
(Fig.~10; for detailed
explanations, check particularly the 
plots obtained at phases 0.233 and
0.710).
In Fig.~11, we display the BFs in the
phase-RV plane. Here one can clearly
see not only the track of the bright region
(hot spot) in second quadrature 
but also several tracks of cold
spots on the primary in
both quadratures.

Accretion
processes between components 
considerably disturb
the RV curve of the secondary, 
with $v_{\gamma}$ as 
obtained from the RV curve of the
secondary 10~km/s smaller
than the $v_{\gamma}$ obtained from the
apparently undisturbed RV 
curve of the primary. We consequently 
determined the spectroscopic mass
ratio of this system using only the
secondary-component RV measurements
obtained at second quadrature
(Fig.~6).  At these phases (Fig.~10,
right panels), the influence of the
well defined bright region can be
easily taken into account 
during the RV
measurment process.

With a
hot spot on the secondary component, 
we expected to obtain a semi-detached
configuration from light curve
modelling. However, as in the case of
CX~Vir, our best solution converged
at the contact configuration with
large $\Delta T$ between components
and a non-negligible (4\%-to-5\%) 
contribution
from a third light (Table~7).

\begin{figure}
\centerline{%
\begin{tabular}{c@{\hspace{1pc}}c@{\vspace{-0.7pc}}}
\includegraphics[width=1.10in,angle=-90]{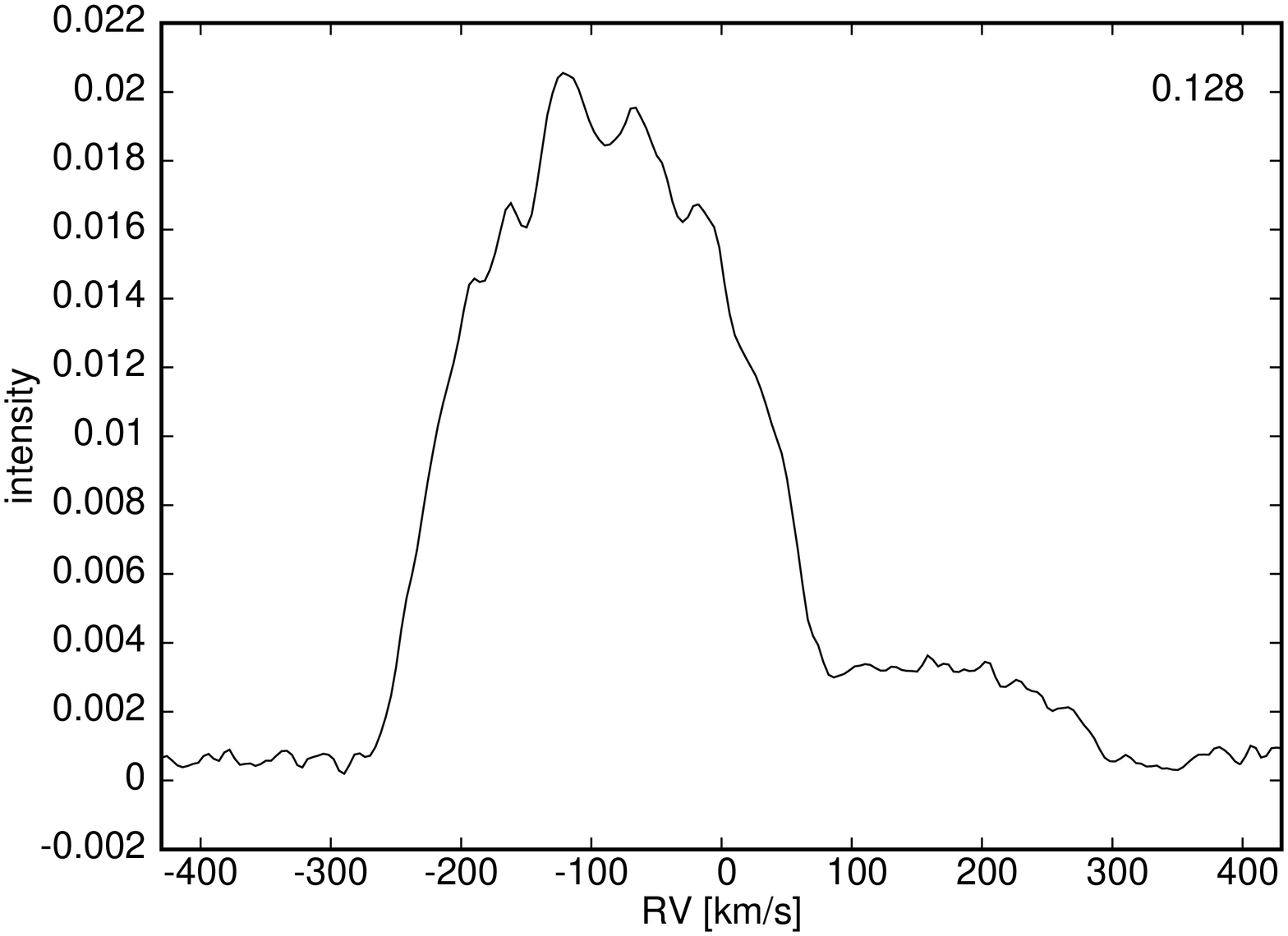}&
\includegraphics[width=1.10in,angle=-90]{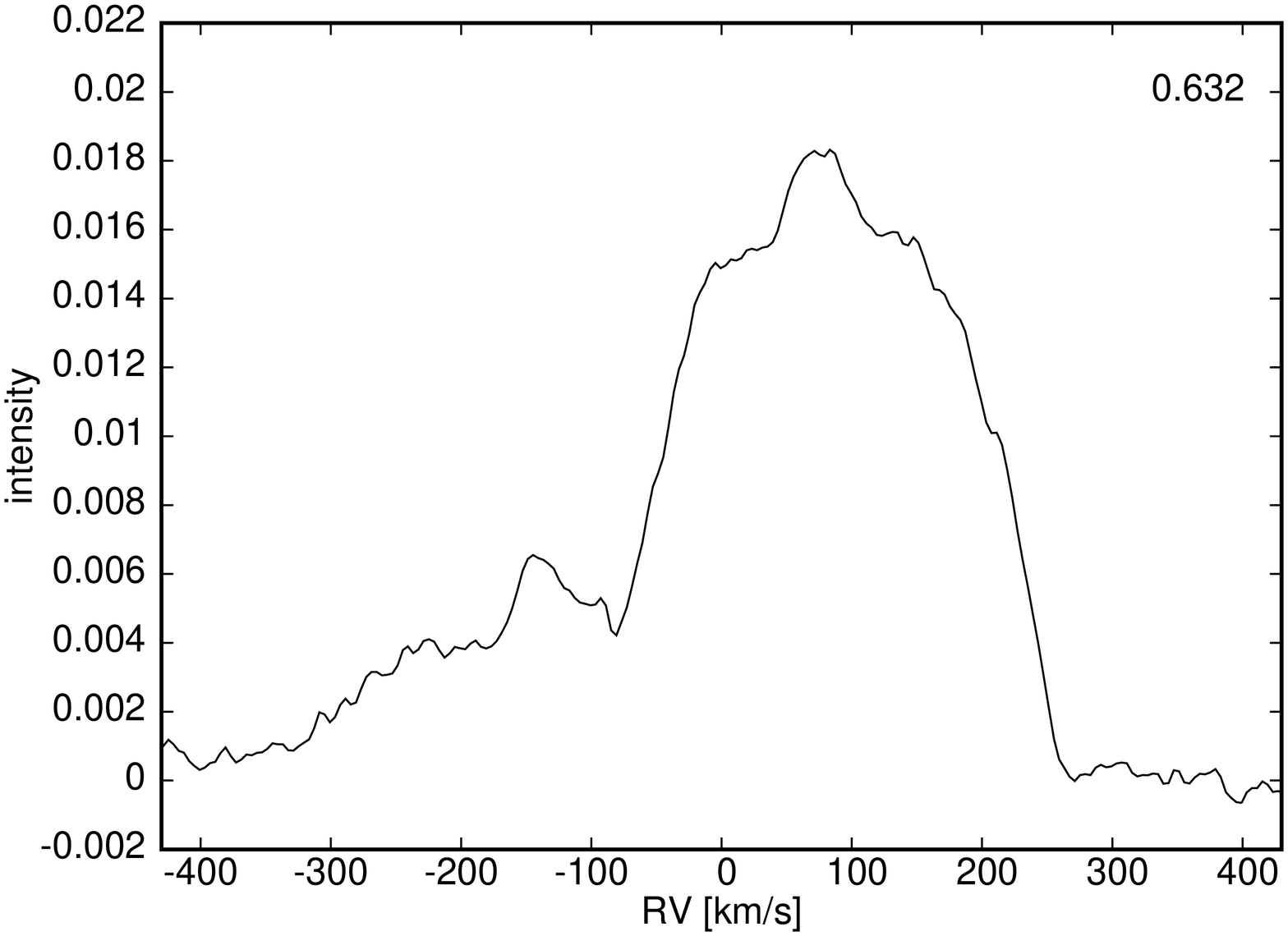}\\
\includegraphics[width=1.10in,angle=-90]{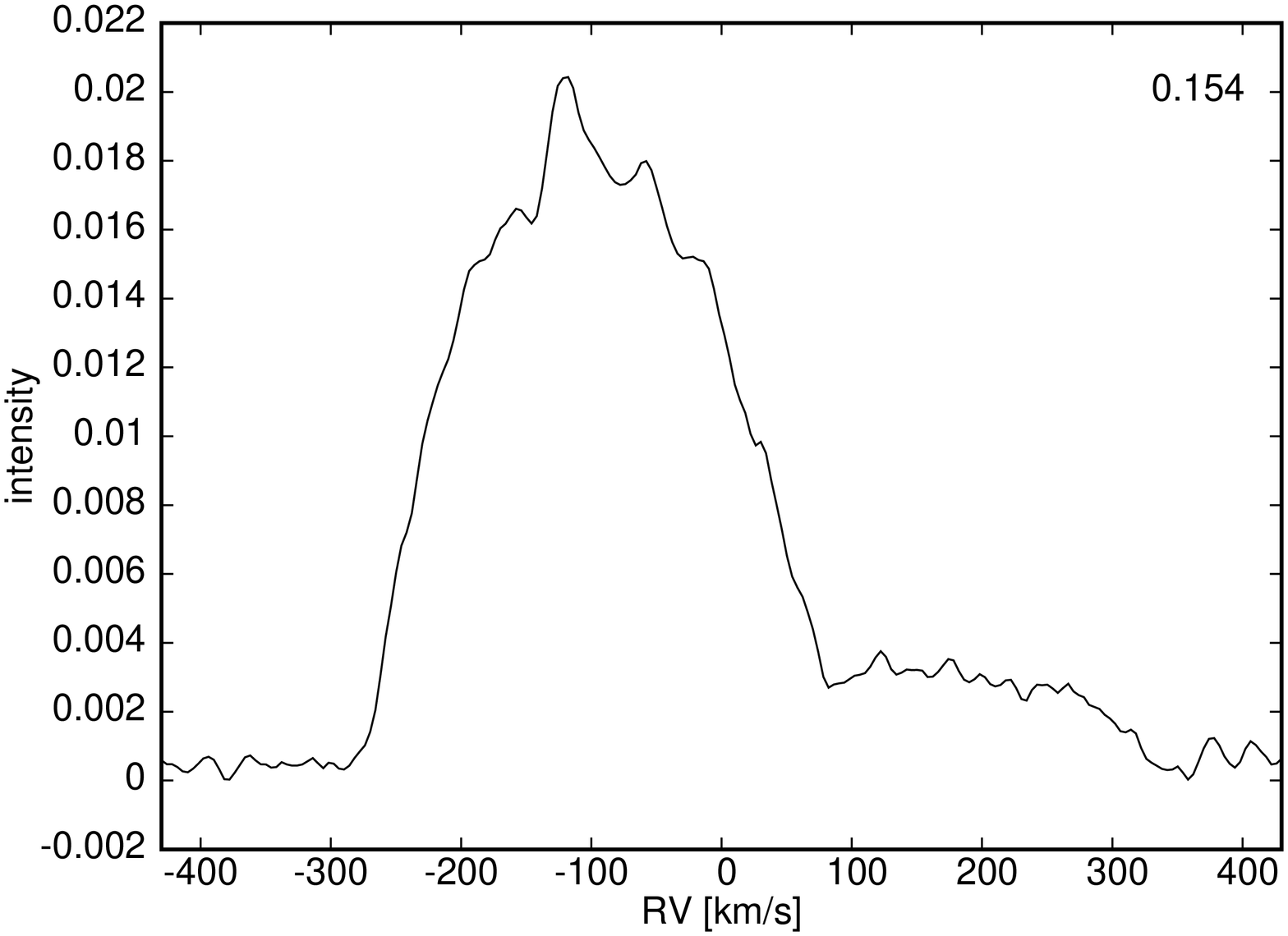}&
\includegraphics[width=1.10in,angle=-90]{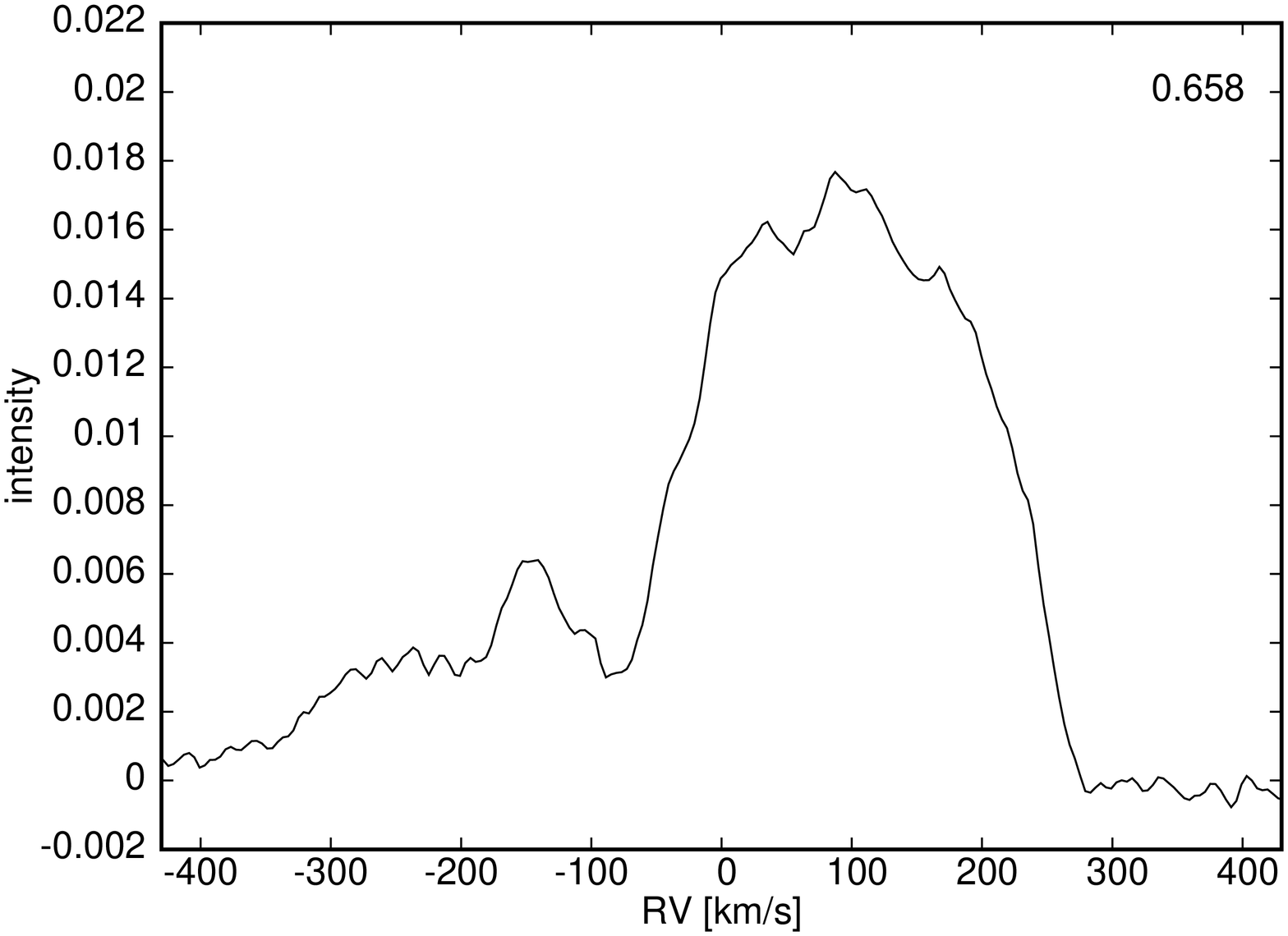}\\
\includegraphics[width=1.10in,angle=-90]{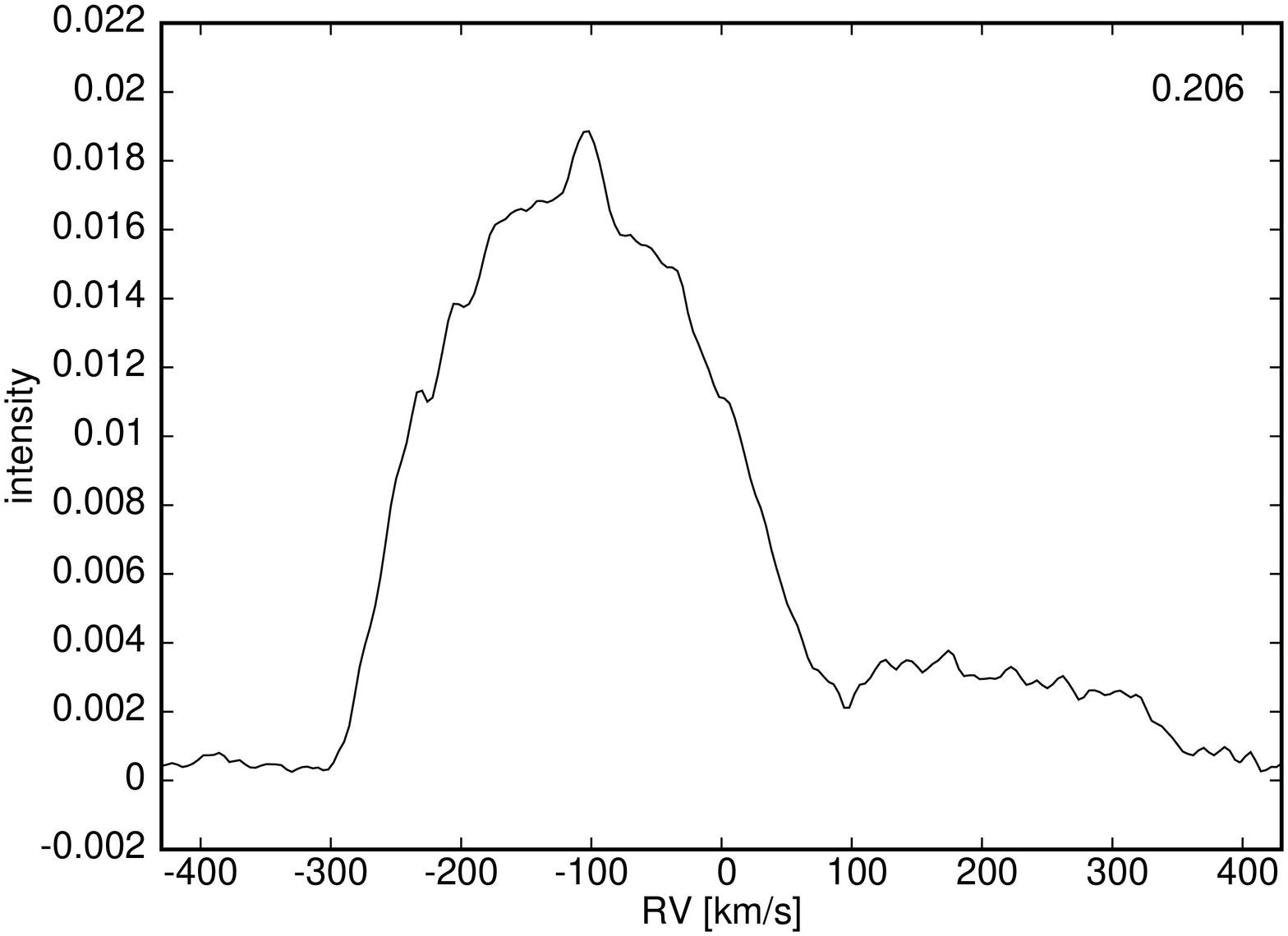}&
\includegraphics[width=1.10in,angle=-90]{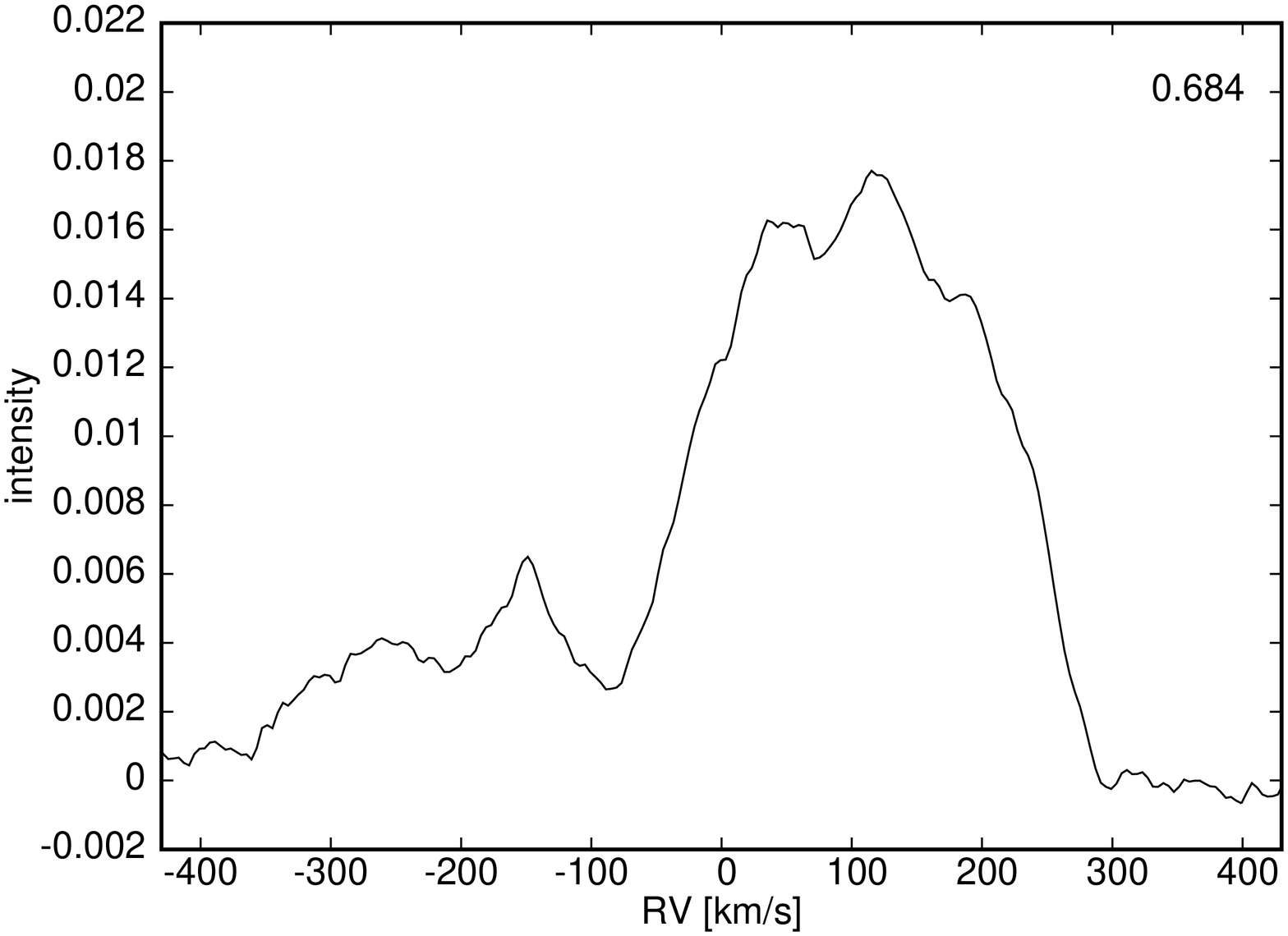}\\
\includegraphics[width=1.10in,angle=-90]{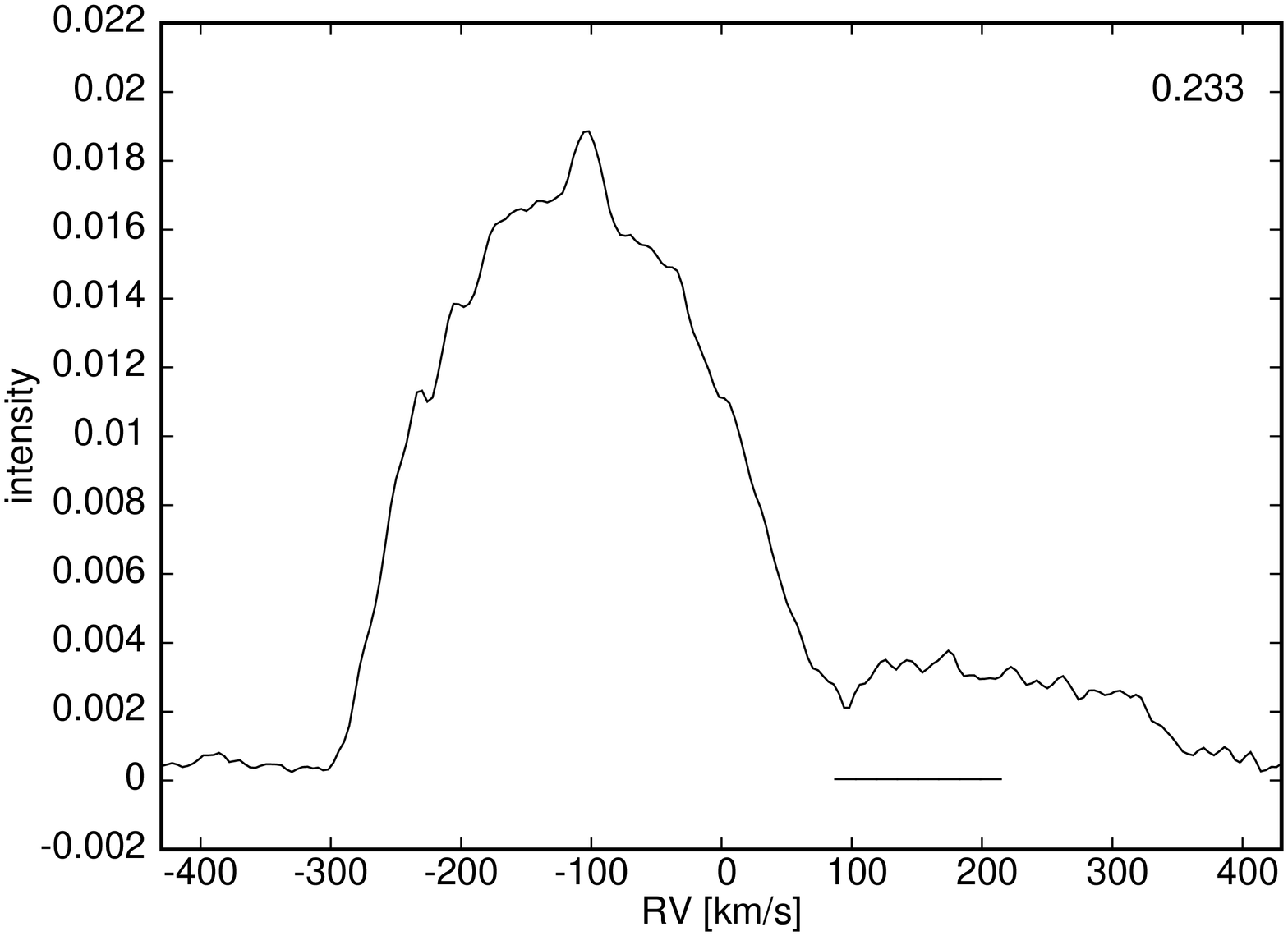}&
\includegraphics[width=1.10in,angle=-90]{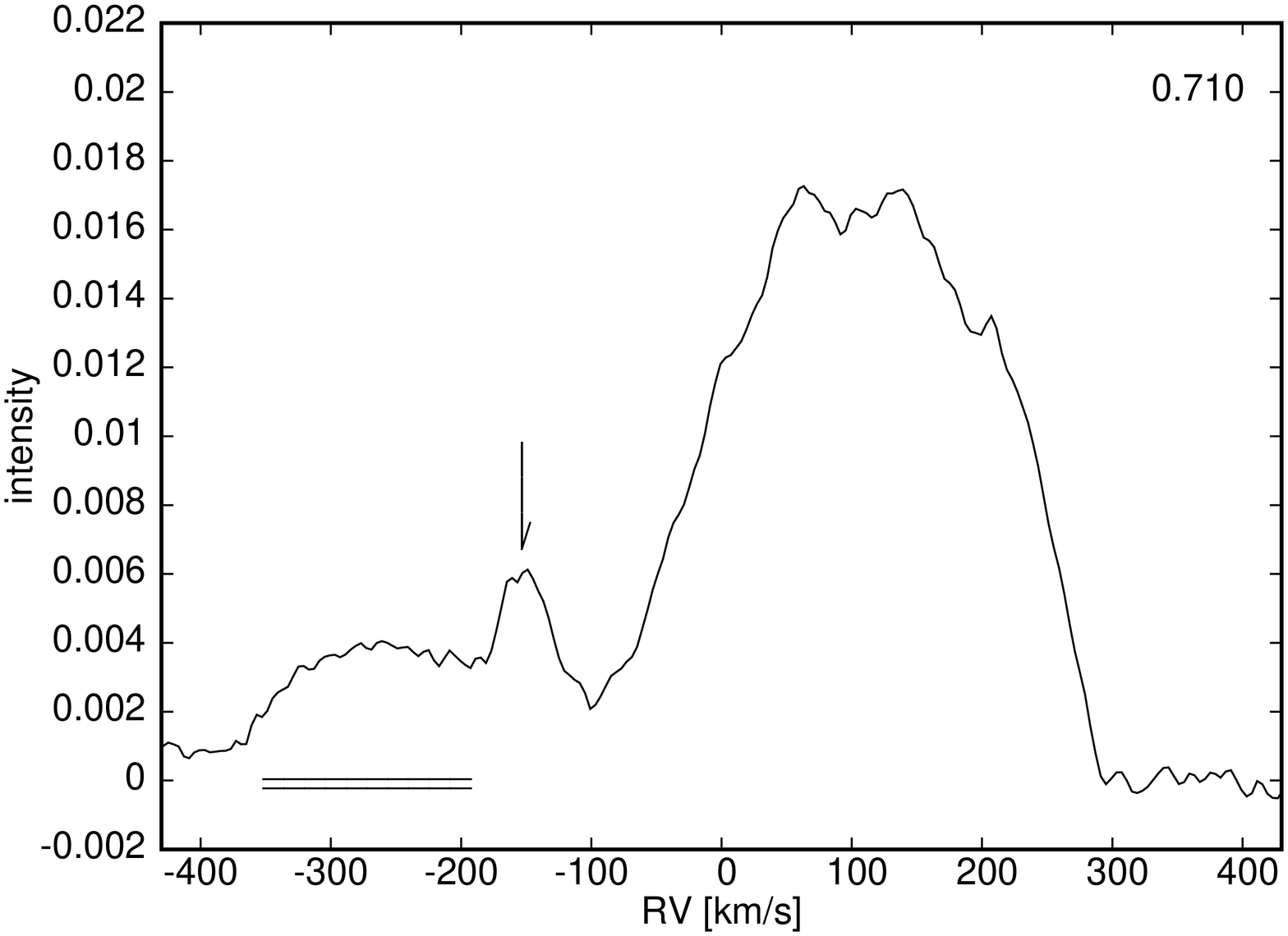}\\
\includegraphics[width=1.10in,angle=-90]{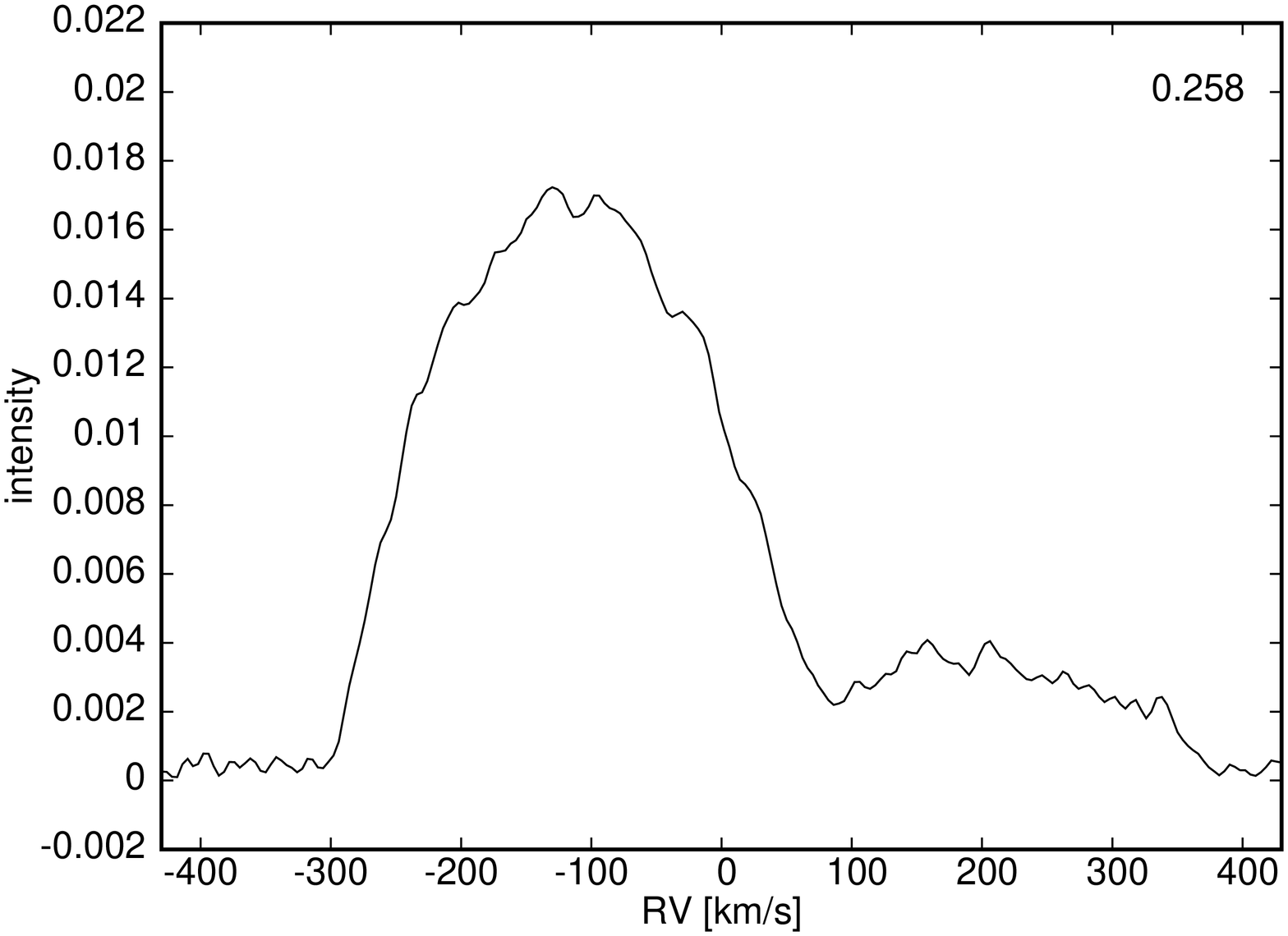}&
\includegraphics[width=1.10in,angle=-90]{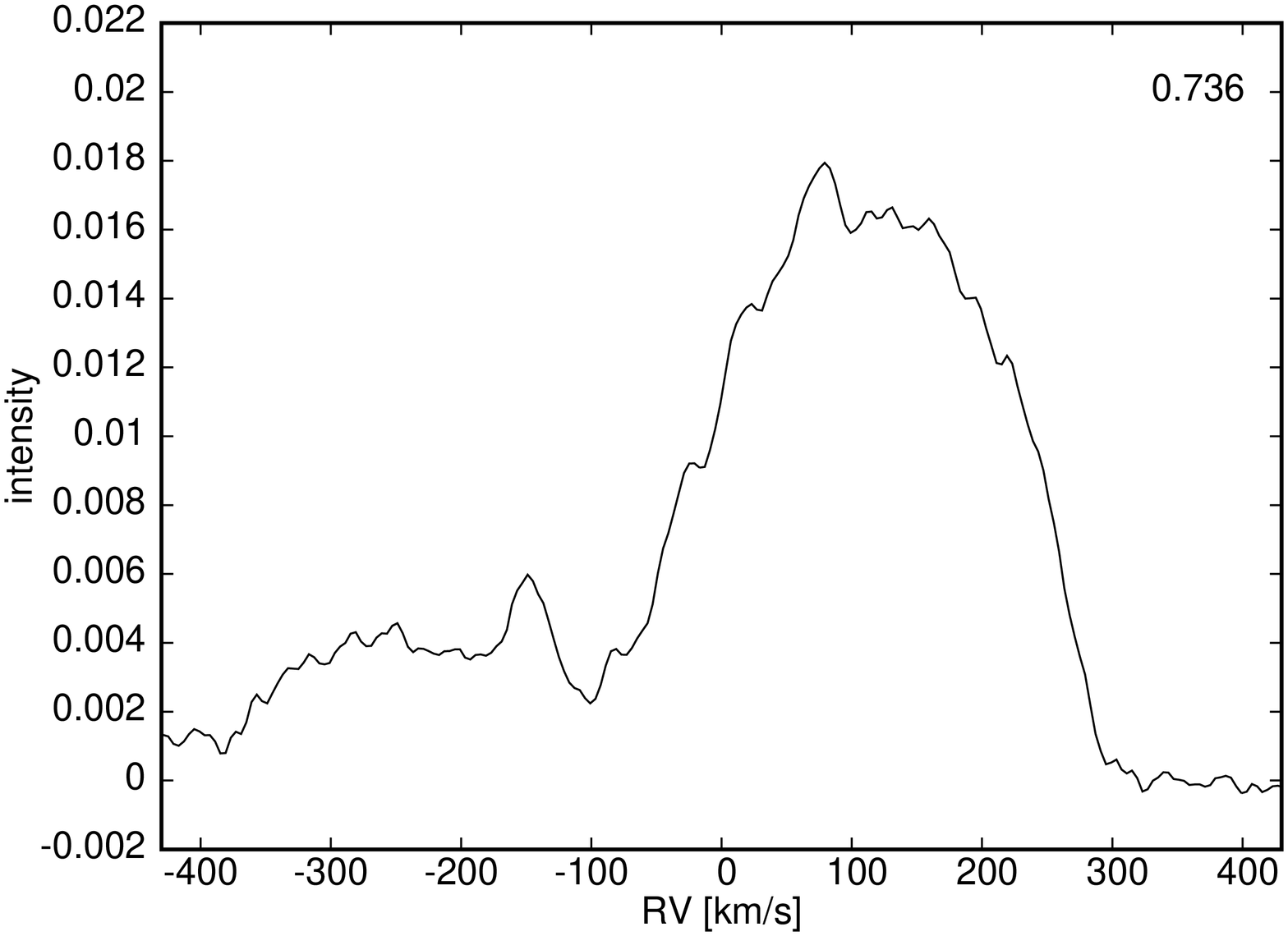}\\
\includegraphics[width=1.10in,angle=-90]{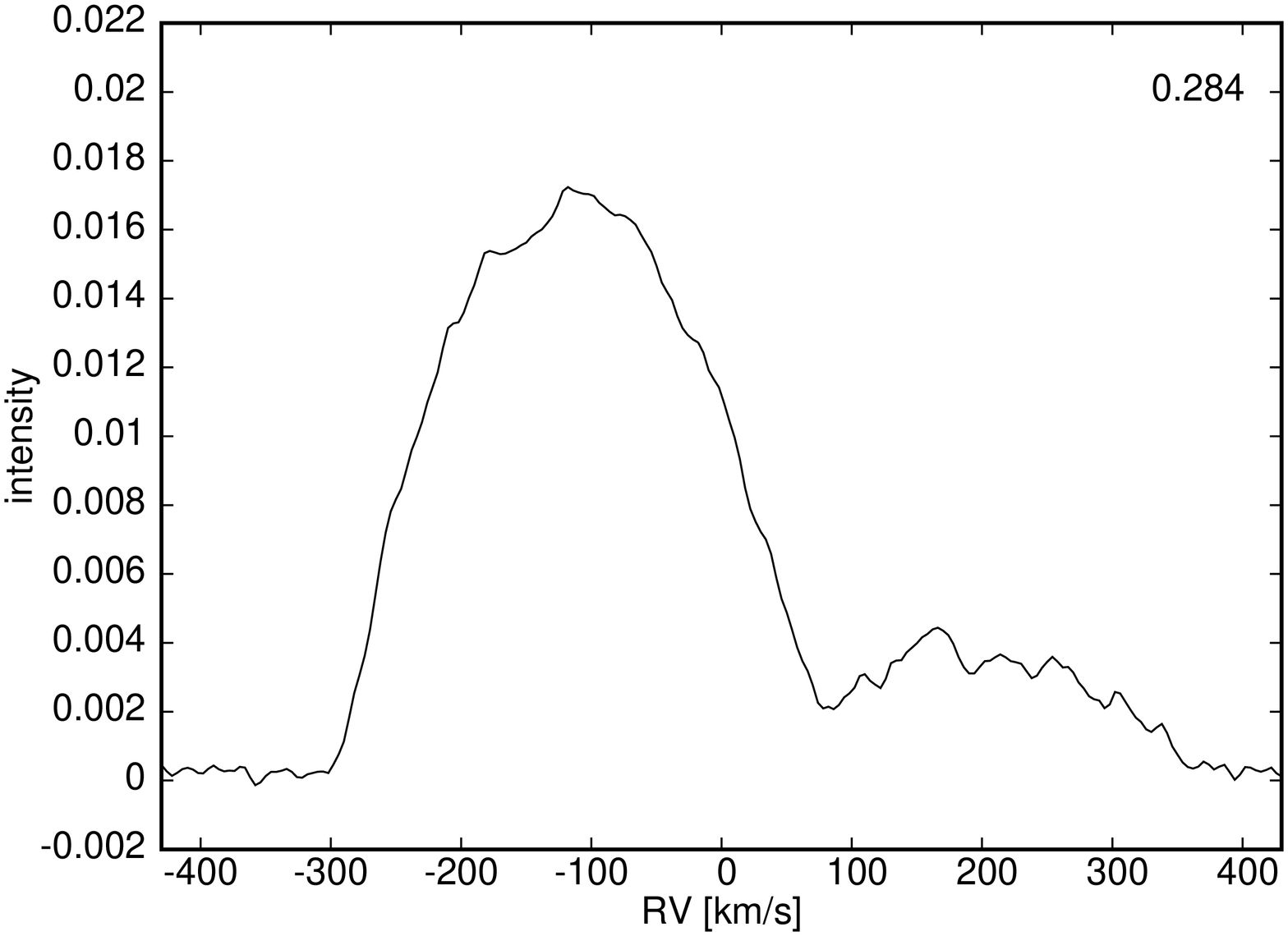}&
\includegraphics[width=1.10in,angle=-90]{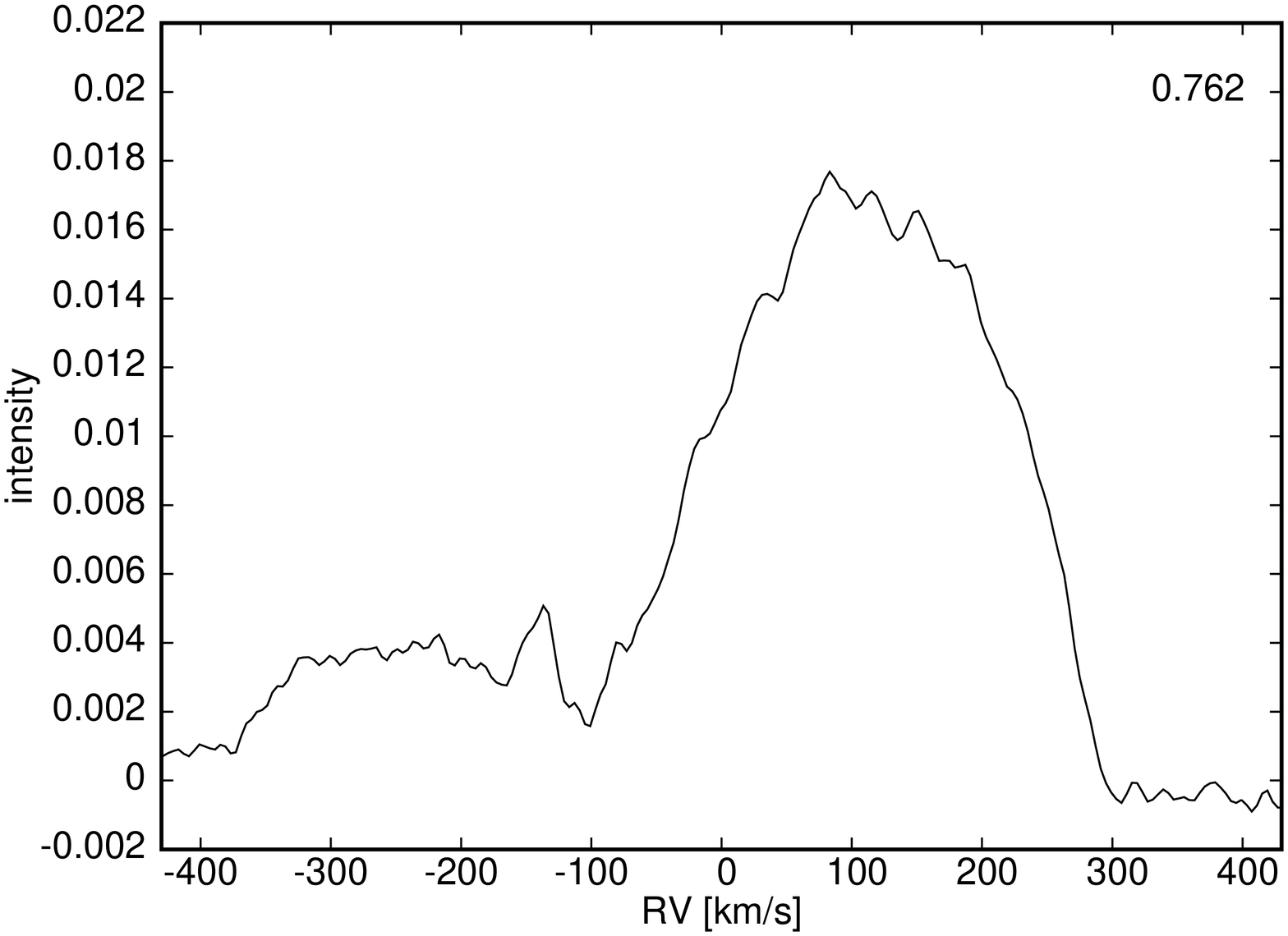}\\
\includegraphics[width=1.10in,angle=-90]{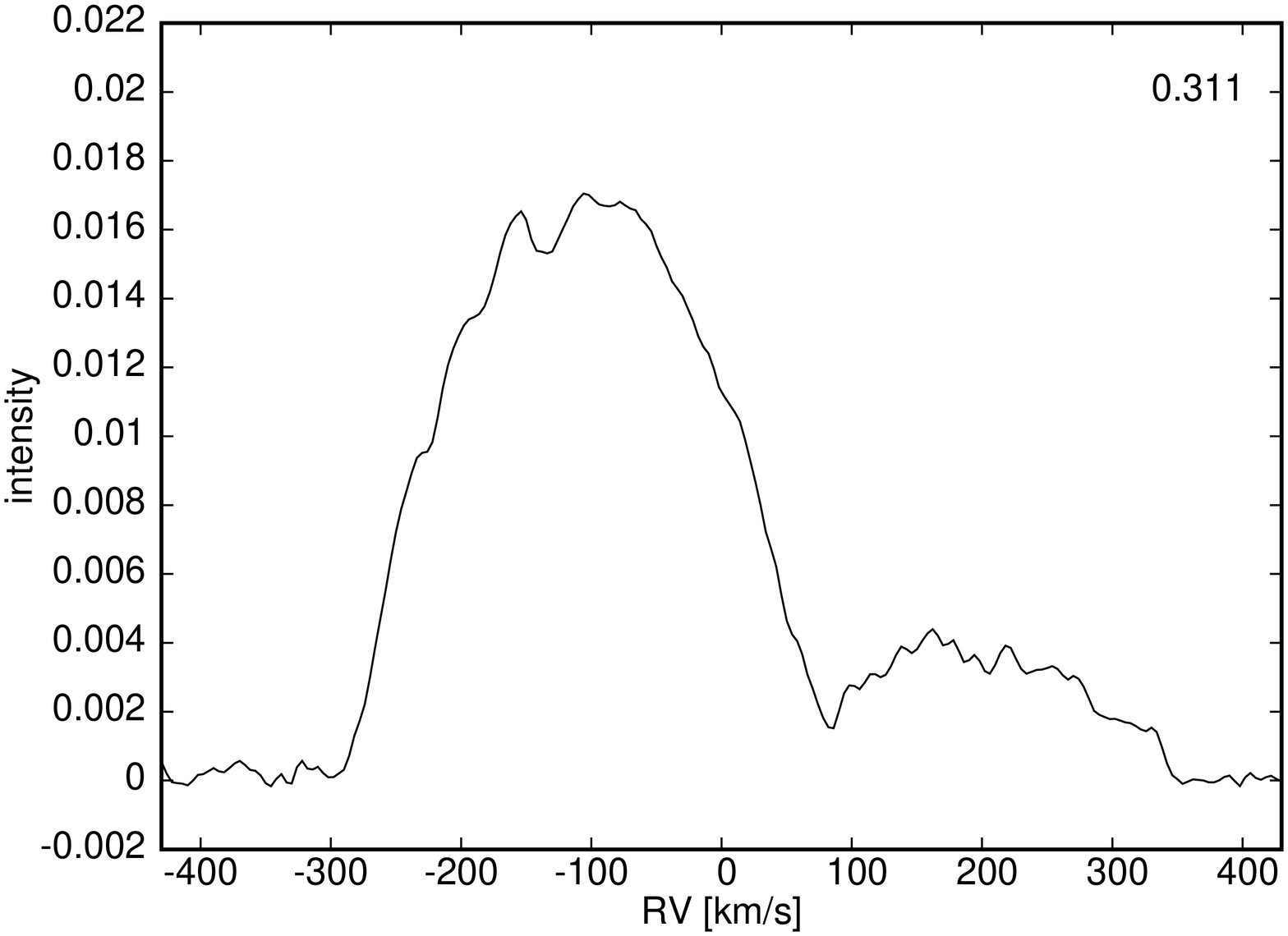}&
\includegraphics[width=1.10in,angle=-90]{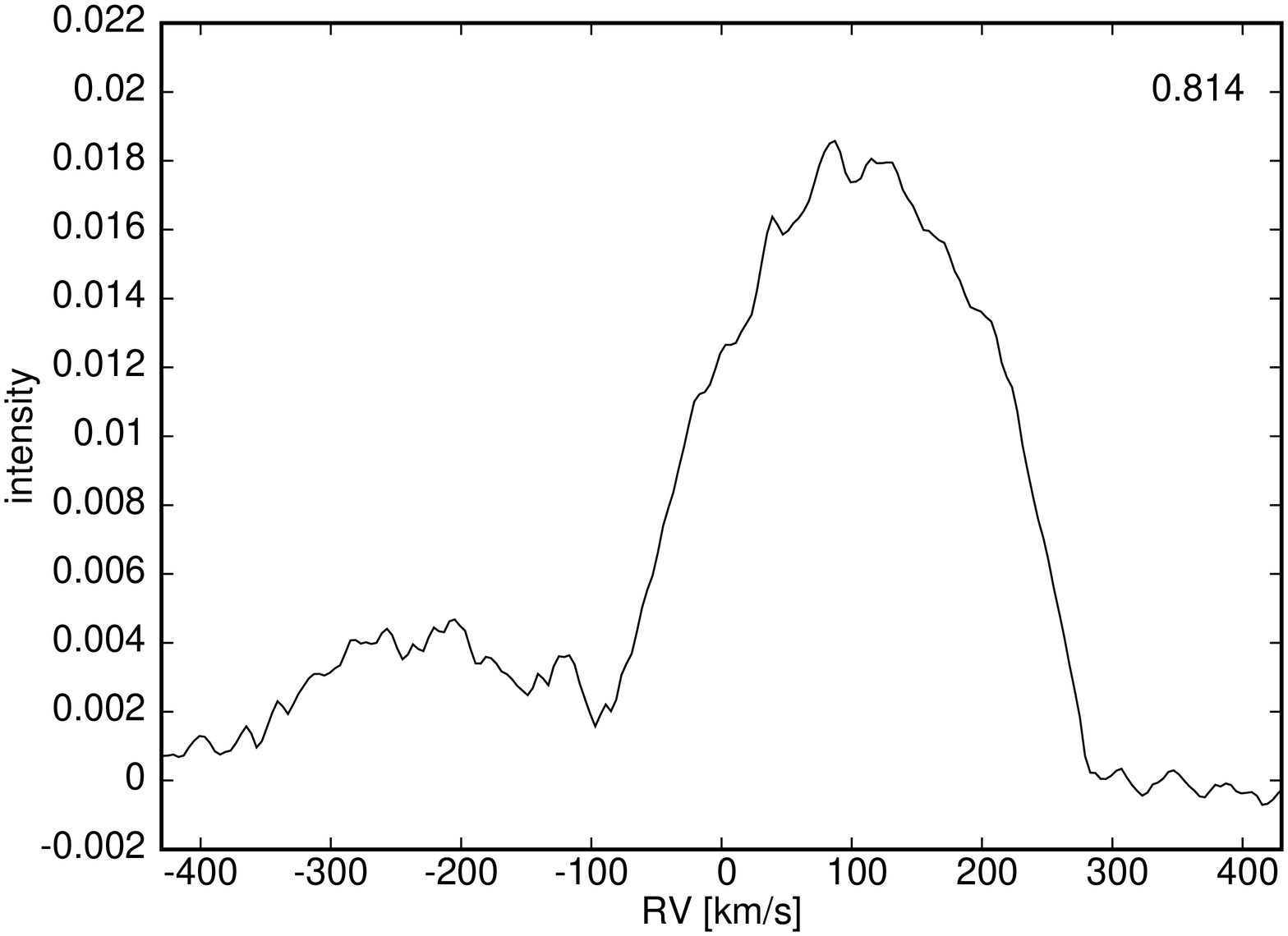}
\end{tabular}}
\vspace{1pc}
\FigCap{BFs of FT~Lup 
obtained by means of F2V RV standard. 
The region of increased intensity
on the secondary component in the
vicinity of the primary component
is underlined in phase 0.233 (see
Section~5.2).  The well defined 
high-intensity region visible in second
quadrature (right panels) is indicated
by an arrow on the BF obtained at phase
0.710, with the apparently pure profile of
the secondary component doubly underlined
on the same plot.}
\end{figure}

\begin{figure}[h]
\centering
\includegraphics[]{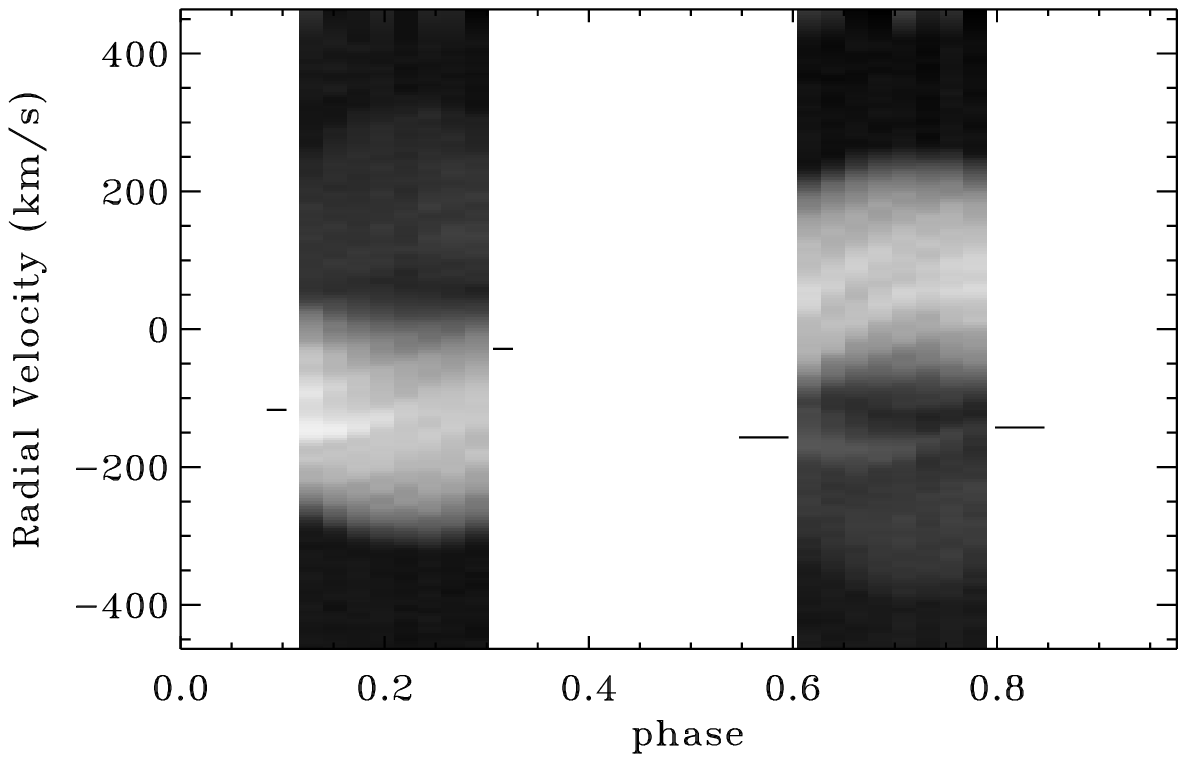}
\FigCap{BFs of FT~Lup in phase-velocity  
representation, rebinned in phase
with a constant increment of 0.025. 
In addition to 
the well defined region of increased
intensity on the secondary component
in second quadrature (indicated by
longer marks), a few cold spots on the
primary are visible. 
(One of these is indicated by shorter
marks in first quadrature.)}
\end{figure}

\subsection{BV Eri}

The BF profiles of BV~Eri again 
reveal a well defined brigter region
on the secondary component (Fig.~12), 
although not as clearly as in 
three systems previously discussed. 
Similarly, in
first quadrature the BF profiles
of the secondary are also
brighter in the vicinity of the 
primary-component profile.

To perform light-curve
modelling, we used the only available 
very good-quality UBV data, published
by Badee et al.~(1983).  In the
model, we assumed a hot spot on
surface of the secondary component
responsible for the O'Connel effect
and for the features discovered in the BFs.
Our computations resulted in a contact
configuration with large $\Delta T$
and with a significant (13\% to 14\%)
contribution from a third light.

\begin{figure}[h]
\centering
\includegraphics[]{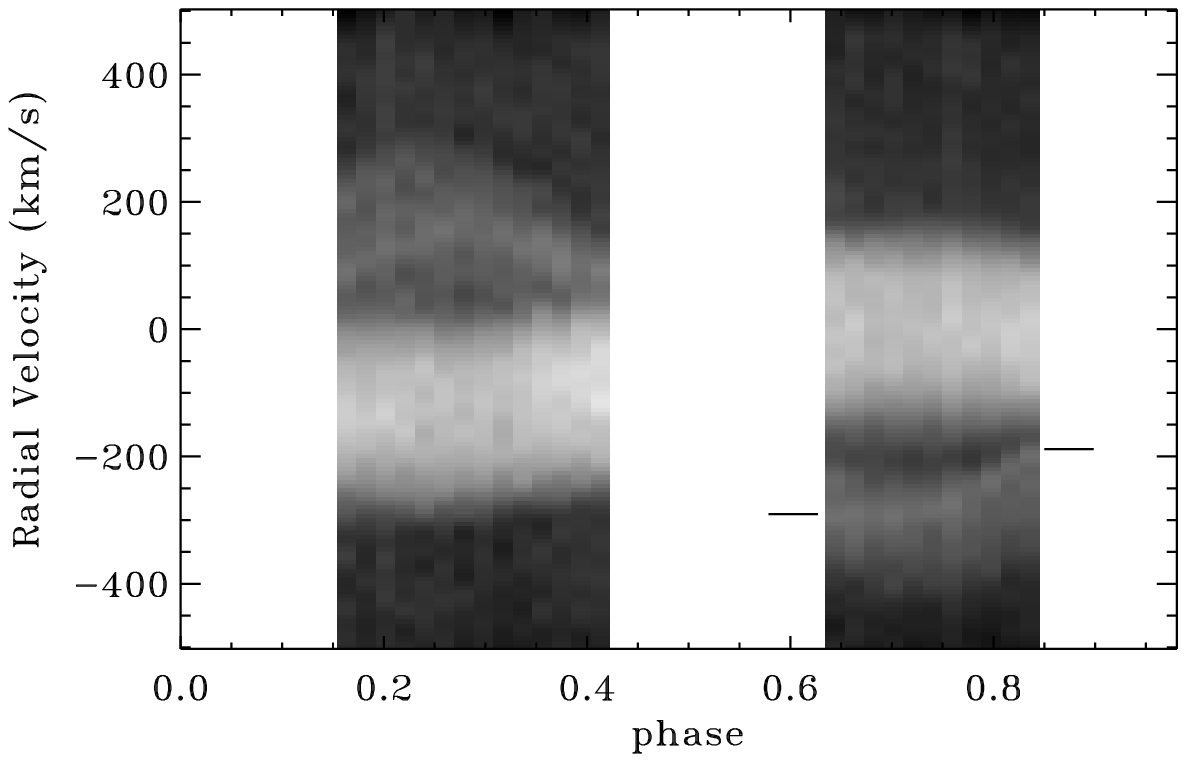}
\FigCap{BFs of BV~Eri
in phase-velocity representation,
rebinned in phase with a constant increment 
of 0.02.  The well defined bright
region of increased intensity on the
secondary component (indicated by
marks) is visible at phases in 
second quadrature.}
\end{figure}

\subsection{FO Hya}

Our new BV$(RI)_c$ data confirmed
the F0V spectral type of the primary
component determined by Candy \& Candy
(1997).  First RV curves
yielded a considerably smaller value
of the mass ratio ($q_\mathrm{spec} = 0.238$, 
in contrast with 
the value $q_\mathrm{phot}=0.552$ 
which Candy \& Candy determined). 
Despite the
relatively poor quality of the spectra
obtained at LCO, we additionally notice
differences between the BF profiles
of the secondary component obtained
in both quadratures: the profiles
appear brighter and are well defined
in first quadrature, being
considerably weaker in 
second quadrature.
The profiles obtained before phase 0.75 
seem to reveal a region of increased
intensity, as with the 
four systems discussed above.

Because the incompletness
of the new light curve
deprived us of evidence 
for an O'Connel effect, we first
attemped to model the light curves
without spots. This first model, however, 
did not yield a good fit. 
In our second attempt,
using information obtained
from BF profiles and following Candy
\& Candy (1997), we introduced a hot
spot on the secondary. 
The model from these
computations,  while yielding a reasonably 
good fit (Fig.~6), additionally
delivered a contact
configuration, a large $\Delta T$ between
the components, and a consirable
contribution from a third light (Table~7). 

\subsection{CN And}

CN~And is a very frequently observed
and analyzed system, and different
configurations were obtained from
its light and RV curve
modelling (Table~1).  As almost all
available archival light curves of
CN~And are of low quality, we decided to
obtain new high-precision, multicolour
(BVRI) data.  Radial velocity curves
were already obtained at DDO, and
the spectroscopic mass ratio was
derived by the BF method
(Rucinski et al.~2000).  In Fig.~13
we display binned BFs of CN~And in 
the phase-velocity plane.  
The BFs, although individually
rather noisy, give information of
better quality when stacked, 
revealing a 
subtle track of a brighter region on
the secondary component at second
quadrature.  We therefore decided to
explain the O'Connel effect apparent
in light curves by assuming one large
hot spot on the secondary component.
We obtained a  marginal-contact 
configuration with large $\Delta T$. 
However, in contrast with 
CX~Vir, FT~Lup, BV~Eri,  
and FO~Hya, it was not necessary to
include a third light to obtain a good
fit.  

We note that Zola et al.~(2005)
obtained a semi-detached configuration. 
Their model, however, assumed two cold spots 
on the primary component, whereas in our
model a hot spot was placed on the
surface of the secondary,
in agreement with information obtained 
from BFs.

\begin{figure}[]
\centering
\includegraphics[]{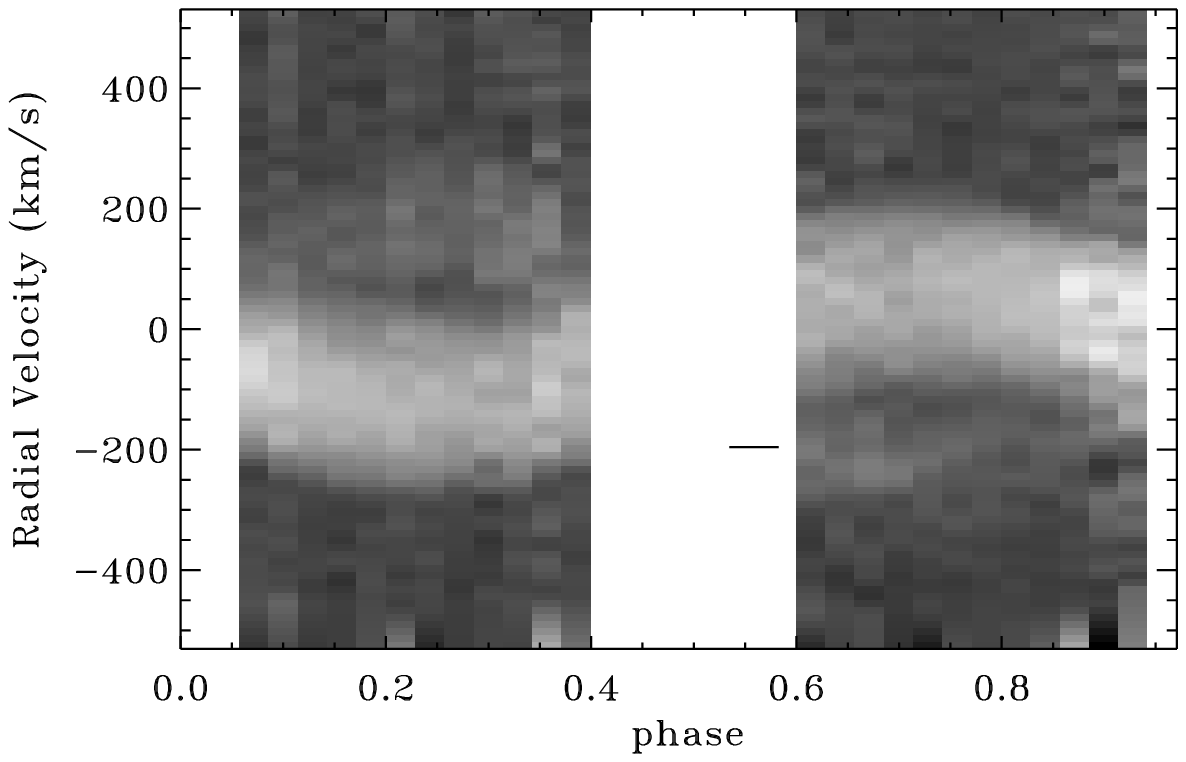}
\FigCap{BFs of CN~And
in phase-velocity representation,
rebinned in phase with a constant increment of
0.03.  The mark indicates the brighter
region on the secondary component
visible at second quadrature in
the phase range 0.6-0.8.} \end{figure}

\subsection{BX And}

At first glance, our new BVRI
light curves appeared to be free from an
O'Connel effect. However, synthetic light
curves obtained within a spotless model
(yielding a contact configuration
with $f=10\%$ and large $\Delta T$
between components) did not fit the
observed curves very well, with the
discrepancy becoming especially pronounced
around the secondary minimum.  A closer
inspection of the region around phase 0.3
reveals that it is slightly brighter
than the one around phase 0.7.
The assumption of a hot spot on the
equator of the secondary component
yielded a marginal-contact 
configuration with $f=4.5\%$ and large
$\Delta T$ (Table~7), but 
with about $40\%$
smaller $\chi^{2}_\mathrm{red}$. 

In order to validate the assumption of a
hot spot in our model, we examined the
available BF profiles obtained from
medium-resolution spectra, but they
did not show any visible tracks of a
hot spot on the secondary component.
This outcome may be due to the small size of
the spot (as obtained from light curve
modelling), with detection perhaps only
possible through BFs derived from
high-resolution spectra. 

\MakeTable{c c c c c c c r c}{12.5cm}{Physical parameters of components derived
from light- and RV-curve modelling.}
{\hline\hline
system   & $M_1[M_{\odot}]$&$M_2[M_{\odot}]$&$R_1[R_{\odot}]$&$R_2[R_{\odot}]$&$T_1[K]$&$T_2[K]$&$L_1[L_{\odot}]$ & $L_2[L_{\odot}]$ \\ \hline
V1010 Oph&1.887(19)&0.887(15)&2.01(3)&1.40 &7500 &5132 &11.45 &1.22  \\
WZ Cyg   &1.589(32)&1.003(37)&1.70(4)&1.37 &6530 &4932 & 4.71 &1.00  \\
VV Cet   &1.042(43)&0.296(20)&1.48(5)&0.79 &8150 &6252 & 8.66 &0.85  \\
DO Cas   &2.130(45)&0.650(25)&2.22(4)&1.29 &8350 &4297 &21.47 &0.51  \\
FS Lup   &1.301(23)&0.611(20)&1.23(3)&0.85 &5860 &5130 & 1.60 &0.45  \\
V747 Cen &1.541(28)&0.493(17)&1.70(3)&0.98 &8150 &4275 &11.42 &0.29  \\ \hline
CX Vir   &1.731(21)&0.594(13)&2.24(3)&1.39 &6450 &4694 & 7.78 &0.84  \\
FT Lup   &1.871(21)&0.823(15)&1.64(2)&1.13 &6700 &4651 & 4.86 &0.54  \\
BV Eri   &1.560(17)&0.430(07)&1.70(2)&0.96 &6700 &5387 & 5.22 &0.70  \\
FO Hya   &1.588(61)&0.378(34)&1.67(3)&0.89 &7000 &4667 & 5.68 &0.33  \\
CN And   &1.433(30)&0.552(20)&1.48(3)&0.95 &6450 &4726 & 3.40 &0.40  \\
BX And   &2.148(52)&0.977(41)&2.01(5)&1.40 &6650 &4758 & 7.08 &0.90  \\ \hline\hline}

\section{Physical parameters of components}

Using the results obtained from light-
and RV-curve modelling
(Table~6 and Table~7), we computed
physical parameters of components
(Table~8).  Because the highly distorted
components of contact binary stars
significantly differ from a sphere, 
their effective radii $R$
were calculated under
the assumption that the surface
area of each component is equal to the
surface area of a spherical star of 
radius $R$.  As the radii are scaled by
their respective distances $a$ 
between centres of
masses of components, $a$-errors
propagate accordingly
on the radii of both components. 
Effective temperatures
of primary components derived from
spectral classification are known with
an accuracy no better than 100--200~K,
and these external errors, much
larger than those obtained 
from light-curve modelling alone (Table~6 
and Table~7), affect also the
effective temperatures of secondary
components. 
These uncertainties
make it even
more difficult to estimate 
errors in the luminosity $L$, 
but $L$ errors are usually 
equal to 10\% to 30\% of $L$ values (Yakut
\& Eggleton, 2005).

Zola et al.~(2006) considered a
sample of contact binary systems
for which physical parameters were
derived from combined light- 
and RV-curve modelling. From the 
available data, they constructed 
mass--radius (M--R), mass--luminosity
luminosity (M--L) and 
temperature--luminosity (T--L) 
diagrams.  The M--L
and T--L diagrams were also constructed
using component physical parameters 
corrected for the effect caused by
energy transfer from the primary to the
secondary, by applying equations
from Mochnacki (1981), with an
ad-hoc modification taking
into account temperature differences
between components (Zola~et~al., 2006).
Theoretical parameters of stars on
ZAMS, TAMS, and hydrogen depletion in
the core  were computed with the 
EZ program (Paxton, 2005), which is
based on the most recent version of
Eggleton's STARS code (Pols~et~al., 1995).

We have appended to the above-mentioned
diagrams the physical parameters
of the systems analyzed in this work
(Table~8). As can be seen in the M--R
diagram (Fig.~14), almost all primary
components of our targets (the only
exception is VV~Cet) have parameters
corresponding to the MS
stars, while secondary components are
oversized by a factor similar to those
of components of W~UMa-type systems.

In Figure~15, we present
two T--L diagrams. The first
is constructed with the physical
parameters from Table~8, while the
second uses parameters corrected
for an effect caused by an energy
transfer between components.  While the
uncorrected parameters of the CLdTs
secondary components (red squares) do not differ from
those of the secondary components of
near-contact systems taken from
literature, in the second plot they
show an obvious tendency to occupy the
region avoided by secondary components
of W~UMa type systems.  This matter
will be discussed more thoroughly in
the next section.

\begin{figure}
\centering
\includegraphics[width=3in, angle=-90]{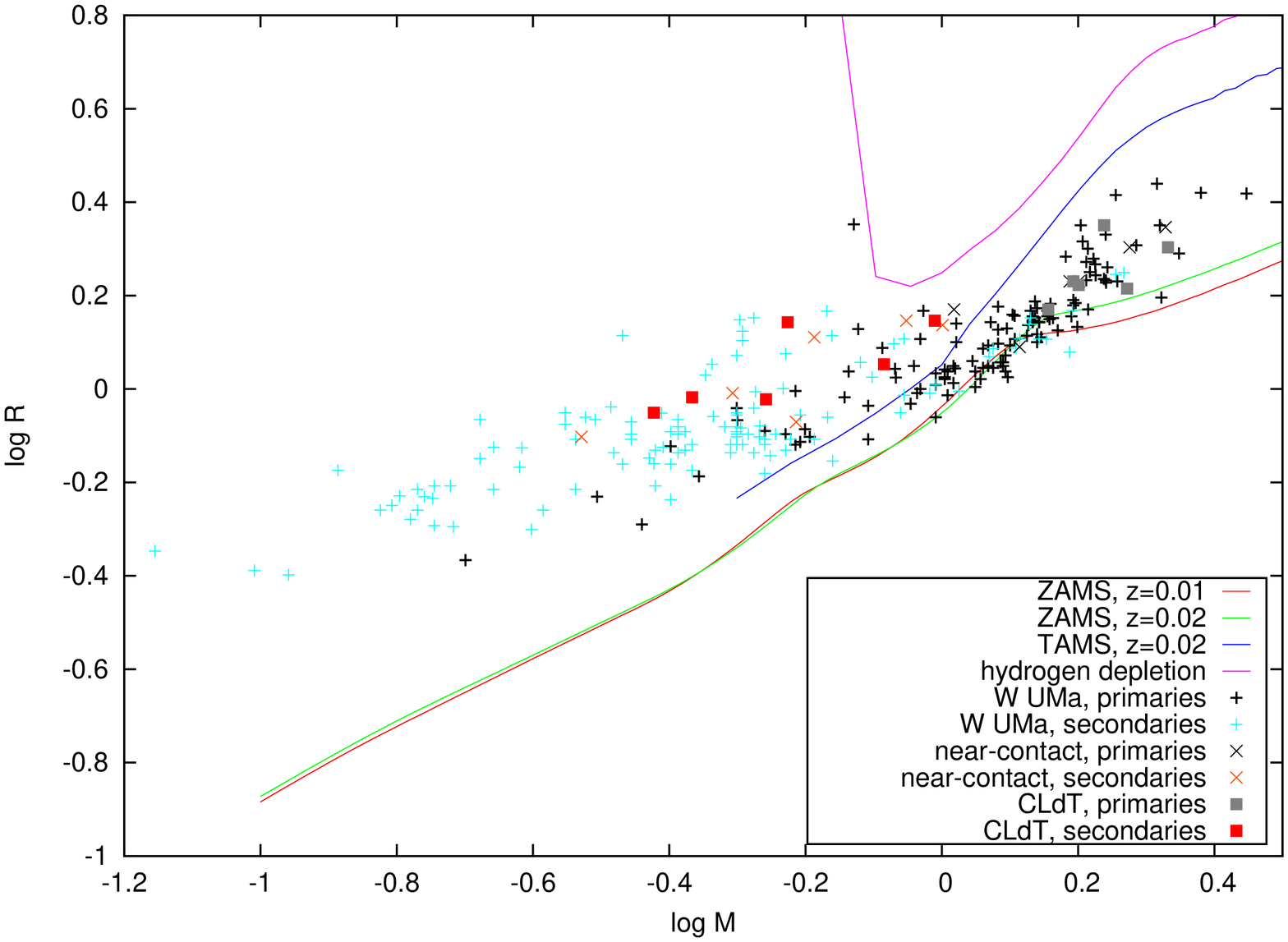}
\vspace{0.5pc}
\FigCap{The M--R diagram for systems analyzed in this work 
on the background 
of the sample of W~UMa-type stars from Zola et al. (2006). 
The continuous line represents the
theoretical lines of ZAMS and TAMS
computed for metallicities of $z=0.01$
and $z=0.02$.  ``Near-contact''
denotes the systems described in Sections~4.1--4.6, 
and ``CLdT'' the systems described
in Sections~5.1--5.6.  Physical
parameters of these systems are listed in Table~8.}
\end{figure}

\begin{figure}
\centerline{%
\begin{tabular}{c@{\hspace{0pc}}c}
\includegraphics[width=1.8in, angle=-90]{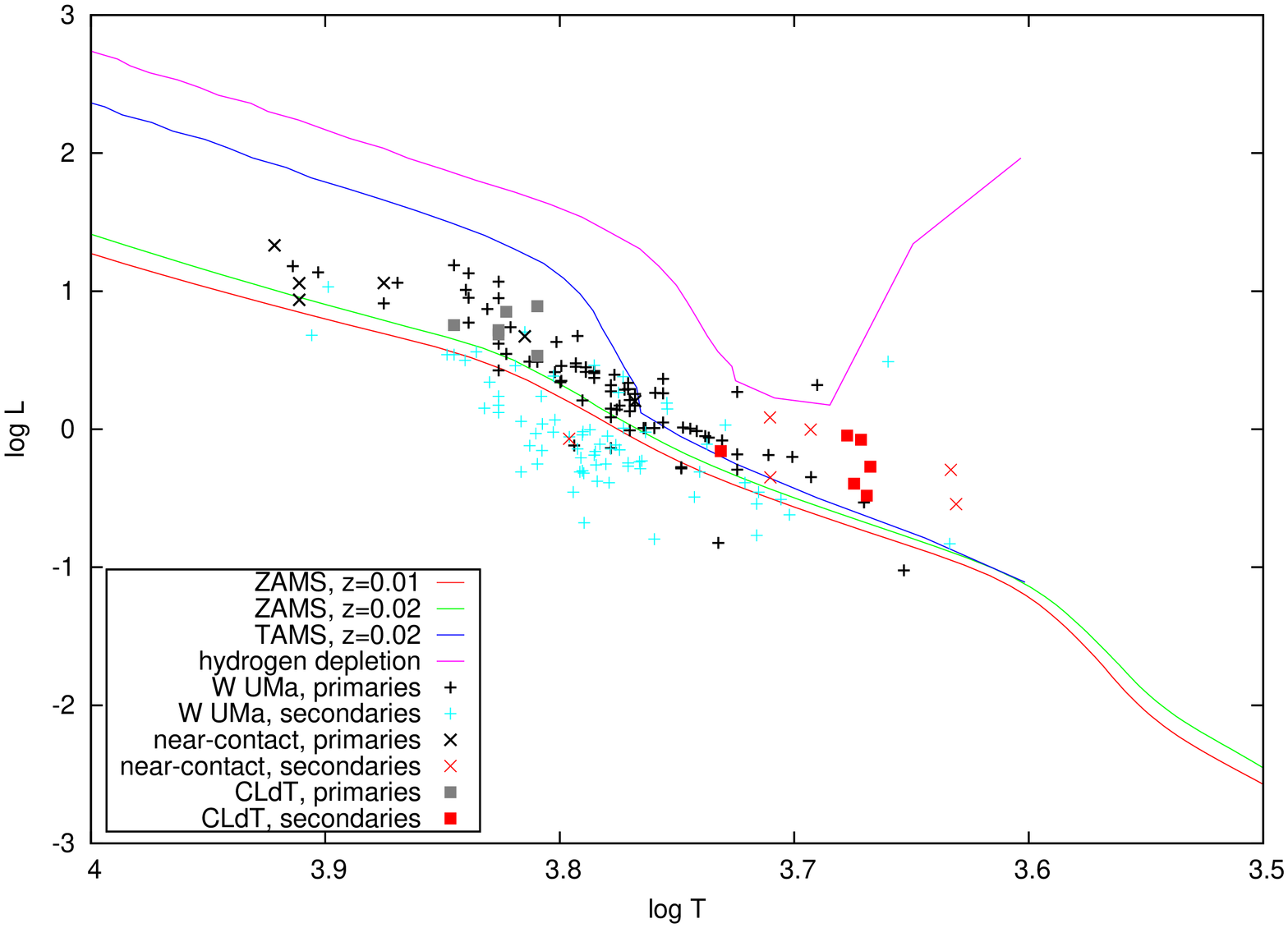}&
\includegraphics[width=1.8in, angle=-90]{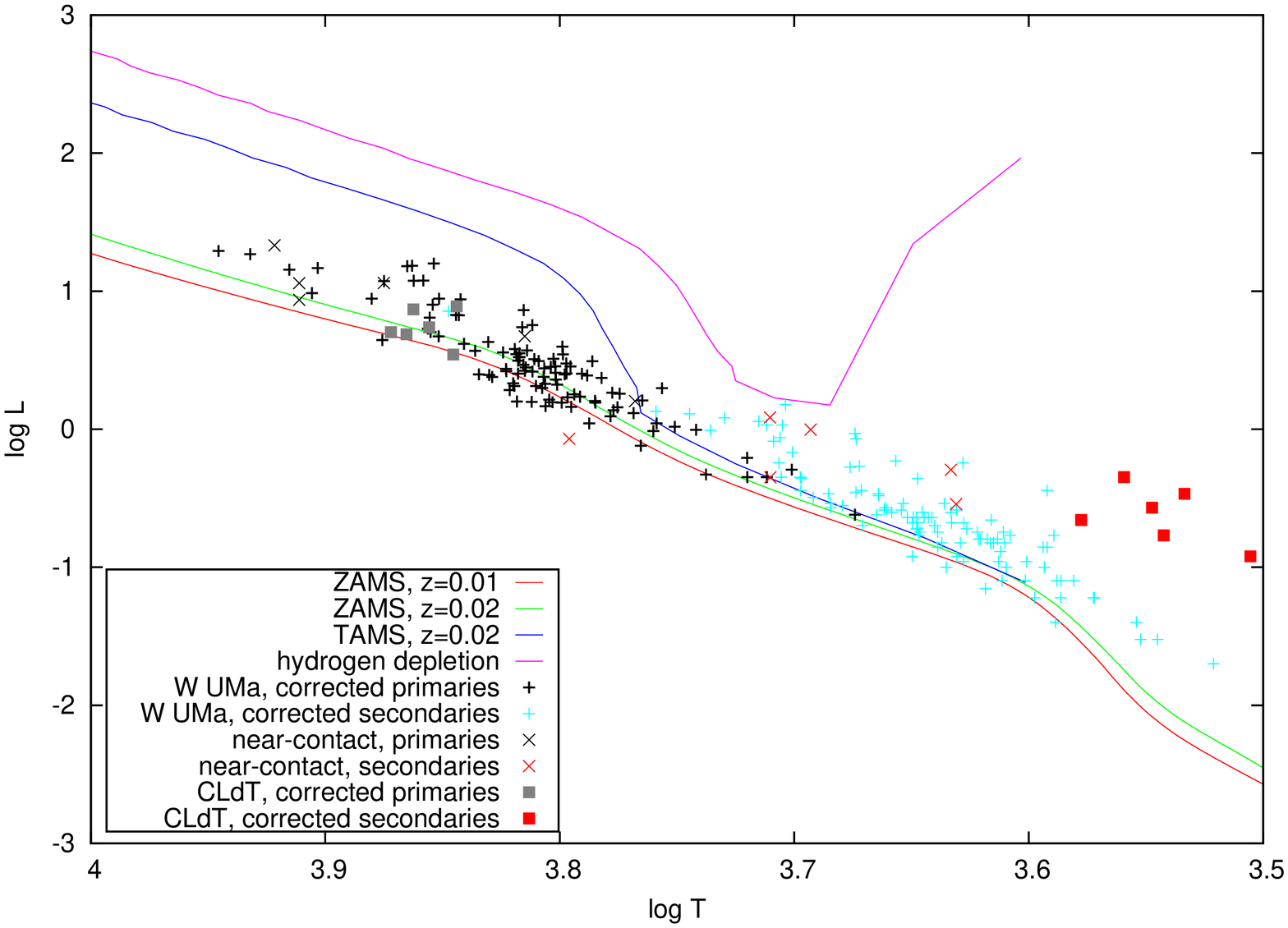}
\end{tabular}}
\vspace{0.5pc}
\FigCap{T--L diagrams for systems
analyzed in this work, without (left panel) 
and with (right panel) provision for 
effects of energy transfer.
Parameters of the W~UMa-type stars 
in the sample from Zola~et~al.~(2006)
are also shown.  The second plot
indicates that secondary components of
CLdT systems lie in the region which
secondary components of W~UMa-type
systems tend to avoid.}
\end{figure}

\section{Discussion and conclusions}

We performed light- and RV-curve 
modelling of twelve close binary
stars, for which contact configurations
with large temperature differences
between components have been reported
at least once in the literature.
In our analysis we applied new
spectroscopic mass ratios, obtained
through the BF
method, determined for the first time
for all systems but CN~And.

We obtained a near-contact
configuration for V1010~Oph, WZ~Cyg,
VV~Cet, DO~Cas, FS~Lup, and V747~Cen.
The primary components of V1010~Oph,
WZ~Cyg, VV~Cet, FS~Lup, and V747~Cen
fill (or nearly fill) their Roche lobes, 
and accordingly their secondaries are
considerably oversized; an ongoing or
past mass transfer can explain their
larger radii according to the mechanism
proposed by Webbink (1976) and Sarna \&
Fedorova (1989).  In this configuration,
an ongoing mass transfer should cause
period shortening, but this
is only confirmed in
the O--C diagram of V1010~Oph (Kreiner
et al., 2001).

Among systems from this group, the
most interesting results were obtained
for V747~Cen and FS~Lup.  The latter
appears to be the second-known
short-period period (0.38~d)
semi-detached system, which next to
V361~Lyr (Hilditch et al., 1997),
is captured in a broken-contact stage
of oscillation, as predicted by the
TRO theory (see Section~1), or during
the phase of a first-time approach to
a contact phase.  In the high-quality
BFs  (i.e., stellar images
in velocity space) of V747~Cen, 
we discovered bright regions on the 
secondary component. The discovery 
is a sign of 
one or more hot
spots, most probably a result of
accretion occuring between the stars.
This picture is in full accordance
with the result obtained from light-
and RV-curve modelling,
which resulted in a semi-detached
configuration for this system.

From the light- and RV-curve 
modelling of CX~Vir, FT~Lup,
BV~Eri, FO~Hya, CN~And, and BX~And,
we obtained contact models with
large temperature differences between
components ($\Delta~T>1000~\mathrm{K}$), 
in contradiction to 
theoretical predictions
for contact systems. 
However, as in the case of V747~Cen, the most
interesting results for CX~Vir,
FT~Lup, BV~Eri, FO~Hya, and CN~And
were obtained from discovery of hot
spots in BFs of their
secondary components.  As the existence
of the hot spots can be explained
only by accretion effects occuring
between components, that discovery
strongly supports semi-detached rather
than contact configurations for these
systems.

The just-stated direct result aside, 
there additionally are 
indirect indications of the possibility 
that (at least some) contact models with
large $\Delta~T$, obtained from
light- and RV-curve modelling, are
wrong. 

(1)~From light-curve modelling of
four systems with the strongest
bright regions, we also obtained
significant values of third light
(Sections 5.1--5.4, and Table~7).
The hypothetical third component is
neither visible in BFs (as a slowly
rotating sharp-line star) nor as
a nearby field star in DSS images
of these systems.  However, without
adding a third light into the model as
adjusted parameter, in all cases flat
secondary minima and depth of primary
minima were not reproduced well by
these models, and as a consequence the 
obtained $\chi^2_\mathrm{red,weigh}$ were at least
a few dozens of percent higher.  It is
also worth noting that the bright
regions are common features not only
for CX~Vir, FT~Lup, BV~Eri, FO~Hya
and CN~And, but also for V747~Cen, 
for which light- and RV-curve 
modelling gave a semi-detached
configuration without any third light,
in full consistency with all the
information inferred from its BFs.
On this empirical basis, one
may deduce that all the systems are in
fact semi-detached, with the duality of
configurations obtained within a pure
Roche-lobe model resulting from accretion
processes taking place on different
levels in the various individual systems.

(2)~The BFs of secondary components
obviously appear different in
the two quadratures. Whereas in    
first quadrature they do possess a   
characteristic smoothly increasing
intensity toward primary components
(this can be explained as a result of
star-stream interaction producing a hot
spot and causing considerable velocity
turbulence), the well defined bright
regions in second quadratures
can be produced by returning streams
encircling secondary components.
If this is a correct picture, then
these systems must
be semi-detached. 
The positive (in
absolute sense) departures of RVs of
secondary components from the Roche model,
observed in the case of V1010~Oph,
DO~Cas (Fig.~1, right panel) and BX~And
(Fig.~7, right panel), in phases close
to both minima, may be a sign of such
high-velocity streams.  Unfortunately,
other systems were observed only during
phases close to quadratures.

(3)~According to the O--C diagrams
(Kreiner~et~al.,~2001), continuous
period shortening of FT~Lup and CN~And
argues for a semi-detached configuration
allowing for mass transfer from
primary to secondary components. 
This empirical fact is consistent
with information extracted from BFs of
these systems, but it is inconsistent
with configurations obtained from 
light-curve modelling.  CX~Vir, BV~Eri,
FO~Hya, and BX~And do not reveal any
obvious signs of a permanent period
decrease or increase. The latter is
expected to be easily observable in O--C
diagrams of contact systems with large
temperature differences, as the mass
transfer rate from secondary to primary
star in common envelope is predicted
to be of order $10^{-8} \mathrm{M}_{\odot}/\mathrm{yr}$
(Flannery (1976), Robertson \& Eggleton
(1977)).  The lack of permanent period
increase in this group of systems
is again in contradiction
to the supposition of contact
configurations.

(4)~The position  on the T--L
diagram of secondary
components of contact systems with
large $\Delta~T$,
constructed from their physical
parameters as corrected for effects of
energy transfer in a common envelope
(Fig.~15, right panel) indicates 
that these stars differ significantly from
the components of typical W~UMa-type
systems.  If their input physical
parameters (Table~8) obtained within
a false contact model from the WD code
are wrong (due to the problems with
unexplained sources of $l_3$), the
corrections for energy transfer may
in turn lead to very different output
values, as observed in Figure~15
(right panel).
 
(5)~The discovery
of bright regions in
BFs of secondary
components, the 
observation of significant
differences in BF profiles
in the two quadratures, and
the discovery of spotted primaries
in (at least)  
CX~Vir and FT~Lup (as described in
Sections~5.1--5.5) suggests that
it may not be possible to derive 
accurate configurations from light-curve
modelling.  This is because smoothly
varying EW- and EB-type light curves
contain only information about changes
of integrated binary-star light with
phase, offering constraints too weak 
to yield a model consistent
with the whole range
of information extracted
from spectroscopy.
The modelling software based on the
Roche model cannot properly handle
all these effects, because
there remains a large set
of free parameters: our
attempts to obtain much more reliable
models of CX~Vir, FT~Lup, and CN~And
did not deliver unique fits.

Although all the above findings 
strongly support the
hypothesis of Rucinski (1986) and
Eggleton (1996), that systems for
which Roche-lobe-based computer
models give contact configurations
with $\Delta~T > 1000~\mathrm{K}$ are in fact
semi-detached (see Section~1), further
observations are necessary
for full confirmation of  
this finding.  
Doppler-imaging observations, as made
by Pribulla \& Rucinski (2008) for
AW~UMa, may be needed.  
The cited observations have revealed 
that the secondary component 
has the aspect of a  
small core-like star surrounded
by optically thick matter. Further, the
recent discovery of a double-peaked
$H_{\alpha}$ line in a short-period (1.6~d) 
SV~Cen (Siwak~et~al., 2009) and 
(0.42~d) TYC~2675-663-1 W~UMa type
systems (Caballero-Garcia~et~al.,
2010) may indicate an existence of
accretion disk around one component,
and a semi-detached rather than a contact
configuration. This is a possibility
previously envisaged by
Zola (1995).  The echelle spectra
of CX~Vir and FT~Lup, covering a wide
spectral range (UV--IR), do not reveal
emission components in hydrogen lines.
Further progress on 
contact-binary systems with large $\Delta~T$
between components may consequently be obtained
from modelling of the available and
the prospective new BFs,
obtained over a whole orbital period.

\Acknow{ The service-mode observations 
made at the ESO La~Silla 
have been financed 
by the Optical Infrared Coordination Network 
(OPTICON), a major international 
collaboration supported by the Research 
Infrastructures Programme 
of the European Commission's Sixth Framework 
Programme.

The observations at the David Dunlap
Observatory and at the South African
Astronomical Observatory have been
financed by research funds of the
Astronomical Observatory of the
Jagiellonian University, the Polish
Foundation for Astronomy, and the Polish
Foundation of SALT.
The observations at Las Campanas
have been financed by
a Canadian Space Agency grant
to Prof.~Slavek M.~Rucinski within
the CSA Space Science
Enhancement Program.

MS and DKW appreciate the help and the hospitality of SAAO 
staff during observations, especially on
the part of  Mr.~Fred Marang, 
Mr.~Francois van Wyk, and Ms.~Melony Spark. 

MS especially acknowledges the following: \newline
- Prof.~Slavek M.~Rucinski, for
access to DDO observing facilities,
for many discussions concerning 
the BF method and modern results in
contact-binaries work, 
and for stimulating discussion
of this paper;\newline
- Prof.~Jerzy Kreiner and 
the Mount Suhora Astronomical Observatory staff, 
for frequent access to observing facilities;\newline
- Dr.~Maria Kurpi{\'n}ska-Winiarska,
Dr.~Maciej Winiarski, Dr. Wac{\l}aw
Waniak, and Dr. Andrzej Baran, for
comments and advice concerning
details of stellar photometry and
spectroscopy;\newline
- the ESO-La Silla technical staff and astronomers,
for executing observing blocks  
during service-mode observations, 
and especially for taking excellent data;\newline  
- Ms.~Heide de~Bond, Mr.~Jim Thompson, and Dr.~Toomas Karmo,
for their
night assistance during observations at DDO;\newline
- the Las Campanas Observatory technical staff, for day work, 
and Herman Olivares and Javier Huentes, 
for night assistance;\newline
- the Canadian Space Agency, for a
post-doctoral grant to Prof.~Slavek M.~Rucinski 
within the CSA Space Science Enhancement Program;\newline
- Dr.~Toomas Karmo and Mr. Bryce Croll for language corrections.
\newline
This paper reports the most important results obtained during the 
preparation of the MS PhD thesis, with SZ as thesis advisor.}


\begin{references}
%
%
%
%
\refitem{Badee, D., Duerbeck, H.W., Karimie, M.T., Yamasaki, A.}{1983}{Ap\&SS}{93}{69}
\refitem{Balona, L.A.}{2000}{XLUCY: Photometer control and data acquisition}{~}{SAAO}
\refitem{Baran A., Zola S., Rucinski S.M., Kreiner J.M., Siwak M., Drozdz M.}{2004}{AcA}{54}{195}
\refitem{Barone, F., Covino, E., Di Fiore, L., Milano, L., Russo, G.}{1991}{Ap\&SS}{183}{117}
\refitem{Barone, F., Di Fiore, L., Milano, L., Pirozzi, L., Russo, G.}{1992}{Ap\&SS}{198}{321}
\refitem{Bell, S.A., Rainiger, P.P., Hill, G., Hilditch, R.W.}{1990}{MNRAS}{244}{328}
\refitem{Caballero-Garc{\'i}a, M.D., Torres, G., Ribas, I., R{\'i}squez, D., Montesinos, B., Mas-Hesse, J.M., 
Domingo, A.}{2010}{A\&A}{514}{36}
\refitem{Candy, M.P., Candy, B.N.}{1997}{MNRAS}{286}{229}
\refitem{Chambliss, C.R.}{1970}{AJ}{75}{731}
\refitem{Claret, A., Diaz-Cordoves, J., Gimenez, A.}{1995}{A\&AS}{114}{247}
\refitem{Corcoran, M.F., Siah, M.J., Guinan, E.F.}{1991}{AJ}{101}{1828}
\refitem{Diaz-Cordoves J., Claret A., Gimenez A.}{1995}{A\&AS}{110}{329}
\refitem{Eggen, O.J.}{1967}{MmRAS}{70}{111}
\refitem{Eggleton, P.P.}{1996}{ASPC}{90}{257}
\refitem{Flannery, B.P.}{1976}{ApJ}{205}{217}
\refitem{Gleim, J.K., Winkler, L.}{1969}{AJ}{74}{1191}
\refitem{Gu, S.}{1999}{A\&A}{346}{437}
\refitem{Gray, R.O.}{2001}{http://phys.appstate.edu/spectrum/spectrum.html}{~}{Department of Physics and Astronomy, Appalachian State University}
{~}
\refitem{Harmanec, P.}{1988}{BAICz}{39}{329}
\refitem{Hilditch, R.W., King, D.J., Hill, G., Poeckert, R.}{1984}{MNRAS}{208}{135}
\refitem{Hilditch, R.W., King, D.J.}{1988}{MNRAS}{231}{397}
\refitem{Hilditch, R.W., Collier Cameron, A., et al.}{1997}{MNRAS}{291}{749}
\refitem{Jassur, D.M.Z., Khodadadi, A.}{2006}{JApA}{27}{47}
\refitem{Ka{\l}u{\.z}ny, J.}{1983}{AcA}{33}{345}
\refitem{Ka{\l}u{\.z}ny, J.}{1985}{AcA}{35}{327}
\refitem{Ka{\l}u{\.z}ny, J.}{1986}{AcA}{36}{113}
\refitem{Karimie, M.T., Duerbeck, H.W.}{1985}{Ap\&SS}{117}{375}
\refitem{K{\"a}hler, H.}{2004}{A\&A}{414}{885}
\refitem{Kreiner, J.M., Kim, Ch.-H., Nha, I.-S.}{2001}{An Atlas of O-C Diagrams of Eclipsing Binary Stars}{~}{Wydawnictwo Naukowe AP, Krakow}
\refitem{Kreiner, J.M., Rucinski, S.M., Zola, S., Niarchos, P., et al.}{2003}{A\&A}{412}{465}
\refitem{Kreiner, J.M.}{2004}{AcA}{54}{207}
\refitem{Kuiper, G.}{1941}{ApJ}{93}{133}
\refitem{Kurucz, R.}{1993}{Atomic data for opacity calculations. Kurucz CD-ROM No. 1. - 18.}{~}{Cambridge Mass. Smithsonian Astrophysical Observatory}
\refitem{Leung, K.-Ch.}{1974}{AJ}{79}{852}
\refitem{Leung, K.-Ch., Wilson, R.E.}{1977}{AJ}{211}{853}
\refitem{Li, L., Han, Z., Zhang, F.}{2004a}{MNRAS}{351}{137}
\refitem{Li, L., Han, Z., Zhang, F.}{2004b}{MNRAS}{355}{1383}
\refitem{Li, L., Han, Z., Zhang, F.}{2005}{MNRAS}{360}{272}
\refitem{Lipari, S.L., Sistero, R.F.}{1986}{MNRAS}{220}{883}
\refitem{Lucy, L.B.}{1967}{Zeitschrift f{\"u}r Astrophysik}{65}{89}
\refitem{Lucy, L.B.}{1968}{AJ}{151}{1123}
\refitem{Lucy, L.B.}{1973}{Ap\&SS}{22}{381}
\refitem{Lucy, L.B.}{1976}{ApJ}{205}{208}
\refitem{Lucy, L.B., Wilson, R.E.}{1979}{ApJ}{231}{502}
\refitem{Mauder, H., Kappelmann, N.}{1982}{MitAG}{55}{72}
\refitem{Milano, L., Russo, G., Terzan, A.}{1987}{A\&A}{183}{265}
\refitem{Mochnacki, S.W.}{1981}{ApJ}{245}{650}
\refitem{Niarchos, P.G., Manimanis, V.N.}{2003}{A\&A}{405}{263}
\refitem{Oh, K.-D., Ahn, Y.-S.}{1992}{Ap\&SS}{187}{261}
\refitem{Paxton, B.}{2005}{airXiv0405130}{~}{~}
\refitem{Pols, O., Tout, Ch.A., Eggleton, P.P., Han, Z.}{1995}{MNRAS}{274}{964}
\refitem{Pribulla, T., Rucinski, S.M.}{2008}{MNRAS}{386}{377}
\refitem{Pribulla, T., Rucinski, S.M., Blake R. M., et al.}{2009}{AJ}{137}{3655}
\refitem{Rahman, A.}{2000}{PASP}{112}{123}
\refitem{Rafert, J.B., Markworth, N.L., Michaels, E.J.}{1985}{PASP}{97}{310}
\refitem{Robertson, J.A., Eggleton, P.P.}{1977}{MNRAS}{179}{359}
\refitem{Rovithis, P., Rovithis-Livaniou, H., Suran, M.D., et al.}{1999}{A\&A}{348}{184}
\refitem{Rucinski, S.M.}{1969}{AcA}{19}{245}
\refitem{Rucinski, S.M.}{1986}{IAUS}{118}{159}
\refitem{Rucinski, S.M.}{1992}{AJ}{104}{1968}
\refitem{Rucinski, S.M.}{1999}{IAU Coll.170, ASPC}{185}{82}
\refitem{Rucinski, S.M., Lu, W.-X., Mochnacki, S.W.}{2000}{AJ}{120}{1133}
\refitem{Rucinski, S.M.}{2002}{AJ}{124}{1746}
\refitem{Samec, R.G., Fuller, R.E., Kaitchuck, R.H.}{1989}{AJ}{97}{1159}
\refitem{Sarna, M.J., Fedorova, A.V.}{1989}{A\&A}{208}{111}
\refitem{Siwak, M., Zola, S., Rucinski S.}{2009}{airXiv0909.5415}{~}{"Binaries - Key to Comprehension of the Universe" Brno, Czech Rep. 8-12 June 2009. To appear in Dec.2010 as ASP-CS435, eds. A. Prsa and M. Zejda}
\refitem{Stefanik, R.P., Latham, D.W., Torres, G.}{1999}{ASPC}{185}{354}
\refitem{Stetson, P.B.}{1987}{PASP}{99}{191}
\refitem{Yakut, K., Eggleton, P.P.}{2005}{ApJ}{629}{1055}
\refitem{Van Hamme, W., Samec, R.G., Gothard, N.W., et al.}{2001}{AJ}{122}{3436}
\refitem{von Ziepel, H.}{1924}{MNRAS}{84}{665}
\refitem{Webbink, R.F.}{1976}{ApJS}{32}{583}
\refitem{Wilson, R.E.}{1996}{Documentation of Eclipsing Binary Computer Model}{~}{~}
\refitem{Zhu, L., Qian, S.}{2006}{MNRAS}{367}{423}
\refitem{Zola, S.}{1995}{A\&A}{294}{525}
\refitem{Zola, S., Kreiner, J.M., Zakrzewski, B., et al.}{2005}{AcA}{55}{389}
\refitem{Zola, S., Gazeas, K., Kreiner J.M., Zakrzewski, B.}{2006}{Ap\&SS}{304}{109}
\refitem{Zola, S., Gazeas, K., Kreiner, J.M., Og{\l}oza W., et al.}{2010}{MNRAS}{408}{464}
\end{references}
\end{document}